\documentclass[12pt,a4paper]{article}   %

\usepackage[latin1]{inputenc}
\usepackage{color}
\usepackage{pstricks}
\usepackage{multicol}   %
\usepackage{anysize}
\usepackage{fancyhdr}
\usepackage{wasysym}
\usepackage{lettrine}
\usepackage{type1cm}

\usepackage{setspace}

\usepackage{graphicx}
\usepackage[small,bf,centerlast]{caption}
\usepackage[sorting=anyt,citestyle=authoryear-comp,maxcitenames=1,bibstyle=authoryear]{biblatex}
\bibliography{Bibliography.bib}  %

\marginsize{1.5cm}{1.5cm}{0.5cm}{2cm}


\definecolor{blue4}{rgb}{0,0,0.5}
\definecolor{indianred3}{rgb}{0.8,0.33,0.33}
\definecolor{gainsboro}{rgb}{0.86,0.86,0.86}
\definecolor{dodgerblue4}{rgb}{0.06,0.31,0.55}



\usepackage{authblk}  %

\title{Frustration in Biomolecules}
\author[1]{Diego U. Ferreiro}
\author[2]{Elizabeth A. Komives}
\author[3]{Peter G. Wolynes}  %

\affil[1]{Protein Physiology Lab, Dep de Qu\'imica Biol\'ogica, Facultad de Ciencias Exactas y Naturales, UBA-CONICET-IQUIBICEN, Buenos Aires, Argentina.}
\affil[2]{Department of Chemistry and Biochemistry, University of California San Diego , La Jolla, California 92093, USA}
\affil[3]{Center for Theoretical Biological Physics and Department of Chemistry, Rice University, Houston, Texas 77005, USA}

\date{}    %

\begin{document}

\maketitle

{\rule[0mm]{170mm}{0.25mm}}

\begin{abstract}
Biomolecules are the prime information processing elements of living matter. Most of these inanimate systems are polymers that compute their own structures and dynamics using as input seemingly random character strings of their sequence, following which they coalesce and perform integrated cellular functions. In large computational systems with a finite interaction-codes, the appearance of conflicting goals is inevitable. Simple conflicting forces can lead to quite complex structures and behaviors, leading to the concept of {\it frustration} in condensed matter. We present here some basic ideas about frustration in biomolecules and how the frustration concept leads to a better appreciation of many aspects of the architecture of biomolecules, and how biomolecular structure connects to function. These ideas are simultaneously both seductively simple and perilously subtle to grasp completely. The energy landscape theory of protein folding provides a framework for quantifying frustration in large systems and has been implemented at many levels of description. We first review the notion of frustration from the areas of abstract logic and its uses in simple condensed matter systems. We discuss then how the frustration concept applies specifically to heteropolymers, testing folding landscape theory in computer simulations of protein models and in experimentally accessible systems. Studying the aspects of frustration averaged over many proteins provides ways to infer energy functions useful for reliable structure prediction. We discuss how frustration affects folding mechanisms. We review here how a large part of the biological functions of proteins are related to subtle local physical frustration effects and how frustration influences the appearance of metastable states, the nature of binding processes, catalysis and allosteric transitions. We hope to illustrate how {\it Frustration} is a fundamental concept in relating function to structural biology.

\end{abstract}

\tableofcontents

\newpage
\section{Perspectives on Frustration}

Life is based on molecular information processing. Thousands of nucleic acids and proteins are organized in cells, cooperate and reproduce themselves as a group and respond to their environment. Most of the biomolecules individually carry out the complex task of decoding their own one dimensional sequences in order to find three dimensional structures (\cite{pmid4124164})  furthermore yielding four dimensional dynamical patterns that permit each molecule to carry out its ``functions'' (\cite{pmid1749933}).  This decoding task, the first step of which is called folding, is a challenging information processing step that only recently has been successfully carried out for some small proteins by computers (\cite{pmid22822217}). This task, even with computer aid, severely challenges the most clever human gamers (\cite{pmid20686574}). Contemplating why this task is so hard both for computers and for people but is apparently not hard for the molecules themselves leads us to consider the concept of {\it Frustration} in biomolecules, the subject of this review.

\begin{figure}
\centering
	\includegraphics[width=0.7\textwidth]{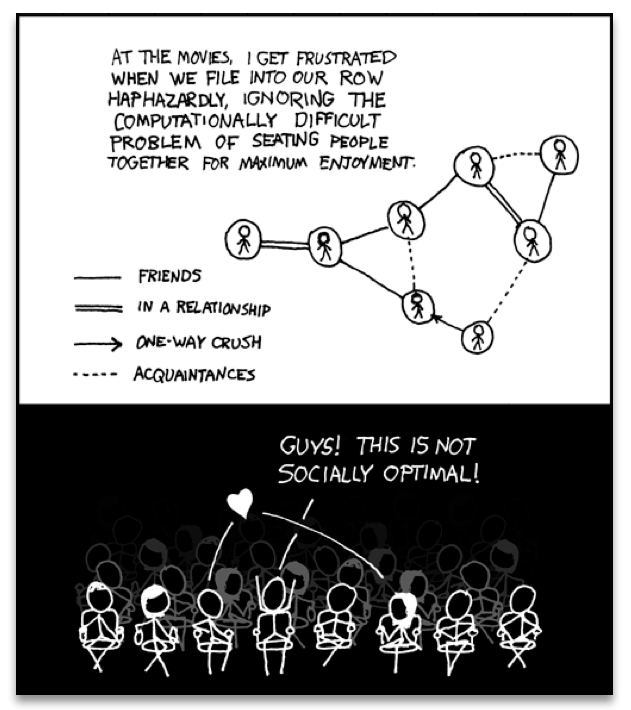}
	\caption{ Frustration is a general feeling and a deep concept. This cartoon presents part of the problem: a group of people that cannot be seated optimally in a single row. The upper panel shows a social graph that takes into account the interactions among people. The lower panel shows the irritation of the character that tries to calculate the arrangement for he and his seven friends, finding out that there will always be unsatisfied agents. Reproduced with permission from http://xkcd.com/173 }	
	\label{fig:xkcd}
\end{figure}

In human psychology frustration is an emotion ``the experience of nonfulfillment of some wish or needs'' (\cite{frustrawiki}). The emotion of frustration according to psychologists, often arises because a person has conflicting goals that interfere with one another (see Fig. \ref{fig:xkcd}). It is this type of internal frustration that can plague information processing rather generally even outside the human context. This sort of frustration is important in formal logic and in computer science and is closely related to the notion of frustration as it is used to describe physical systems, such as disordered magnets (\cite{neelnobel}, \cite{Vannimenus1970}, \cite{Wannier1950}, \cite{nordblad2013competing}, \cite{SchifferRamirez96}) or biomolecules (\cite{pmid3478708}). For these inanimate systems, the ``goals'' are set by evolution and are reflected in the systems' own laws of development and motion given the genetic information. These laws are usually encoded in an energy function or ``hamiltonian'', which in turn determines the kinetics and thermodynamics at the macroscopic level.

\medskip
\medskip

Frustration slows the performance of computer algorithms and indeed can make some tasks impossible for a computer to carry out in a timely fashion. For physical systems, frustration leads to complex dynamics that often spans many orders of magnitude in time scale. Some disordered magnets, called spin glasses, can never seem to settle into any equilibrium state, no matter how long one waits. This behavior resembles the way some liquids do not crystallize into their lowest energy structure when cooled below their freezing point but instead become trapped in a glassy state. Proteins have some of the characteristics of glasses at low temperatures (\cite{pmid1749933}, \cite{pmid3478708}) but at physiological temperatures the functionally relevant motions are generally fast enough to reliably take place in a cell (\cite{frauenfeldernyas87}). Glassy dynamics of a condensed matter system or of a biomolecule resembles a computer code still in the process of crunching the problem - generally when a system is glassy the actual system's state depends on its history and kinetics. While kinetic control of pathways is paramount at the cellular and network level in biology, the physics of individual biomolecules is largely controlled thermodynamically and is independent of the history of the macromolecular system. We shall see the reason for this is that Nature has evolved biomolecules where the frustration of protein dynamics is minimal. Nevertheless some frustration is present in evolved biomolecules and is crucial for many aspects of their function. 

The purpose of this article is to explain the basic ideas about frustration in biomolecules and how these ideas let us understand many aspects of the architecture of biomolecules and how that architecture connects with their motions. These ideas are simultaneously both seductively simple and perilously subtle to understand completely. To do so without a mathematical framework is impossible. Such a framework is provided by the energy landscape theory of protein folding which indeed gives criteria for quantifying frustration globally in large systems (\cite{pmid15664893}, \cite{goldstein1992optimal}, \cite{pmid3478708}). Formally such criteria would be rigorous for infinite sized proteins, so that they can only be applied using approximations to finite systems. The global criterion for quantifying frustration can be implemented at several levels of mathematical sophistication and has led to concrete and useful protein structure prediction algorithms (\cite{pmid22545654}). A complete mathematical description is quite complicated and this review will only sketch some of the basics (\cite{Plotkin:2002ve}, \cite{pmid7784423}).  But this summary will allow us to concentrate on reviewing how a local notion of frustration can also be quantified (\cite{Ferreiro:2007bh}) and also how localizing frustration can be used to look at protein structures and motion both in a survey sense and to understand specific examples. The local notion of frustration is not rigorous and the approximate form we use could be severely criticized by statistical mechanics purists. Nevertheless we have found in several investigations, that a very simple estimate of local frustration can give surprising insights into how proteins fold and misfold, how proteins move once they are folded and how they may have evolved to carry out their functions.

The structure of this review is as follows. First, we develop the notion of frustration seeing how it enters the area of abstract logic and how it is used for the simplest condensed matter systems which are magnets. We then discuss how the frustration concept applies specifically to heteropolymers and discuss the role of frustration in the energy landscape theory of protein folding. We further review tests of folding landscape theory in the stylized computer simulation world provided by lattice models of proteins. We then briefly review how understanding frustration globally for real proteins provides a ``license to do bioinformatics'', enabling us to extract physical information about molecular forces from structural databases. Studying frustration thus provides practical tools for protein structure prediction and leads to an appreciation of the complexity of the protein folding code. Some illustrations of how frustration occurs in evolved natural proteins and in artificially designed proteins are then discussed. Following this we review how residual frustration in some biomolecules affects their folding kinetics focusing on some anecdotal examples. {\it En passant} we discuss broader uses of the frustration concept in folding, describing topological frustration, chemical frustration as it occurs in metalloproteins and the role of symmetry, which surprisingly can frustrate the folding process. The larger part of the review then goes beyond the study of folding to seeing how the notion of frustration help understanding biomolecular function by examining the role of frustration in metastable protein states, binding processes, enzymatic catalysis and allosteric transitions. We also discuss frustration in protein aggregation and in protein-DNA assemblies.

\subsection{Frustration, Logic and Magnets}

To get a feeling for frustration it is fun to start with some old, seemingly simple, logic puzzles. Medieval philosophers developed abstract logic using such puzzles anticipating many modern ideas in metamathematics and computational complexity. They are known as ``sophismata" (\cite{heytesbury1494regule}, \cite{kretzmann1982syncategoremata}). Consider these two sentences:

	1. All men are donkeys or men and donkeys are donkeys.
	
	2. The sentence ``Socrates says something false'' in the case where Socrates says nothing other than ``Socrates says something false''.
	
Are these statements true or false? How do you feel when you try to figure this out? Frustrated? Figuring these puzzles out in the original Latin in which they were written is even more challenging than it is in English translation. Modern students of logic can do better than most students did centuries ago by systematically using the idea that one must first disambiguate the sentences (with appropriate punctuation in English, like parentheses and commas, not used in classical Latin) and then judge the truth or falsity of the elemental parts e.g., we are all pretty sure that the proposition ``All men are donkeys'' is false etc. After generating a ``truth table'' of possibilities a ``simple'' search of all the entries tells us the answer. Using more cumbersome argumentation, Albert of Saxony concluded that two different renditions of the truth or falsity of sentence 1 are possible. Sentence two is one of the forms of the ``liar paradox'' with which even twentieth century philosophers like Whitehead and Russell continued to grapple. In modern computer science, much longer strings of logical statements are needed to answer practical problems such as how to detect objects in a scene that is digitally encoded (\cite{Field01091999}). Are there methods that allow a computer to decide the truth or falsity of very complex, long statements? Can a computer find consistent solutions of any set of abstract assertions? Such questions are known as constraint satisfaction problems (\cite{mezard2009constraint}, \cite{monasson1999determining}). In general a huge number of possibilities must be tested to see if the logical constraints in the sophism can be satisfied in some way and the computations can become quite lengthy.

That solving a particular problem must lead to frustration is easiest to see when the choices for the subparts of the problem are few -the binary choice of whether an individual proposition is true or false being the simplest. Binary choice is the ``atom" of logic and computer science. The consequences of conflicts for problem solving are most profound, however, when many such multiple choices must be made to specify the solution. This is because the search space through the truth table becomes exponentially large with the number of choices needed to specify the answer. We quickly see the liar paradox has no solution but for people, the task of deciding whether the longer strings of logical statements in typical computer programs actually lead to solutions or whether they do not is much, much harder. As a practical matter we usually just run the programs and check whether they stop or whether they yield a unique solution after trying out all possible inputs.

In physics, frustration entered the lexicon in the study of magnets (\cite{Vannimenus1970}, \cite{anderson1950}). Like the logical puzzles we have just discussed, the elementary units in magnets, the spins, have a limited number of choices - classically they can ``choose'' to be up or they can ``choose'' to be down. Actually in some magnets the quantum mechanical aspect of the spin can give the spins the opportunity not to choose at all but to remain in the limbo of a quantum superposition of up and down, like a resonance hybrid in organic chemistry, but this complication corresponding to non-Aristotelian logic can often be ignored. The analog of the logical-grammatical constructions for the spins in a magnet are the energies of interactions between the spins. The simplest energy describing the interactions between two spins, one on the $i$th atom, $S_i$, and on another, usually neighboring atom $j$th, $S_j$, has the form $E_{ij}= -J_{ij} S_i S_j $  where the two choices for $S_i$ and $S_j$ are +1 i.e. up or down. If $J_{ij}$ is positive, the low energy state aligns the two spins in the same direction; if $J_{ij}$ is negative the spins prefer to point in opposite directions.

\begin{figure}
\centering
	\includegraphics[width=0.9\textwidth]{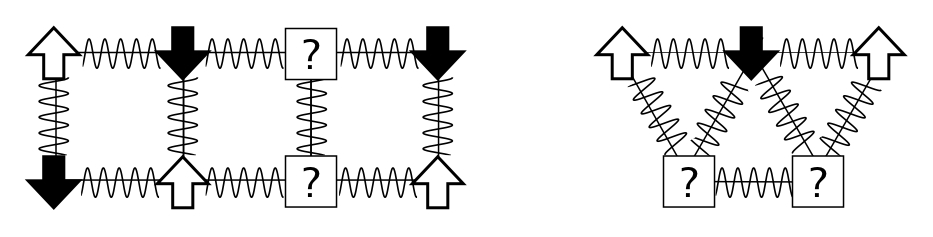}
	\caption{ Frustration entered the physics lexicon in the study of magnets. The arrows represent spins that can be in any two states: {\it up} or {\it down}. Favorable antiferromagnetic interactions between spins are represented by squiggly lines. The potentially frustrating problem is: what is the lowest energy arrangement of the up and down spins that can be reached dynamically? On the left, the particles are arranged on a rectangular lattice. How would you set the spins such that all local interactions are satisfied? What happens if the particles are arranged on a triangular lattice, as shown on the right? There is no way to arrange the spins so that every component interaction can be minimized. The triangle is ``frustrated'' }	
	\label{fig:lattice}
\end{figure}

If we have a large magnet the energy can be taken as the sum of the individual (pair) interactions $E=-\sum_{i\neq j} J_{ij} S_i S_j $ .  The potentially frustrating problem to solve then is, at low temperature, what is the lowest energy arrangement of the up and down spins that can be reached dynamically? (See Fig \ref{fig:lattice}) This problem is ``solved'' by the system when it is cooled and random thermal motions allows it to try out various configurations or arrangements of the spins, according to their energy.

\medskip
\medskip

If all the $J_{ij}$ are positive it is easy even for people to see the answer: the lowest energy state has all the spins pointing in the same direction -they are either all up or all down. What if all (or even some) of the  $J_{ij}$'s are negative? Now things seem to be tricky, but sometimes people can figure out the answer pretty quickly for certain situations. Consider first what turns out to be an easy problem: suppose the atoms bearing the spins are situated on a simple square lattice and all the $J_{ij}$ are negative. In this case an interlaced checkerboard of up and down spins has the lowest energy and minimizes simultaneously every local interaction $E_{ij}$. The possible conflict between different local interactions can be resolved (See Fig \ref{fig:lattice}). The resulting pattern is called an anti-ferromagnet. This solution while it can be quickly checked by anyone, was not really that easy to see {\it a priori}, and indeed (along with other insights) figuring out this pattern netted its discoverer, Louis Neel, a Nobel Prize in Physics (\cite{neelnobel}). Although it takes at least a few moments for humans to see the answer for the square lattice antiferromagnet the system itself has a much easier time figuring out what to do. When an antiferromagnet is cooled below a certain temperature it orders spontaneously and fairly quickly.

But antiferromagnets on other lattices can be more complex than the uniform square lattice antiferromagnet case. Imagine the spins reside on atoms situated on a triangular lattice with all antiferromagnetic interactions ( Fig \ref{fig:lattice} -b). Already a single triangle of the lattice shows the problem. There is no way to arrange the spins so that every component interaction can simultaneously be minimized; always at least one interaction must be still in a high energy state. We would say the individual triangle is ``frustrated''.

What happens for a big lattice of frustrated triangles?  In the strictly two-dimensional system one finds a huge degeneracy of possibilities. There is no unique ground state. This would lead to an apparent contradiction to the Third Law of Thermodynamics, which requires entropy to vanish at absolute zero, implying a unique ground state. Wannier showed that $T=0$ was a critical point for the triangular antiferromagnet and thus there would be large length scale fluctuations even at low temperature (\cite{wannier1950antiferromagnetism}). In real three dimensional systems, weaker interactions between planes of spins restore the 3rd Law and lead to ordering to a unique lowest energy state, but the energies required to excite the system from the ground state now depend only on the weak interactions and so are much smaller than you would otherwise expect. The kind of ordering found in such three dimensional systems is also very complex and generally hard to predict with precision (\cite{collins1997review}).

\medskip
	\begin{figure}
\centering
	\includegraphics[width=0.9\textwidth]{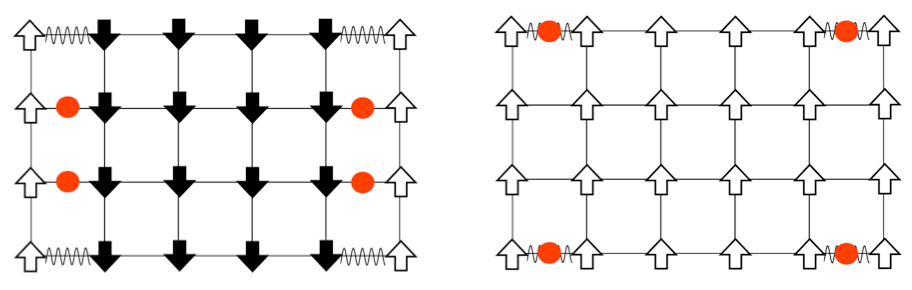}
	\caption{ Frustration leads to barriers in the energy landscape. The arrows represent spins that can be in any two states on a rectangular lattice. Favorable ferromagnetic interactions are represented by straight lines, antiferromagnetic interactions by squiggly lines. In this case most of the bonds are ferromagnetic so at low temperature most of the spins will be parallel. Two choices of spin assignments are shown. These have similar energies, are relatively stable, but they are globally dissimilar. At left most spins are down, at right most spins are up. In both cases the same number of interactions remain locally unsatisfied (red dots). There are two ways of arranging the ``misaligned'' spins. To rearrange from one to the other a very complicated set of spin changes must be made. This rearrangement is entropically disfavored at high temperature and energetically disfavored at low temperature. Many large scale moves must be made for the system to find out which arrangement or set of arrangements are most stable. Although there is a ground state it can be hard to find because of the very slow dynamics that emerges.}	
	\label{fig:square_red}
\end{figure}
\medskip

You might think the triangular antiferromagnet case is special because of the exact symmetry of the frustration and indeed it is. Without symmetry, frustration and degeneracy still go together but manifest themselves in a very different way: very slow dynamics emerges that gives the system difficulty reaching equilibrium at low temperatures. Although there is a ground state it can be hard to find. Such a system is called a ``spin glass''. The models that are simplest to describe place spins on a simple lattice and at random assign some bonds to be ferromagnetic and others to be antiferromagnetic (Fig \ref{fig:square_red}). A particular assignment of bonds is analogous to a sophism having a particular combination of logical connectors (AND, NOT, OR ... ). While there are numerous approaches for finding the low energy states of such a magnet, in the general case all known algorithms take a very long time. This time scales up exponentially with the size of the magnet. The reason for this slowness is there are many choices of spin assignments that have similar energies that are relatively stable but that are globally quite dissimilar. It is difficult for the system to carry out such rearrangements between degenerate states. How does frustration lead to barriers? In fact the locality of moves makes this hard to understand but you can get some intuition from a weakly frustrated spin system with only a few frustrated interactions like the example shown in Fig. \ref{fig:square_red}. Since most of the bonds in Fig \ref{fig:square_red}A are ferromagnetic, at low temperature most of the spins will be parallel, let's say oriented up. Yet some interactions must be unsatisfied locally. In the example shown there are two ways of arranging these ``misaligned'' spins. To rearrange from one to the other a very complicated set of spin changes must be made. This rearrangement is entropically disfavored at high temperature and energetically disfavored at low temperature. Many large scale moves must be made for the system to find out which arrangement or set of arrangements are most stable.

When frustration is widespread the system resembles a glass-the exact arrangement of spins found at any one time depends on the history of the system, just as a liquid cooled into a glass has properties that depend on its annealing schedule. This is the feature of frustration most noted in the condensed matter physics of magnets. The global effect of widespread frustration is that it leads to such completely new phases of matter. The illustrative example, however, focuses attention on a point of special relevance to biochemistry: even a little frustration only locally may lead to new degeneracies that could have dynamical consequences.

Nearly degenerate states will interconvert slowly not only because the frustrated parts of the system have to be moved about, but possibly also because other parts of the system that are not frustrated have to move in order to allow the rearrangement to occur. This entails breaking up strong interactions and is energetically costly and thus difficult to be achieved by random thermal motions.

\begin{figure}
\centering
	\includegraphics[width=0.7\textwidth]{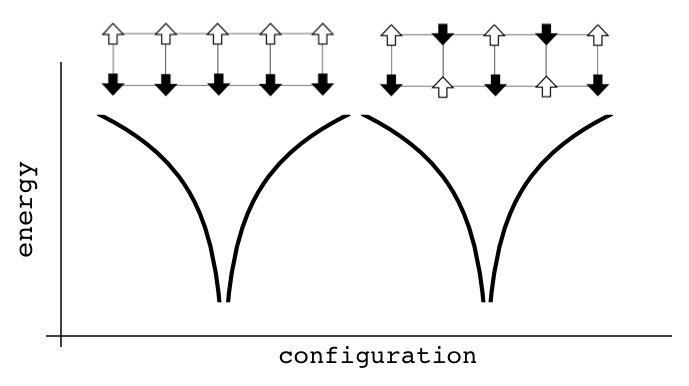}
	\caption{ Patterns can be encoded within magnets. The magnet is composed of different patterns of binary spins arranged on a rectangular lattice. Configurations are sorted by similarity to each other in a natural reaction coordinate measuring their overlap. The energy of each configuration is given as a function of the local interactions. In this case only two patterns are encoded, allowing two globally different ground states. Configurations close to any one of them are fairly low in energy. If there are not too many interferences between the patterns, the existence of one pattern only contributes a small random energy that doesn't destabilize the other energy minimum much. If the magnet starts in a configuration resembling one of the individual patterns, at low temperature the spins will rearrange and recollect the complete encoded pattern. These patterns remain retrievable even if many bonds of the lattice (but not too many!) become modified. The memory patterns are thus stored globally, not just locally. There is a finite capacity in the maximum number of storable patterns, as too many of them make the system behave like a random spin glass and robust recollection of the associated input memories becomes impossible.}	
	\label{fig:enespins}
\end{figure}

While completely random magnets and weakly, locally frustrated magnets are not very familiar real-world systems the binary choices of spins in magnets have made magnetism a convenient playground to understand degeneracies in many other nonmagnetic systems. While spin glasses have huge numbers of near-degenerate states, ferromagnets or the simple antiferromagnets have only a few low energy patterns. We can imagine magnets that exhibit intermediate levels of degeneracy. The most conceptually important of such mildly complex landscapes are models that encode a modest number of spin patterns. Such magnets would reliably order to a specific input pattern if only a fraction of the spins were initially organized to resemble the template. Such magnetic models thus represent a very simple model of how one could construct an associative memory using very simple components. William Little (\cite{little1974existence}) and later John Hopfield (\cite{hopfield1982neural}) began the study of such associative memory models. The energy function of such a magnet has the form $E=-\sum_{ij} \sum_{\mu=1}^M \xi_i^\mu \xi_j^\mu S_i S_j $ where there are $M$ patterns $\xi_i^\mu$ of spins $S_i$ and the sum goes over all pairs. If there were only one such pattern $\xi_i^\mu$, the magnet would not be at all frustrated but is really just a disguised ferromagnet. Its ground state corresponds to each spin $S_i$ taking on the value $\xi_i$. If only a few patterns are encoded you can see that patterns close to any one of them remain as fairly low energy states (See Fig \ref{fig:enespins}) and the existence and encoding of the other patterns just adds some random energy contributions that probably don't completely destabilize the state, in any event, if there are not too many interferences. If the magnet starts in a configuration resembling one of the individual patterns at low temperature, as the spins reconfigure they will rearrange to resemble the complete encoded ``memorized" pattern. Such an easily reachable pattern of spins would resemble the native structure of a protein which can be reached quickly and is fairly robust to environmental changes (Fig \ref{fig:enespins}). Like a human brain which can retrieve memories after losing many cells, if many bonds of the lattice (but not too many!) are modified the pattern still remains retrievable. The ``memory'' pattern is stored globally not just locally. This is quite different from a typical digital computer today where memories are put into specific addresses, locally, and recall can be far from robust.

\medskip
\medskip

Associative memory magnets can be studied very thoroughly using modern statistical mechanics. A somewhat surprising property of associative memory models is that they have a finite ``capacity'' to store information: the maximum number of storable patterns scales only linearly with system size. If there are too many patterns encoded, robust recollection becomes impossible. When there are too many patterns, the magnet instead behaves like a random spin glass and becomes ``confused'' in trying to recall the encoded patterns. The situation is like that recounted in the joke about the old zoology professor who, when meeting a new faculty member at the Fall faculty party, says, ``No, no. Please don't tell me your name. Every time I learn a new name I forget a fish!''

Frustration enters into protein biochemistry in ways very analogous to what happens for magnets and for these neural network models of associative memory. Frustration leads to degeneracy. Finite degeneracy allows interesting and controlled emergent behaviors. Too much frustration and degeneracy leads to qualitatively different dynamics in which searching through the degeneracies cannot be carried out in a meaningfully short time. A little frustration can be good, a lot can be disastrous.

\subsection{Frustration in Heteropolymers}

Ferromagnetic interactions or antiferromagnetic; spins up or down, on a fixed lattice- the choices are stark and it is easy to see the origin of frustration in trying to find the lowest energy state for magnets. Similarly we may ask, what is the root of the difficulty in predicting the three dimensional folds of proteins? How do we specify the choices of structures and how do we judge their quality? The choice space is a bit harder to visualize for proteins than it is for magnets and there are several ways to represent choice space, none perfect. Pairs of peptide bonds are restricted in their relative orientations and side chain orientations are restricted too, so local dihedral angles seem like a sensible choice. The global fold is however a very nonlocal function of the individual backbone dihedral conformational selections -wild swings of portions of the chain are induced even by changing a single backbone dihedral by itself, so interactions (contacts) between the residues generally become completely reorganized by local moves.

Many such simple individual local reconfigurations of the backbone would be impeded by steric clashes and of course the chain connectivity must be preserved at all costs. The allowed configurations of a chain with excluded volume, possibly entangled, therefore require a much more complex encoding than a simple set of binary spins on a lattice. Connectivity and excluded volume by themselves already make computer searching through the allowed configuration space nontrivial (just think about the myriad of patterns in books on knitting or crocheting, or about modern day problems in mathematical knot theory). Yet our understanding and study of dense polymer melts suggest the entanglement effect should actually be small for chains of the length of typical protein domains. Knots first begin to appear in random homopolymer chains typically only after hundreds of persistence lengths. Indeed knots appear rather rarely in the natural protein database, perhaps occurring at the 1\% level (\cite{sulkowska2012energy}). They are more important for DNA.

Frustration due to topology alone is thus only a small part of the story of the difficulty of finding the global free energy minimum for proteins (but as we shall see ``topological frustration'' does play a role in affecting {\it in vitro} protein folding kinetics). The problems of moving the chain around that are encountered in making a single dihedral angle change can often be overcome by a local combination of several such moves, if the chain is not too compact, a so called ``bicycle pedal move". Thus an alternative encoding of the protein configuration in terms of the contact map that specifies which residues are near each other in three dimensional space (and that therefore contribute to the energy) can be more useful than the backbone encoding. Such an encoding emphasizes the idea that non-local van der Waals or solvent-mediated interactions determine the energy (or scoring function) for proteins. While more complex than the interactions in magnets, these interactions still are local in space. There are 20 genetically encoded amino acids, so there are at least 200 possible interaction types compared to the $+/-$ choice for magnets. Still amino acid side chains can be classified into groups whose members share similar chemical properties -big, small, acidic, basic, hydrophobic, hydrophilic, etc. Since Kauzmann, the hydrophobicity of the side chains has been known to be the dominant effect in determining protein architecture (\cite{kauzmann1959forces}). The other physicochemical aspects of the amino acids, as we shall see, are also relevant but this caricature of the interaction as being simply hydrophobic versus hydrophilic already makes it clear how frustration might arise in heteropolymers. Frustration arises from the combination of the constraints of fixed chain connectivity for a molecule with a defined sequence along with the heterogeneous energies for contact interactions that specify the greater attraction that hydrophobic residues have for each other than they have for hydrophilic residues. A problem of frustrating choice thus can arise when for example a single hydrophilic residue is surrounded in sequence by hydrophobes. The hydrophobes can lower their free energy by becoming buried, but then their nearby hydrophile must become buried too. If this lone hydrophile doesn't then find another hydrophilic mate in the interior (to form a salt bridge for example), its burial will be costly in free energy terms. The energy of the hydrophile could be alternatively chosen to be minimized by allowing the hydrophile to rise to the surface but then if the backbones are too rigid, this choice will bring also the hydrophobes nearby in sequence up to the surface as well. We see this reconfiguration will be costly for the surrounding hydrophobes. Not everyone can be satisfied at the same time by an arrangement. In general then low energy states of a random heterpolymer with no specific sequential pattern of hydrophobes and hydrophiles will be the result of compromises. In addition, excluded volume and chain connectivity make it likely that the other different compromised structures that have low energy will actually be quite different in arrangement or topology. To avoid this frustration special patterns of hydrophobicity are needed like those seen in the helical wheel representation of helices (\cite{schiffer1967use}).

\medskip
	\begin{figure}
\centering
	\includegraphics[width=1.0\textwidth]{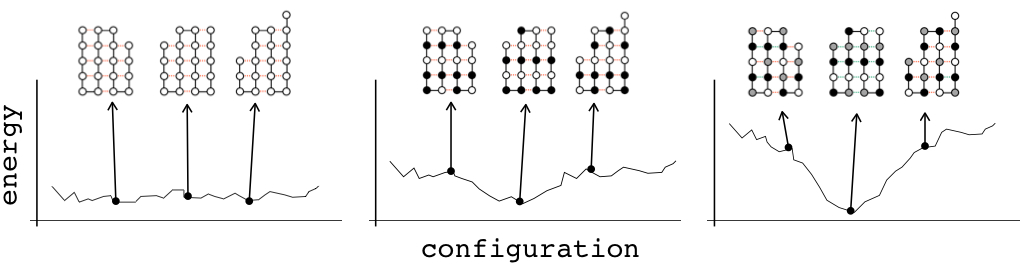}
	\caption{ Specific folding becomes possible as sequence complexity increases. Two-dimensional lattice polymers and its associated energy landscapes are shown. The leftmost is an homopolymer. Many configurational states with nearly similar energy can be formed, for example by shifting the ``strands''. The energy landscape is very rugged with local minima and many barriers. If the temperature is higher than a typical barrier, the polymer will be fluid; at lower temperature the system becomes trapped in many configurationally disparate states and the dynamics gets glassy. At the center a two-letter heteropolymer. The shifting of strands is more difficult as the register can be partially determined by the sequence composition. The energy landscape is still rugged but an overall shape appears. The barriers between any two low energy configurations are high as energetically favorable contacts must be broken up in order to move. At rightmost a polymer with three types of monomers. The energy of the configurations can be better specified by the sequence and the energy landscape becomes funneled shaped. Adventitious trapping in high energy states is reduced and folding to the ground state is robust to both sequence and environmental perturbations. Redrawn with permission from (\cite{Wolynes:1997tg})}	
	\label{fig:simplecode}
\end{figure}
\medskip

Energetic frustration of the type just discussed is easiest to see and to quantify in the world of ``lattice polymers''. Lattice polymers have been much studied by theorists because the simplicity in visualization and ease of computation of such models brings intellectual benefits that overwhelm our desire for realism. A concrete example from the lattice polymer world illustrates the power of simplicity, at least in pedagogy. In Fig. \ref{fig:simplecode} some low energy configurations of a polymer chain are laid out on a two dimensional checkerboard lattice (\cite{Wolynes:1997tg}). The left panel shows configurations of a homopolymer. If every pair interaction has the same value, each of these structures is nearly degenerate in energy. Only surface costs differentiate between the structures energetically. (Of course in a very small system there is always a big surface to volume ratio!) Artificial homopolymers in bulk indeed choose amongst these degenerate states in a kinetically controlled way, making the materials science of solid polymers a challenging topic. We may say, with modest exaggeration that homopolymers slip and slide on a nearly featureless energy surface like that shown schematically below the homopolymer structure. This situation for the homopolymer is a little like what we found for the triangular antiferromagnet where we have an extremely degenerate landscape even near absolute zero temperature. The middle panel of Fig. \ref{fig:simplecode} shows configurations of a binarily encoded heteropolymer. The energies of the various structures now vary a lot as they reconfigure, even their bulk energy contributions are pinned by the heterogeneity of contact energy. The landscape is rugged and two topologically very different configurations can compete at being possible ground states. The barriers between any two low energy configurations are high since now many energetically favorable contacts must be broken up in order to move from one low energy state to the other low energy state. The frustration of finding the lowest ground state comes from the conflict of many reasonably similar, but nevertheless highly compromised, choices. The energy landscape of a random heteropolymer is much like that for the magnetic spin glass (\cite{pmid3478708}, \cite{pmid7784423}).

Even when there are more than two interaction types, frustration, degeneracy and slow dynamics remain the norm for complex heteropolymers. But having more choices in the interactions allows molecular evolution to pre-select landscapes to fold better. A canny ``choice'' (read successful but rare evolutionary outcome!) makes it possible for a single structure to clearly emerge as ground state winner and no longer have significant competitors. An example of such pre-selection is the hydrophobic patterning of $\alpha$-helices and $\beta$-strands that we mentioned earlier. In the lattice world such a pre-selected example is shown in the third panel (Fig \ref{fig:simplecode}). The energy landscape for the ``evolved" sequence plotted below this structure now has a larger overall energy scale than for the random heteropolymer but the landscape is locally a bit smoother and at a low enough temperature only a single structure will dominate. This is much like the disguised ferromagnetic order found at low temperatures in the spin models of associative memory.

\medskip
	\begin{figure}
\centering
	\includegraphics[width=0.8\textwidth]{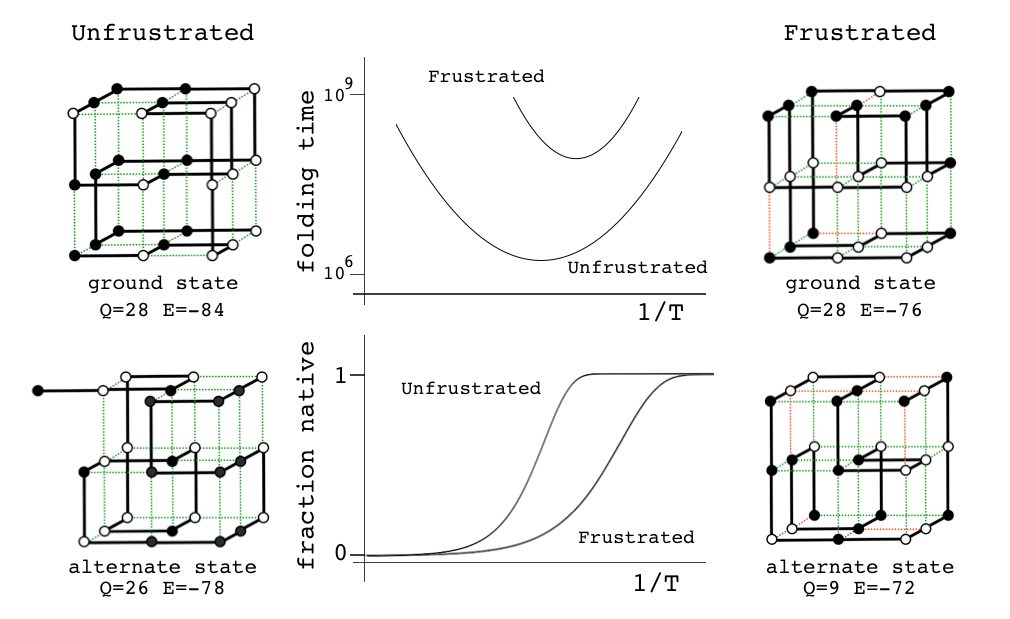}
	\caption{ Frustration in the lattice polymer world. Different degrees of frustration lead to distinct kinetic and thermodynamic behaviors. Here the 27-mer polymers fold into 3x3x3 cubic lattices. If two monomers are adjacent in space there is an attractive interaction between them. This interaction is strong (-3 energy units, green) if the monomers are of the same type and weak (-1 energy units, red) if they are of different types. The most compact configurations are cube-like and have 28 contacts. The structures of the ground state of two sequences are shown at the top. At left an unfrustrated sequence where all interactions are optimal is in the lowest energy state. At right a sequence for which it is impossible to optimize all interactions in any structure, making it necessary to compromise and have weak contacts in the ground energy state. $Q$ measures the similarity to the ground state as the number of native contacts, $E$ is the pair-wise added energy. For the unfrustrated sequences, most of the configurations with energy just above the ground state are very similar to the ground state configuration. For frustrated sequences however there are configurations with energy just above the ground state that are very different from the ground state configuration. When the system gets trapped in one of these low energy states, it takes a long time to completely reconfigure before it can try to fold again, as schematized. The middle panel sketches the results from various folding simulations in the lattice world (\cite{Onuchic:1997hc}). At top the kinetics measured as the number of MC steps required to reach the native structure. Below the thermodynamics of the same systems. Even at its fastest folding temperature the frustrated sequence is not thermodynamically stable in the folded state but takes up other configurations. The unfrustrated sequences fold fast even at temperatures where the folded state wins out thermodynamically over the panoply of other possibilities. When there is a single ground state (as is true when there is little frustration) the global minimum of the lattice polymer can be accessed fairly quickly at low temperature. For the typical frustrated sequence there are several near degenerate low energy states so the folding becomes thermodynamically reliable only at very low temperature. This degeneracy leads to kinetic trapping and the folding time course becomes non-exponential. }
	\label{fig:bosw95}
\end{figure}
\medskip

Different degrees of frustration at the global level lead to different kinetic and thermodynamic behaviors for the heteropolymer. These different behaviors show up in direct simulations of heteropolymers in the lattice world (\cite{Onuchic:1997hc}). In Fig. \ref{fig:bosw95} we show the results of Onuchic {\it et al.} for the folding times of two minimally frustrated lattice polymers (one with a two letter code, the other a three letter code) and one for a rather frustrated but typical two letter code polymer. These are small systems and folding times of random heteropolymers scale exponentially with size so the contrasts in folding time would be bigger for longer sequences but you can see the systems with less frustration are able to achieve faster folding under ideal conditions. The contrast is made even clearer when we simultaneously examine the lattice polymers' thermodynamics: even at the temperature of its fastest folding the typical 27-mer is not thermodynamically stable in the folded state but instead takes up a myriad of unfolded configurations. The minimally frustrated sequences however still fold fast even at temperatures where their folded state wins out thermodynamically over the panoply of other possibilities. When there is a single ground state (as is true when there is little frustration) the global minimum of the lattice polymer can be accessed fairly quickly at low temperature, perhaps with a modest nucleation barrier largely entropy controlled still remaining that can slow folding. In this case a single exponential rate characterizes the folding process. Folding occurs as a Poisson distributed random event in what would appear to be a single step in a classic mechanistic investigation. For the typical sequence as shown in Fig. \ref{fig:bosw95} there are several near degenerate low energy states. This degeneracy leads to kinetic trapping and the time course of folding therefore exhibits several distinct exponential phases. These phases signal the distributed escape from traps. Non-exponentiality is always a sign of a multiplicity of states and can also occur for minimally frustrated polymers. But for fast, minimally frustrated folders the non-exponentiality coming from intermediate states is not the result of frustration but reflects high energy states involved in direct diffusion to the ground state (down-hill folding) which typically occur on the microsecond timescale. For slower, but minimally frustrated folders with nucleation, longer lived intermediate ensembles involve species where only part of the chain (a set of ``foldons'') has become organized and the remainder remains disorganized (\cite{pmid8700876}). When there is partial frustration these intermediates that represent entropically stabilized traps can be further stabilized by allowing the formation of non-native but nevertheless stabilizing contacts. Not only does frustration influence the net stability but also it changes the character of the thermodynamic transition. From the thermodynamic viewpoint when there is little frustration, the folding process appears very cooperative-a single very rapid transition occurs sharply with changing the temperature or solvent conditions as seen for the two or three letter minimally frustrated chains (\cite{Wolynes:1997tg}). With growing frustration, the formation of populated intermediates corresponding to the near degenerate ground states with non-native contacts causes the transition to be spread out over a range of temperatures so the change appears less cooperative. This is manifested in significant shifting baselines in the observable ensemble averages sometimes seen in equilibrium denaturation experiments. Unfrustrated heteropolymers, in other words, exhibit first-order transitions, resembling the freezing of a liquid into a crystal while highly frustrated heteropolymers, in contrast, exhibit thermodynamically continuous transitions that resemble the transition from a viscous liquid to a glass. 

\begin{figure}
\centering
	\includegraphics[width=0.5\textwidth]{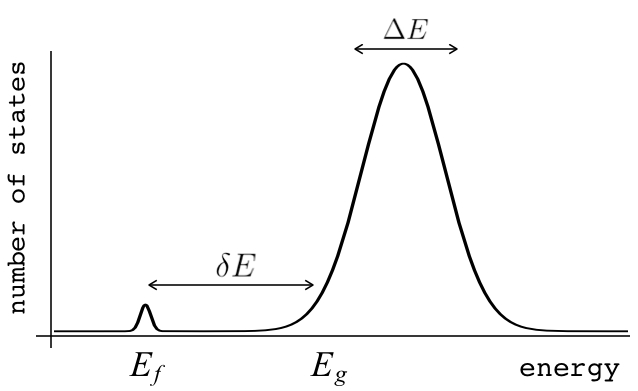}
	\caption{ The global effects of frustration can be quantified with energy landscape theory. A frustrated heteropolymer will have the energy distribution drawn from a random energy landscape, a Gaussian with some mean $\overline{E}$ and a standard derivation $\Delta{E}$. The energy of the lowest energy states $E_g$ can be estimated given the size of the configuration space and the variance of the energy distribution. A large protein whose energy is just $E_g$ will have quite a few kinetic traps of nearly the same stability. Therefore, if a protein is to fold robustly, its ground state energy must be substantially below $E_g$. If the energy of the completely folded state $E_f$ is substantially below this estimate we can predict that frustration effects will be minimal for this sequence (\cite{pmid3478708}). The condition that $E_f$ is below $E_g$ is known as the {\it gap} condition for folding. Comparing $E_f$ and $E_g$ and requiring them to be well separated ($\delta{E}$) provides one way of quantitatively stating the Principle of Minimal Frustration.}	
	\label{fig:denstates}
\end{figure}

Energy landscape theory allows us to quantify the global effects of frustration. In a completely random energy landscape a simple argument (made first in a more abstract setting by Derrida, (\cite{derrida1981random})  allows us to estimate the energy of the lowest energy states by combining knowledge of the size of the configuration space to be searched with knowledge of the statistical variation of energy between substantially different configurations (the variance of the energy). Let's call this typical ground state energy $E_g$.  If the energy of the completely folded state, for the particular sequence being studied, $E_f$ is substantially below this estimate we can predict that frustration effects will be minimal for this sequence. On the other hand if $E_f \approx E_g$ there will likely be at least a few competitors for the ground state and folding will be slow, accounting for the time needed to escape from traps becomes necessary and the ensemble will change more continuously with thermodynamic conditions so folding will not seem very cooperative in equilibrium experiments. Comparing $E_f$ and $E_g$, requiring them to be well separated (Fig \ref{fig:denstates}), then provides one way of quantitatively stating the Principle of Minimal Frustration (\cite{pmid3478708}). 

\medskip
\medskip

It is worth reprising how one makes the random energy model estimate of the globally frustrated ground state energy here, because later we will see how this estimate can be used to give a meaning to the notion of having specific locations for frustration in proteins later. The random energy landscape argument is simplest for a large system. For a large frustrated heteropolymer we expect there to be, for any given configuration, both a large number of stabilizing interactions and a large number of destabilizing interactions. We can thus invoke the central limit theorem of probability to conclude that the energy of a given configuration will be chosen from a Gaussian distribution having some mean  $\overline{E}$ and a standard derivation $\Delta{E}$. Thus if we choose a configuration at random, the energy of that chosen specific state is a random variable having a probability distribution

\begin{equation}
	P(E)= \frac {1} {\sqrt{2 \pi \Delta{E}^2}} . exp - \frac{[E-\overline{E}]^2} {2\Delta{E}^2}
	\label{eq:random}
\end{equation}

The mean $\overline{E}$ and standard deviation $\Delta{E}$ may depend on properties that characterize globally the sampled ensemble of structures such as their degree of collapse and average secondary structure content. Both these statistical quantities can be estimated from an incomplete sampling of the states so long as it is a fair sampling. This makes the statistical characterization of the landscape feasible even while the exact enumeration of all possibilities would be impractical or even impossible owing to the exponentially large configurational space. We can also understand how these statistical properties vary in different parts of the energy landscape. For example if contact pair interactions are the dominant energy contribution, the extended configurations (as measured by radius of gyration) will have very few contributing terms, while compact configurations will have many specific pairs in contact to be summed over. In this case the mean energy $\overline{E}$ depends on the expected number of pairs in that configuration, which in turn depends on the degree of collapse as does $\Delta{E}^2$.

\begin{equation}
	\overline{E} = \overline{\varepsilon} \rho N
\end{equation}

\begin{equation}
	\Delta{E}^2 = \Delta{\varepsilon}^2 \rho N
\end{equation}

If compact configurations are generally stable, $\overline{\varepsilon}$ is negative and mostly depends on the average hydrophobicity. In fact compactness can be guessed quite well from knowledge of sequence composition alone. Intrinsically disordered proteins often are known not to be compact merely on the basis of their composition which is generally insufficiently hydrophobic. Like $\overline{\varepsilon}$, $\Delta{\varepsilon}^2 $ also depends on composition because it depends on how many energetically distinct pairs of interactions there are. Therefore also $\Delta{\varepsilon}^2$ is a function of protein composition. $\Delta{\varepsilon}^2 =0$ for a homopolymer, for example. Generally $\Delta{\varepsilon}^2$ grows also with the number of different kinds of amino acids used to encode the protein.

The Gaussian distribution for the energies of states at first looks like it would allow the energy of a state to be arbitrarily low, but because there is a finite, albeit exponentially large, number of protein conformations, an infinitely low energy can't be realized: there must always be some particular lowest energy state to satisfy the third law of thermodynamics. What is the lowest energy we would expect then? The actual number of states that have energy $E$ is a product of the total number of states regardless of energy (called $\Omega$) times the probability of a specific energy $E$, $P(E)$. Thus,

\begin{equation}
	W(E)= \Omega \frac {1} {\sqrt{2 \pi \Delta{E}^2}} . exp - \frac{[E-\overline{E}]^2} {2\Delta{E}^2}
	\label{eq:random}
\end{equation}

$\Omega$ depends exponentially on the chain entropy $\Omega = e^{S_0 / k_B}$ . Since entropy is an extensive quantity, $S_0 = sN$, $\Omega$ is exponential in protein size. This exponential scaling is the essence of the Levinthal calculation of the difficulty of folding on a flat energy landscape (\cite{levinthal1968there}).  Again we can account for how the number of chain configurations varies with several measures that stratify the energy landscape of the heteropolymers such as its overall compactness, average amount of secondary structure, etc. A lot of interesting polymer physics is involved in such calculations (\cite{luthey1995helix} \cite{plotkin1996correlated} \cite{koretke1996self}). Since both factors in $W$ vary exponentially in protein size, for large proteins, we see that depending on the energy $E$, $W(E)$ is either exponentially large in $N$, suggesting that every finite part of the chain still has some finite number of energetically equivalent configurations or else $W$ is exponentially small in $N$ - there would not be expected to be a single viable configuration for any segment of the chain with this energy. The expected ground state energy $E_g$ then satisfies the equation $W(E_g) \cong 1$ , or explicitly:

\begin{equation}
	E_g = \overline{\varepsilon} \rho N - \sqrt{\rho} . \frac{\Delta{\varepsilon}}{\sqrt{2}} \sqrt{S_0} N
	\label{eq:ground}
\end{equation}

From this formula we see the expected ground state of random heteropolymers becomes deeper in energy, the higher the specific entropy of the chain and the larger the variety that is allowed in the choice of the interactions.

At the value of the energy $E_g$ given by equation ( \ref{eq:ground} ), the number of states is not, in fact, actually precisely one. This is because we have only kept the dominant terms in the exponentials. The number of traps at precisely $E_g$ is known to scale polynomially in polymer length $N$  i.e. $N(E_g) \approx N^{\alpha}$. A large protein whose energy is just $E_g$ thus still will have quite a few kinetic traps of nearly the same stability. Therefore, in order for a protein to fold robustly, its ground state energy must be substantially below $E_g$ on a per residue basis, otherwise trapping will become a problem. Without this gap in energies the actual ground state will be degenerate but perhaps only in short (intrinsically disordered?) segments. The condition that $E_f - \overline{E}$ is below  $E_g - \overline{E}$ is sometimes known as the {\it gap} condition for folding.

The ratio $(E_f - E_g) / \sqrt{\Delta{E}}$ thus indicates how confident we can be, based on the statistics of the random energy landscape that there are no competing traps for the ground state.  Since $E_g - \overline{E}$ is proportional to $\sqrt{\Delta{E}}$ , this ratio is also related to the $Z$-score used in statistics, $Z= \frac{(E_f - \overline{E})}{\sqrt{\Delta{E}}}$. Quantitatively we can say the effects of frustration are minimal if the Z-score is very large. See Fig  \ref{fig:denstates} for a picture of the distribution of energy levels for a minimally frustrated protein. The number of standard deviations below the mean that are needed in order to get a gap depends on the specific entropy $S_0$ and chain length. $S_0$ can be estimated for compact polymers with secondary structure (\cite{luthey1995helix}, \cite{koretke1996self}). Notice that if there are further a priori constraints lowering $S_0$, one can get by with a lower $Z$-score.

The minimal frustration principle can also be stated in terms of the characteristic temperatures at which states with the energies $E_f$ and $E_g$ would be accessed at equilibrium. If there is a large stability gap the folded state will be accessed in a highly cooperative first order transition at $T_f$. At this ``folding temperature" the entropy of the disordered chain is balanced by the deep ground state energy. Clearly then $T_f$ is roughly $(E_f - \overline{E})/S_0$. If the heteropolymer is frustrated, on the other hand, the lowest energy trapped states can only be accessed at a temperature $T_g$ which is roughly $\Delta{E}/\sqrt{S_0}$. In terms of these characteristic temperatures we can say a heteropolymer is minimally frustrated if $T_f / T_g > 1$. We see that $T_f / T_g$ also depends monotonically on the Z-score. One advantage of this way of stating the criterion for minimal frustration is that this criterion doesn't explicitly involve chain length. Also this criterion in terms of temperatures turns out to be fairly robust to there being correlations in the energy landscape which are ignored in the rough argument we have just given (\cite{plotkin1996correlated}). The correlations between energy levels predicted by polymer theory certainly exist as they allow states to be grouped together into basins of attraction which can interconvert without significant barriers. The number of such basins is much smaller than the number of individual configurations. The criterion in terms of $T_f / T_g$ resembles very much the criteria that have long been used in metallurgy to distinguish those systems that easily crystallize from those alloys that form glasses (\cite{greet1967glass}, \cite{chaudhari1978structure}, \cite{stevenson2010ultimate}). The success of the idea in the macroscopic context of material science gives us more confidence in its use at the nanoscale of protein folding.

\begin{figure}
\centering
	\includegraphics[width=0.4\textwidth]{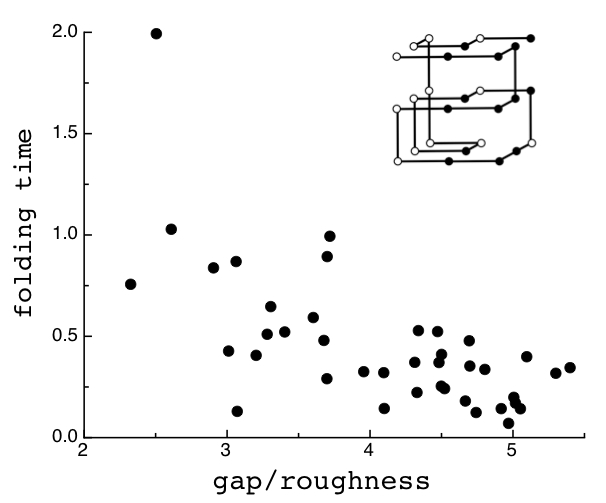}
	\caption{ Quantifying frustration in lattice world. Different degrees of frustration lead to distinct kinetic and thermodynamic behaviors. Folding simulations in enumerable systems have confirmed the criteria for achieving non-degenerate ground states and robust rapid folding. The configurational space of a two-letter polymer folding on a 3x3x3 lattice can be exhaustively analyzed, so that the ground state energy $E_f$ can always be found, the energy distribution all other compact structures can be computed explicitly and the relevant parameters of the energy landscape calculated. Each dot in the figure corresponds to a different sequence. The ratio of the energy gap $\delta{E} = E_f - E_g$ to the roughness $\overline{E} / \Delta{E}^2$ define the relevant temperatures $T_f$ and $T_g$. The average folding time at constant temperature for each sequence is shown. At the same time gap/rough correlates with the cooperativity of the thermodynamic folding transition as measured by its width. For a lattice world heteropolymer to behave in a protein-like way it must conform the minimal frustration principle requirement of having a sufficiently large $T_f / T_g$ ratio as measured by the Z-score statistics. Seeing that this principle works in the controlled world of lattice proteins with known force laws gives us confidence in applying the minimal frustration principle to proteins in the real world, where the force laws are still uncertain. (Redrawn with permission from (\cite{melin1999designability}) }	
	\label{fig:melin}
\end{figure}

The criteria for achieving non-degenerate ground states and also rapid, reliable folding have also been checked many times by carrying out simulations in the lattice polymer world (\cite{sali1994kinetics}). The results of a very comprehensive study by Mellin {\it et al.} (\cite{melin1999designability}) are shown in  Fig. \ref{fig:melin}. The sequences in this study use a two letter code with a reasonably well chosen interaction threshold so that ground state structures are collapsed 27-mer cubes. Competing, perfectly collapsed structures can be completely enumerated explicitly (there are 103346 of them) (\cite{shakhnovich1990enumeration})
so that the ground state can always be found (and checked to see that there is no competing non-compact structures) and the average $\overline{E}$ and $\Delta{E}^2$ can be computed explicitly. The ratio $\Delta= (E_f - E_g)/ \sqrt{\Delta{E}}$ can then be computed explicitly. We can see in Fig.  \ref{fig:melin} that the folding time at constant temperature correlates well with $\Delta$ and therefore $T_f / T_g$. The contrast in rates would be even larger at temperatures where the low stability sequences with low $\Delta$ would be thermodynamically stable. The dramatic slowing of folding of low $T_f / T_g$ with cooling has been explicitly confirmed then in the lattice world. At the same time $\Delta$ correlates with the cooperativity of the thermodynamic folding transition as measured by its width $\delta$. Clearly for a lattice model heteropolymer to behave like proteins do in the laboratory it must obey the minimal frustration principle requirement of having a sufficiently large $T_f / T_g$ ratio or high Z-score. Seeing that this principle works in the controlled world of lattice proteins with known force laws gives us confidence in applying the minimal frustration principle to describe how proteins behave in the real world, where the force laws are still uncertain.

\medskip
\medskip

\section{Frustration and the energy landscape of real proteins}

\subsection{The energy landscape of real proteins}

The principle of minimal frustration has some very direct consequences for understanding folding mechanisms and kinetics in the laboratory. The most important consequence is that thermodynamic stability and folding kinetics should be correlated for real proteins of similar structure in much the same fashion as these thermodynamic and rate quantities are correlated for lattice models. This correlation between rates and equilibria while not completely perfect in the real world is an especial surprise since our introductory chemistry classes have drilled into our heads the idea that kinetics and thermodynamics are supposed to be completely distinct aspects of chemical reactions. Indeed in small molecule chemistry the relations between changes in rates and in equilibria are often unpredictable, but this is not true for the folding reactions of proteins because the overall folding reaction is itself a concatenation of many elementary steps each step being mostly near to local equilibrium. The smooth funnel-like landscape of completely minimally frustrated proteins thus implies a rate-stability correlation because the landscapes of minimally frustrated proteins should have few energetically generated traps. If there were many significant kinetic traps on the landscape the changes in the stabilities of the trap states when the protein is mutated would change the rate unpredictably and would thus interfere with finding a simple correlation of the overall rates with the stability of the ground state configuration alone. In a minimally frustrated protein, the final native structure of the molecule itself is generally directly correlated with the energy landscape throughout the thermally accessed configuration space because the native-like interactions are stronger than random and thus are most common along folding paths. This structure-energy correlation is what we mean by there being a ``funneled'' energy landscape for a minimally frustrated protein. We should remember, though, that the degree of rate/equilibrium correlation will not be uniform across the protein sequence. This nonuniformity of correlation arises even on a perfectly funneled landscape because some parts of the protein make few contacts or if they make many contacts a significant fraction of these contacts are distant in sequence so that forming them becomes entropically costly and thus unfavorable in the transitions state for folding. At the same time other parts of the protein may have the bulk of their contacts rather close in sequence and thus these parts could easily form with lower entropy loss and thus will tend to form early. Studying the structural features of the rate/equilibrium correlation is called  $\phi$-value analysis (\cite{Fersht:1992pi}, \cite{Matthews:1987ff}). $\phi$-value analysis reveals what are the most critical parts of the protein for the rate limiting steps of the folding process. 

The success of $\phi$-value analysis confirms the basic tenet of landscape theory, the minimal frustration principle. This along with the controlled behavior seen in lattice model simulations gives us confidence (at a global level) that we can use the minimal frustration principle to think about natural proteins. These successes of energy landscape theory are partial, however. They don't tell us precisely what the physical interactions actually are that have allowed such an apparently low level of frustration to have been achieved through evolution. Also we would like to know how perfect is the funneling of the landscape of real proteins. The level of agreement between purely structure based landscape models and experiment is so good that we can't tell directly from the existing kinetic experiments how much energetic residual frustration if any actually remains in most real proteins. The data show the overall frustration level is small for most proteins, but how small precisely? Basically, the glassy effects on the thermodynamics and rates of most folding reactions are insufficient to provide an accurate value of the glass transition temperature $T_g$. A quantitative scale for the energetic ruggedness cannot be easily found using the measured folding kinetics in the laboratory alone. Nevertheless some inferences on ruggedness, quantified by $T_g$ and perhaps more important, the energy of the typical competing near-ground states characterized by $E_g$, can be made by using indirect experiments on structure and dynamics of molten globules in concert with the random energy model and polymer physics (\cite{Onuchic:1995mi}). The key quantity that we need to know is the configurational entropy to be associated with molten globules. In other words how numerous are collapsed structures that already have their secondary structure partially formed? Such ``molten globule'' structures should make up the bulk of the decoy structures with which typical final native structures have had to compete energetically. Using the theory of helix coil transition in collapsed states ( \cite{luthey1995helix}), the combination of collapse and helix formation has been shown to reduce the configurational entropy to about 0.6 $k_B$ per residue from a nominal value of 2-3 $k_B$ that would be available for an amino acid chain in an expanded random coil. Thus using the random energy model we can conclude that the ground state energy of a configuration that might compete with the native structure is approximately $E_g=\overline{E} - \Delta{E} \sqrt 0.6$ .  If the energy of the folded state $E_f$ is lower than this value, the native structure will have few competitors. If, on the other hand, $E_f$ is comparable to $E_g$, traps should appear. To quantify the possibility of there being a multiplicity of states i.e. to measure frustration quantitatively, the ruggedness energy scale itself thus needs to be determined. This scale can be computed once a reliable energy function is given. We will use this approach throughout this review but it is worth noting that as an alternative we can try to estimate the roughness of the energy landscape from experiment by looking at fluctuation rates in molten globules, as was done rather early on. Motions in molten globule proteins can be measured via NMR. Such dynamical studies have been carried out for surprisingly few systems (\cite{pervushin2007structure}, \cite{jane2002insights}, \cite{bruylants200915n}). Using some measured millisecond fluctuation rates in molten globules in this way Onuchic {\it et al.} arrived at an estimate for the $T_f$ to $T_g$ ratio for typical helical proteins. They obtained a value of $T_f$/$T_g$ of about 1.6. Thus for a typical protein, glassy effects would only show up strongly at temperatures well below the freezing point of water, explaining why pure funnel models work so well in practice. This estimate of $T_f$/$T_g$ must be considered quite crude but it is nevertheless higher than what one sees in the simplest lattice models using a two letter hydrophobic-hydrophilic code. Such lattice simulations therefore are plagued by trapping effects that are actually less common for real proteins. The low level of frustration suggested by this analysis for real proteins can be achieved for lattice proteins however by using a maximally distinct three-letter residue folding code. The pedagogically popular two flavor model (hydrophobic / hydrophilic) of lattice proteins is more frustrated, giving a lower $T_f / T_g$ ratio but the three letter code may be qualitatively fine for thinking about real proteins. 

\subsection{A license to learn via bioinformatics: Structure prediction and energy functions}

Given the difficulty that direct experiments have in quantifying frustration, a purely theoretical approach to estimating ruggedness is desired. A sufficiently reliable energy function, of course, would allow us to assess computationally how funnel-like a protein landscape is and would allow us to quantify the degree of frustration that real proteins have by determining the free energies of a sampled variety of conformations. One is tempted to develop a computational approach to the ruggedness then by employing the all-atom energy functions that are routinely used to simulate the dynamics of folded proteins (\cite{Karplus:1983zt}). Such a direct assault however faces several difficulties today. The first difficulty is that simulating completely atomistic models in their aqueous environment still takes a lot of computing time. Only in the last few years have the design of special purpose machines and the development of algorithms made it possible to carry out fully atomistic simulations in solvent covering the millisecond time scale motions. This is the time scale for accessing truly distinct conformational minima as inferred from NMR ( \cite{Zerbetto:2013lh}). Mere sampling of the structures of molten globules is also insufficient: it is necessary to quantify the solvent averaged free energies (not merely energies!) of the individually sampled configurations-since we know hydrophobic effects arising from solvent averaging are important and these have a large solvent entropy component. Finally even when sampling and solvent averaging can be carried out, this direct approach to quantify ruggedness relies on the accuracy of the input all-atom force field. Alas, the accuracy of atomistic force fields while well documented to be good for configurations near to the folded state is, again for reasons of time scale, incompletely tested very far away from the native configurations (until recently!)( \cite{pmid22822217}, \cite{Piana:2011fk}). There are indeed many hints that many of the all-atom molecular force fields used today fail to properly describe the stabilities of incompletely folded states. One hint that there is a problem comes from a meta-analysis of folding simulations made by different research groups: there has been a distinct lack of agreement between the specific results obtained from different simulations of the early stages of folding carried out on the same proteins (\cite{Wolynes:2004uq}). There is often a tendency to ascribe these differences between results from different groups to the sampling protocols each different simulator uses but this is probably not the main source of error. Instead what we are seeing in most of these disagreements is that the landscapes encoded by the popular all-atom force fields (like the more obviously stylized two letter code lattice models) are generally all more frustrated than the actual protein energy landscapes really are. Thus each landscape's particular set of traps is quite sensitive to model details much as would be expected for a random heteropolymer. The conclusion we may draw from these observations is that many current atomistic models are still too frustrated to be completely realistic to describe folding in detail. This conclusion is buttressed more directly by a recent study by the Shaw group of the thermodynamics of folding for a variety of all-atom force field models. They have shown that some tuning is needed in order to get reliable folding to the correct native structure across the spectrum of small proteins and that the better tuned models were more cooperative in their thermodynamic behavior than are earlier, less well-tuned, models (used however by many other groups) (\cite{pmid22822217}). Lower cooperativity correlates as we have seen with there being greater frustration and more intermediates. Even the most successful energy function studied by the Shaw group still gives less cooperative folding transitions than are found via laboratory calorimetry (\cite{Henry:2013fu}). Thus we can infer there probably remains still some room for improving all atom force field models. We also conclude that the direction of improvement will involve further reducing the frustration of the landscape. So today's atomistic simulations probably show larger frustration effects on kinetic pathways than really occur in nature. Recently Best {\it et al.} have shown that the Shaw simulations of folding do conform with a minimally frustrated landscape (\cite{best2013native}).

Many of the difficulties of quantifying frustration using all-atom modeling would disappear if a coarse-grained but transferable energy model were available that was also sufficiently accurate. There is a tendency to think coarse-grained models intrinsically must be less accurate than all-atom models because they are clearly less complete in having fewer degrees of freedom explicitly treated. But, when it comes to the issue of having a realistic level of frustration of the landscapes this reasoning may be flawed. As we have seen the funneled nature of the landscape must emerge from evolution, but evolution itself has had only coarse-grained abilities to reduce frustration: mutations can change only a single amino acid at a time. In contrast, mutations are not able to change single atoms in proteins one at a time! If fluoro-alanine would be the best fitting residue in a protein structure, natural evolution is out of luck if it has to find a way to try to put it in -there is no atomic level mutation facility in the cell. (Well, we might imagine post translational modifications like phosphorylation are attempts at doing atomic level mutation.) This evolutionary meta-argument suggests therefore that a sufficiently complicated but still coarse-grained energy function should be able to capture at least the bulk of the evolved low frustration level. 

One of the best ways to find a good transferable coarse-grained energy function is to use the minimal frustration principle itself. The correlation between structure and energy landscape envisioned in the minimal frustration principle gives a strategy for learning energetics from protein structural databases. The minimal frustration principle provides a ``license to do structural bioinformatics" since it implies there is a non-random part of the sequence-structure correlation. Beyond merely providing such a license, the minimal frustration principle, in mathematical form, motivates a number of specific algorithms for optimally learning about energetics from known structures (\cite{goldstein1992optimal}). These machine learning algorithms try to find energy functions that make the landscape of structurally well characterized proteins as funnel-like as possible (\cite{eastwood2001evaluating}). These algorithms start by surveying the $T_f / T_g$ ratios using various levels of statistical mechanical approximation to estimate the ratio. This must be done for many examples in the database of proteins with known structures but using a transferable energy function having a common set of adjustable parameters that encode the interactions between the residues which are assumed universal. Then the typical $T_f / T_g$ ratios for the various different force fields can be compared. The more funneled potentials should have higher $T_f / T_g$. Thus the relative quality of two different coarse grained force field models can be accessed provided the $T_f / T_g$ ratio can be estimated computationally with sufficient rapidity. The computer can furthermore carry out an optimization of the parameters used in the coarse-grained potential energy function to find the force fields that are most likely to fold properly. In this way optimal protein folding codes have been found ( \cite{goldstein1992optimal}). In the lattice world, this learning strategy has been explicitly tested in a clever way. By first simulating lattice folding with a known energy function one is able first to construct an artificial database of lattice structures that can be folded with the assumed energy function. These structures can then be used as input to the learning optimization algorithm (\cite{Yue:1995uq}). This test shows the learning procedure does reproduce to rather good accuracy the original energy function that was used to evolve (design) the structures. This kind of study has established that the mathematics of the learning algorithm is sound.

Even before the lattice model tests of the learning procedures based on the minimal frustration principle were made, the idea was used in a practical way to find transferable potentials for natural proteins by Goldstein {\it et al.}(\cite{Goldstein:1992ye}, \cite{goldstein1992optimal}). The strategy was first applied to associative memory hamiltonians (\cite{friedrichs1989toward}). These are energy functions whose form resembles the associative memory magnet models that we discussed in Section II. The associative memory function consists of a sequence independent backbone interaction term along with a sequence dependent ``memory" term that encodes particular structure-sequence correlations. 
$E=-\sum_{memories, \mu} \gamma (S_i S_j ; S_i^\mu S_j^\mu) \Theta (r_i - r_j^\mu) $
The transferable coefficients $\gamma$ depend on the sequence of the target protein to be folded $S_i$ and the sequences of memory protein $S_j$ which are aligned to the target sequence.

\medskip
	\begin{figure}
\centering
	\includegraphics[width=0.8\textwidth]{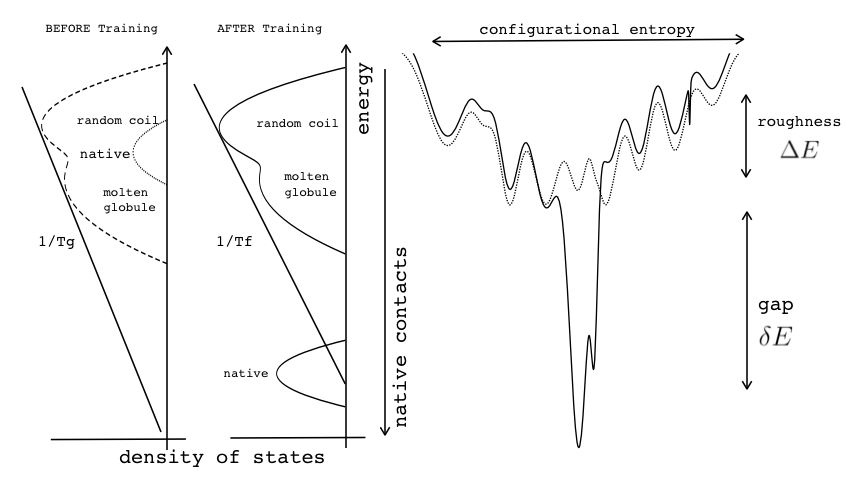}
	\caption{ Reliable energy functions can be learned applying Energy Landscape Theory. The correlation between structure and energy envisioned in the minimal frustration principle gives a strategy for learning energetics from protein structural databases. Machine learning algorithms find energy functions that make the landscape of natural proteins as funnel-like as possible. The more funneled potentials should have higher $T_f / T_g$. The parameter of a coarse-grained potential energy function are optimized self-consistently to find the ones most likely to fold properly. In this way optimal protein folding codes are found (\cite{goldstein1992optimal}). On the right a representation of a realistic protein folding landscape of a globular domain. There are a large number of configurations at the top of the funnel that are nearly random coils, with few nonlocal contacts. As contacts are made the energy on the average decreases. Nonspecific contacts can be sufficiently favorable, a collapsed but fluid set of configurations becomes thermodynamically relevant and indeed may be a separate phase. At left a sketch of the histograms of the energies of the configurations before and after training the energy function. The folded configurations are well separated from these disordered configurations by a stability gap from the molten globule.}	
	\label{fig:koretke}
\end{figure}
\medskip

Some of the alignments used in the memory set may be correct, but they may represent assignments in the ``twilight zone" where sequence identity by itself is so low that we cannot be sure if a supposed homolog is actually evolutionarily related since completely structurally unrelated proteins can also be similarly well-aligned based on their sequences alone. Other memories used in the associative memory model are simply misalignments to proteins that are incongruent in structure to the target but which are not recognizable {\it a priori} as being wrong simply based on sequence. The pairwise distances in the example proteins or memories $(r_i - r_j^\mu)$ encode tertiary contacts, of course but they also encode more distant spatial relationships such as the distances involved in extended helices, strands or turns. The minimal frustration principle then is used first to find the optimal $\gamma$ 's for a training set of proteins. The resulting $\gamma$ 's which are supposed to be transferable are then used to construct hamiltonians for proteins not in the training set. Free energy profile analysis reveals that the resulting transferable energy landscapes were funneled to near native structures. Folding simulations with these associative memory hamiltonians perform well, showing that an algorithm based on learning the $\gamma$  coefficients could ultimately recognize correct alignments even when the alignment by itself was clearly in the twilight zone.

The optimization learning strategy based on the minimum frustration principle has also been applied to off-lattice hamiltonians with simple contact hamiltonians (\cite{koretke1996self}). The interaction parameters learned in this way turn out to be quite adequate to recognize correct structures from candidate homologues in the twilight zone by direct sequence-structure alignment (\cite{Goldstein:1992ye}). Alignment recognition is much easier than molecular dynamics based recognition because the search space for alignment (hundreds of base structures) is smaller. In energy landscape terms, the smaller entropy of possible structures means that a bigger gap results between correct alignments (the ``folded state") and the best decoy alignment than would result from searching through all possible polymeric configurations. Still developing alignment hamiltonians pinpointed the need to estimate better the $T_f / T_g$ ratios in a self-consistent manner; the most troublesome decoys are obviously those that are produced as minima by the same hamiltonian. This notion of self-consistent optimization is illustrated in Fig. \ref{fig:koretke} (\cite{koretke1996self}). This makes the learning process used in the search algorithm iterative and nonlinear but the learning process is still practicable since the development of parameters needs only to be done once. Such optimized but simple contact based hamiltonians also work surprisingly well in off-lattice Monte Carlo structure searches. The server developed by Takada's group (\cite{Jin:2003qo}), predicts structures quite well for small proteins. It is based on such a landscape optimized contact energy function but uses molecular dynamics and fragment assembly as search algorithms.

\medskip
	\begin{figure}
\centering
	\includegraphics[width=0.7\textwidth]{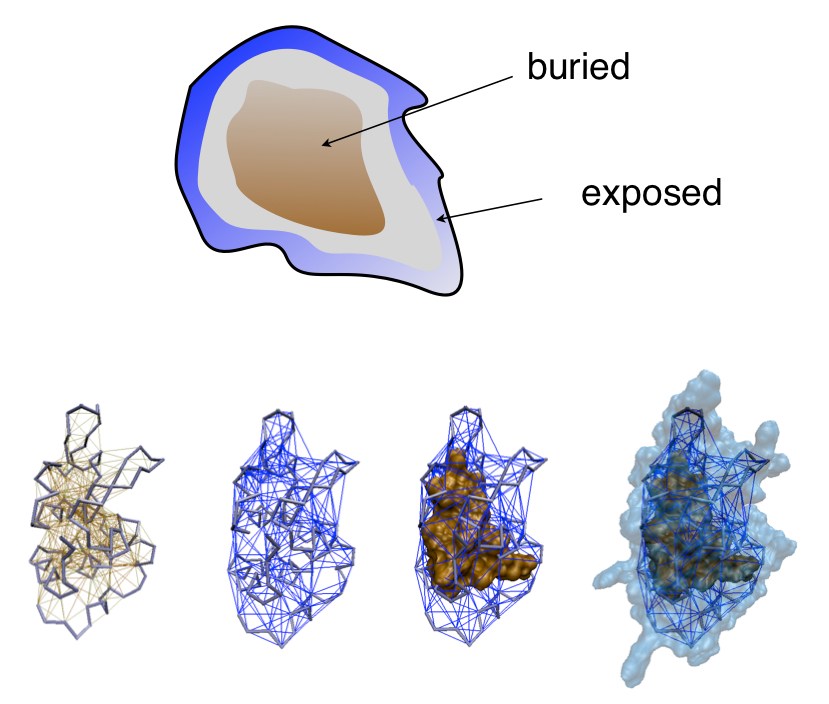}
	\caption{ Proteins {\it live} in water. Amino acid chains interact strongly with their environment, that modulates the intrachain forces and conditions the global protein dynamics (\cite{Frauenfelder:2009tw}). Accurate and reliable force fields must include the solvent, at least implicitly. A sketch of a globular domain is shown on top, where the protein is partitioned in layers, conceptualizing different interaction modes. An implementation of these ideas was realized with a transferable associative memory hamiltonian energy function (\cite{Papoian:2004zt}). The different layers were calculated and projected on a real protein structure (Lysozyme). The backbone chain is shown as a continuous trace, the lines correspond to interactions between amino acids. At left only the direct interactions are drawn (brown), next to only the water-mediated interactions (blue). At right an overlay of the molecular surface of the buried (brown) or exposed (transparent blue) residues is shown.}	
	\label{fig:waterpotential}
\end{figure}
\medskip

\begin{figure}
\centering
	\includegraphics[width=0.9\textwidth]{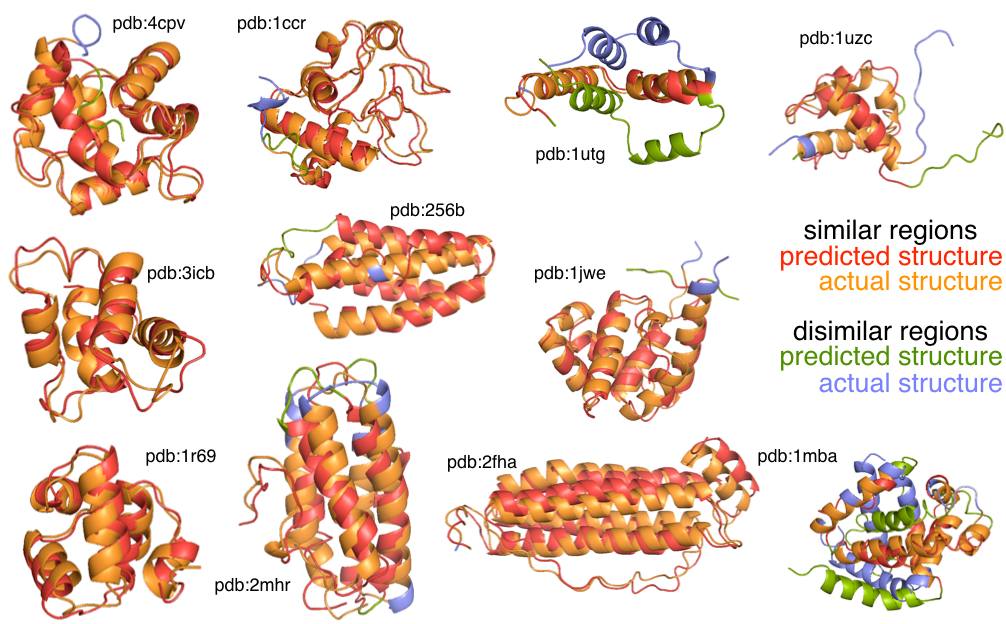}
	\caption{ Protein structure predictions. Energy landscape theory provides a mathematical framework to learn energy functions. The associative memory, water mediated, structure and energy model (AWSEM) is a coarse-grained protein force field optimized with energy landscape theory (\cite{pmid22545654}). AWSEM contains physically motivated terms, such as hydrogen bonding, as well as a bioinformatically based local structure biasing term, which efficiently takes into account many-body effects modulated by the local sequence composition. With appropriate memories taken from local or global alignments, AWSEM can be used to perform {\it de novo} protein structure prediction. The figure shows structural alignments of the best ranking predictions for sequences of different sizes in which no a priori homology information is used. The predicted structures are shown in blue, and the actual experimental structure in green. The regions where the structures are highly similar as judged quantitatively via structural similarity of the contact maps are colored red for the predicted and orange for the actual structure (\cite{Sippl:2012gb}). Regions that are not accurate are shown in blue or green. While not perfect, the predictions are reasonably robust and quite good for moderate size proteins. The same model and energy function can also be used to predict the native interfaces of protein complexes (\cite{Zheng:2012ys}) thus it amounts to a flexible docking algorithm.}	
	\label{fig:water3}
\end{figure}

Seeking greater reliability and accuracy in structure prediction has shown that things are not so simple however. When studying larger protein complexes Papoian, Ulander and Wolynes discovered that the simplest contact-based energy functions while being adequate for some interfaces, were inadequate for describing the binding landscape of many interfaces between proteins (\cite{Papoian:2003il}). They found it was important to allow more distant water-mediated interactions between solvent separated residue pairs to contribute to the energetics  (Fig \ref{fig:waterpotential}). The same parameter learning strategy used for contact models can still be applied to this style of energy function. Structure prediction based on the resulting water mediated energy function while not perfect, turns out to be reasonably robust and quite good for moderate size proteins (Fig:  \ref{fig:water3}). This coarse-grained energy function used for de-novo structure prediction still appears to be more frustrated than the real folding energy landscape in so far as misfolded structures are still sometimes found via molecular dynamics, but the extra frustration largely comes from the ambiguities of the local-in-sequence structure information which is used as input (secondary structure). Such information is encoded via associative memory terms. When a correct native secondary structure alone is used as the only memory input, very few intermediates are found in folding simulations and for most proteins folding is nearly as cooperative as folding processes in the laboratory. In this review we will use this rather sophisticated but still coarse grained energy function to quantify frustration. It is worth remembering therefore that the coarse-grained model we use to illustrate frustration effects in natural proteins still probably errs on the side of exhibiting greater frustration than is present in reality. This being the case, one should take most seriously then indications of very high levels of frustration from this model but probably one should not ascribe too much significance when there are only modest variations or changes in frustration level as measured by this energy function.

\medskip
\medskip

\subsection{Localizing and Visualizing Frustration in Natural Proteins}

Like elementary landscape theory, the traditional phenomenological picture of folding kinetics using single rate coefficients provides only a global description. By global we mean that the protein as a whole is treated as one single entity having no localized substructure. This global phenomenological description works well for small, well-folded proteins but, even phenomenologically, the global picture begins to break down for larger proteins which are usually made up of identifiable domains that can fold independently as kinetic units. In mutidomain proteins, rather often, the individual domains fold in separate steps rather than all at once. The same multi-step folding behavior is seen in long repeat proteins (\cite{Werbeck:2008cq}). These observations impel us to the recognition that the forces in biomolecular energy landscapes act locally even though they have emergent global consequences for the kinetics. How deeply does the locally correlated nature of the dynamics go? The $\phi$-value analysis of transition state ensembles of even small proteins suggests that the more highly folded residues tend to be close to each other in the native state. A simple model of local folding on a minimally frustrated landscape, called the capillarity model (\cite{pmid9177189}, \cite{Galzitskaya:1999kh}, \cite{Ferreiro:2008rq}) likens folding to crystallization of a small droplet. This (oversimplified) analysis suggests that a compact region encompassing about 4/27 $\approx$ 15 percent of a single domain protein must come to a near native configuration before folding can proceed downhill to the native state in a free energy sense (\cite{pmid9177189}). Typically for small proteins such a compact core contains about 20 residues. Local excursions away from the native state can be monitored by hydrogen-deuterium exchange. These unfolding elements are of a similar size to the folding core (\cite{Weinkam:2005if}). The size scale of both folding nuclei and unfolding units is then comparable to the protein segments encoded in the exons of eukaryotic genes (\cite{gilbert1987exon}). It is plausible then that regions of this size are ``foldons" (\cite{pmid8700876}) i.e. minimal folding units. These arguments suggest we can meaningfully discuss the level of frustration of the energy landscape of individual foldons rather than consider only the protein, globally, as a whole. Scanning contiguous regions of sequence using the Z-score criterion for the energy in the native configurations can thus be used to assign levels of frustration locally. By doing this, determining the Z-scores of contiguous segments, Panchenko {\it et al.} showed there often is a rough correspondence between those regions that were minimally frustrated based on their internal interactions and thus would be capable of selecting a unique structure on their own and the regions encoded by genetic exons (\cite{pmid8700876}).  

It is hard to refrain from going still further down in length scale in characterizing the energy landscape in order to ask whether some particular physicochemical interaction between two specific residues in a protein is actually what gives rise to frustration, just as we did for the magnets, where individual plaquettes could be said to be frustrated or not depending on whether as isolated units these clusters of spins would have simple ground states. One must bear in mind though that such an analysis cannot be completely rigorous in proteins any more than it was for magnets: the global structure is determined by the cooperation of many elementary subunits; the clusters are not isolated completely. At the same time we also know that individual interactions in a protein cannot be tuned by mutation -instead several interactions change at once when a mutation is made at the amino acid level, all the interactions involving a given residue change when that residue is mutated. Still this degree of unavoidable spatial coarse graining suggests frustration can be defined at a pretty fine level. Recognizing the mutational limit on the length scale where minimal frustration can be expected suggests a reasonable, but by no means unique strategy for finding and locating the origins of frustration in particular proteins: we can computationally examine the free energy changes that would be made upon introducing mutations and compare the energy of altered choices with what occurs in the original protein. Such a procedure actually generates a series of measures of localized frustration (\cite{Ferreiro:2007bh}). The approximate localization of frustration via this approach has proved to be helpful in visualizing regions of high frustration. The localized frustration analysis, while not being rigorous makes vivid the widespread, delocalized nature of the minimally frustrated interactions that do, in the end, allow proteins to behave globally as minimally frustrated systems with smooth funnels.

When trying to assess frustration locally we are ineluctably breaking up the full energy function of the protein into parts. As we have said, this step has some arbitrariness in it. Nevertheless the breakup of the energy function seems straightforward conceptually when the force field is written in the first place explicitly as a sum over pairs. One can then imagine assigning contacts as being frustrated or not just as we did for the few letter lattice models of heteropolymers that we discussed earlier. The best coarse grained energy functions are not pair-wise additive, however. For non-pairwise additive force fields local frustration indices can be obtained using the {\it in silico} mutational strategy postulated above. A local frustration index should quantify how much a residue or residue pair contributes to the energy in a given structure compared to what it would contribute in a typical decoy or molten globule configuration. Making numerous mutations and changing local environments lets us assess the mean and variance of the energies of molten globule configurations (decoys) relative to the native. By normalizing the difference between the native and the average decoy by the contribution of the same residue (or pair of residues) to the variance of the energies of the decoys we thereby can get an idea whether that contribution to the energy is typical of what would be expected in a minimally frustrated protein or more like that in a random unevolved heteropolymer. In the minimally frustrated protein it is necessary that globally  $E_{Total}$ is less than  $E_g$. Requiring each part of the interaction energy to satisfy a similar constraint $E_{ij}^N > E_{ij,G}^{random}$ is of course overly stringent since only the total energy needs to be low, so we would expect (and find) that only a fraction of individual sites or of the interacting pairs would be minimally frustrated by themselves when we do not consider their cooperation with their neighbors.

At the level of a single residue, a {\it frustration index} can be assigned to each residue via such a set of mutations as 

\begin{equation}
	F_i = \frac{E^{T,N}_i-<E^{T,U}_{i'}>}{\sqrt {1/N \sum^n_{k=1} (E^{T,U}_{i'}-<E^{T,U}_{i'}>)^2}}
	\label{eq:ts}
\end{equation}

Here $E^{T,N}$ is the total energy of the protein in the native configuration, taken as $E^{T,N}=\sum^N_{k\not=i}{(E^{i;k}_{contact}+E^{i;k}_{water})+E^{i}_{burial}}$ according to the tertiary interaction terms of the AMW energy function (\cite{Papoian:2003il}). This energy considers all the interactions that residue $i$ makes with residues $k$, either in a direct contact $E^{i;k}_{contact}$ or in a water-mediated interaction $E^{i;k}_{water}$ and via a single-body burial energy term $E^{i}_{burial}$. The average energy of the decoys $<E^{T,U}_{i'}>$ is computed by mutating residue $i$ to every other possible residue. As the 20 genetically coded amino acids are not all equally probable, the decoy energy is calculated with weights according to the amino acid composition of the chain. These mutations are evaluated from the sequence-specific contact and burial terms from the AMW force field with parameters $\lambda_i, r_{i,k}, \rho_i$ that correspond to the amino acid identity, interaction distance, and density respectively (\cite{Papoian:2003il}). Similar recipes could be used for other coarse grained energy functions.

In the case of pairs of residues we ask: how favorable is the actual native pair relative to other possible interactions? To compute the frustration index for interacting pairs of amino acids $i,j$ simultaneous mutations on residues $i$ and $j$ are made. We have proposed two related but complementary ways for localizing frustration at the pairwise contact level. These ways differ in how the set of decoys is constructed. In one choice, the decoy set is made by randomizing only the identities of the interacting amino acids $i,j$, keeping all other interaction parameters at their native value. This scheme effectively evaluates every possible mutation of the amino acid pair that forms a particular contact in a robustly fixed structure. We call the resulting index the ``mutational frustration": 

\begin{equation}
	F_{ij}^m = \frac{E^{T,N}_{i,j}-<E^{T,U}_{i',j'}>}{\sqrt {1/N \sum^n_{k=1} (E^{T,U}_{i',j'}-<E^{T,U}_{i',j'}>)^2}}
	\label{eq:ts}
\end{equation}

The decoy energy distribution is calculated by randomly selecting amino acid identities from the protein composition and fixing the density $\rho_i$ and the pairwise distances $r_{i,j}$ to the native conformation. It is worth noting that the energy change upon pair mutation not only comes directly from the particular contact probed but also changes through interactions of each residue with other residues not in the pair, as those contributions may also vary upon mutation. One advantage of the mutational frustration index is that, in principle, this local measure of frustration also could be experimentally determined in the laboratory by combinatorial protein engineering. 

A second way of quantifying pairwise local frustration imagines that the residues are not only changed in identity but also can be displaced in location: how favorable is the native interaction between two residues in the native structure relative to other interactions these residues could form in globally different distinct compact structures? The energy variance thus reflects contributions from the energies of molten globule conformations of the same polypeptide chain. For this index, specially suitable for examining alternative tertiary structures, the decoy set involves randomizing not just the identities but also the distances $r_{i,j}$ and densities $\rho_i$ of the interacting amino acids. Given a protein sequence and structure this strategy compares the energy of each native contact with the distribution of a randomly interacting set.

\begin{equation}
	F_{ij}^c = \frac{E^{N}_{i,j}-<E^{U}_{i',j'}>}{\sqrt {1/N \sum^n_{k=1} (E^{U}_{i',j'}-<E^{U}_{i',j'}>)^2}}
	\label{eq:ts}
\end{equation}

When $E^{N}_{i,j} = <E^{U}_{i',j'}>$ the native energy would not be discriminated from the typical energy of a random interaction in the molten globule and $F_{ij}^c\approx0$ . This scheme effectively evaluates the native pair with respect to a set of structural decoys that might be encountered in the folding process. We call the frustration index computed in this way the ``configurational frustration". 

\subsection{Patterns of Frustration in Globular Proteins}

\medskip
	\begin{figure}
\centering
	\includegraphics[width=0.6\textwidth]{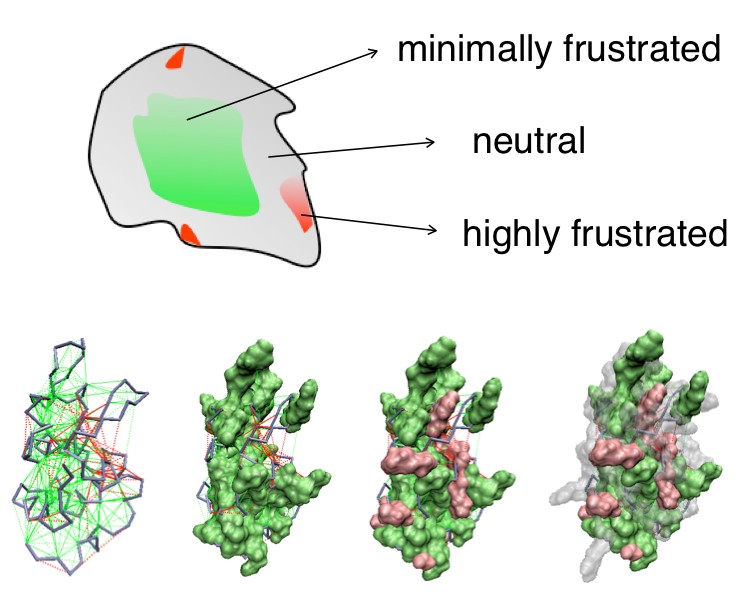}
	\caption{ Proteins can be locally frustrated. A sketch of a globular domain is shown on top, with the protein partitioned in regions conceptualizing local frustration patterns. An implementation of this idea can be quantified with a local Frustration Index, calculated on real protein structures with the AMW force field (\cite{Papoian:2003il}). Below the schematic, the results obtained with the prototypical protein Lysozyme. The backbone is shown as a continuous trace. The protein is networked by a connected set of minimally frustrated contacts (green) while there are two patches of highly frustrated contacts (red). Neutral interactions are now drawn. Direct interactions are shown with continuous lines, and water-mediated interactions with dotted lines. The surfaces of the amino acids are overlaid, according to their frustration index calculated at the single-residue level. Local frustration patterns were calculated with the frustratometer.tk server (\cite{Jenik:2012oq})}	
	\label{fig:local1}
\end{figure}
\medskip

In visualizing frustration in proteins it is useful to simplify the information of the frustration analysis into a discrete form like that used for magnets or lattice models where an interaction can be said to be frustrated or not. Given the continuous nature of the frustration index, physically motivated but ultimately non-unique choices of cutoffs must be made. It is particularly interesting to highlight those interactions that are by themselves minimally frustrated -that is those whose energy is deeper than we would expect in the ground state of a random heteropolymer. Using the entropy estimate for molten states of 0.6 $k_B$ alluded to earlier, minimal frustration corresponds to having an $F_i$ index larger than 0.78 (\cite{Ferreiro:2007bh}). Many of the remaining contacts are neutral, that is they lie near the center of the distribution of possible energies in compact decoy states. While these are important interactions they are so numerous that they would obscure images of the deeper stabilizing contacts. On average if they are consistent they can reinforce the funneling of the landscape but weakly. On the other hand some contacts appear highly frustrated, that is they are not only somewhat destabilizing but are, in fact, unusually destabilizing in the native structure. We define highly frustrated contacts to be those having high energies 1 $\sigma$ above the mean. Such interactions would seem to contribute little to the specificity of the folding to the native structure. So why are they there?

To get a feeling for the patterns of frustration, in Fig. \ref{fig:local1} we show images of the frustration pattern in the prototypical protein Lysozyme. One can see that the protein is networked by a connected set of contacts all of which are minimally frustrated (green) while there are two patches of highly frustrated contacts (red) along with other neutral interactions that are not shown. The frustrated contacts are near to the active site of this protein, which functions as a glycoside hydrolase enzyme. 

\medskip
	\begin{figure}
\centering
	\includegraphics[width=0.8\textwidth]{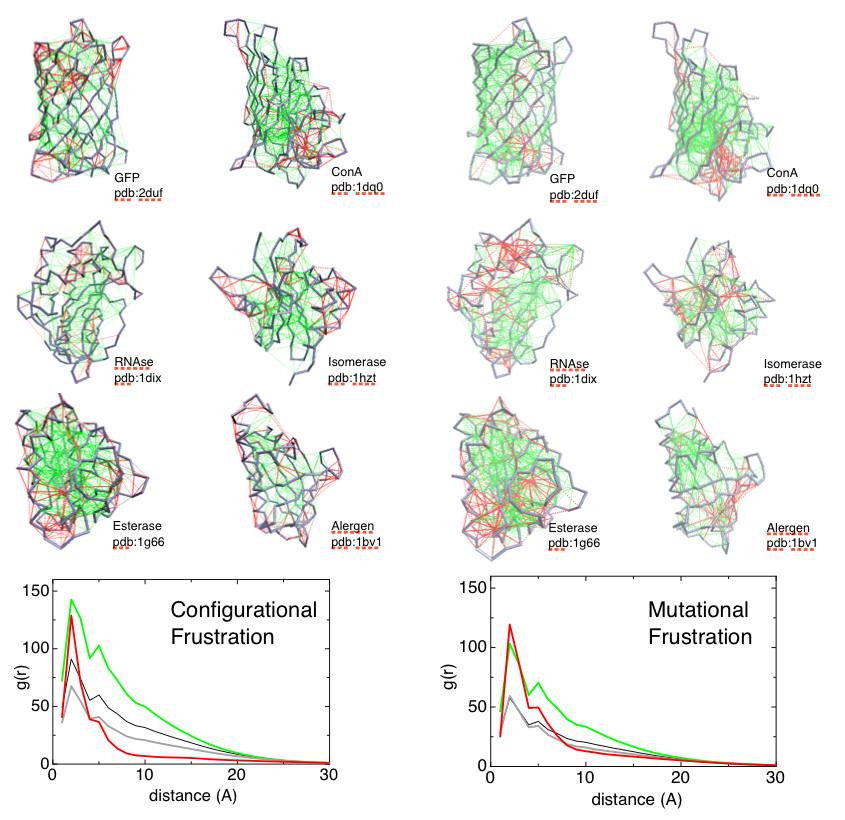}
	\caption{ Local frustration patterns in globular proteins. Examples of {\it frustratograms} of natural proteins. The structural coordinates from protein models were taken the protein data bank (pdb)(\cite{berman2000protein}. The chain backbone is shown as a continuous trace and interactions between residues are represented by thin lines. Minimally frustrated contacts are green, highly frustrated contacts are red, and water-mediated interactions are shown with dotted lines. Two complementary ways of localized frustration are shown for each protein. At right the patterns are calculated with the {\it mutational frustration} definition: how favorable is a native interaction relative to the interactions other residues would form in those locations? At left the patterns are calculated with the {\it configurational frustration} definition: how favorable is the native interaction relative to the interactions these residues would form in other compact structures? A statistical survey was performed with a non-redundant sample of the pdb (\cite{Ferreiro:2007bh}). The pair distribution functions between the centers of mass of the contacts are shown below for the minimally frustrated (green), neutral (gray), or highly frustrated (red) groups. Neutral interactions follow the distribution of all contacts (black); highly frustrated interactions cluster at short distances; and minimally frustrated interactions form a tight network that percolates the structure. }	
	\label{fig:local2}
\end{figure}
\medskip

The bulk of the rest of this review will focus on how frustration patterns of folded proteins can be correlated with folding dynamics and functional aspects of protein dynamics for many specific systems. Before studying particular cases however it is worth making some general observations about the prevalence of frustration when surveying the proteomic database synoptically. First a surprising number of direct contact interactions in native protein structures are individually minimally frustrated. On average about 48\% of the direct interactions are individually minimally frustrated and another 46\% are neutral. It is easy to see then why proteins as a whole behave as if they are minimally frustrated globally. This also makes clear why the notion used in structure-based folding simulations (that nearly all native contacts are stabilizing) is a good first approximation to an evolved protein's energy landscape. Only about 6\% of the direct contacts turn out to be highly frustrated. The indirect water mediated interactions are seemingly less important in guiding the protein to its native state, only 26\% of these are minimally frustrated by themselves while a much larger 61\% are neutral, leaving 13\% highly frustrated. The contribution of water-mediated interactions to global landscape funneling is far from negligible, however. Minimally frustrated residues are generally more buried than highly frustrated residues and are more likely to be found in $\alpha$ helices or $\beta$ sheets rather than in turns or bends. It is possible that some turns actually relieve their frustration in more complex ways not visible in a frustratogram. Side chain-backbone contacts which are commonly made in ``turn signals" may relieve frustration in a way not captured by the coarse grained energy function (\cite{Ferreiro:2007bh}).

The pair distribution functions of minimally frustrated and highly frustrated residues are different from each other and are distinct from the pair distribution of all pairs irrespective of frustration level. As can be seen in Fig. \ref{fig:local2} the minimally frustrated residues are more strongly correlated at large distance than are residues in general. This long range correlation allows the minimally frustrated contacts to provide the main rigidifying elements of the structure. In contrast highly frustrated contacts are strongly clustered and localized, not spread out: whole localities exhibit a high level of frustration: the question of why these frustrated regions exist takes up much of the rest of this review.

\section{Frustration in the evolution of natural proteins and in protein design}
\subsection{Frustration and the complexity of Nature's amino acid code}

Theoreticians aptly explain many aspects of the fundamentals of protein folding by studying simplified models. Sometimes seen as surreal ``spherical cow'' models, these simplifications yield testable and communicable hypotheses such as those uncovered in the lattice worlds discussed above or that one finds using structure-based models (\cite{Clementi:2008kc}). These are thought of as exaggerations because such extreme simplifications are difficult to construct in the laboratory, although some amusing literally macroscopic models of proteins have been built (\cite{Reches:2009hs}). In working with real proteins though, one obvious simplification route would be to somehow reduce the apparent complexity of the amino acid sequences. Twenty amino acid types seems like a lot, at least to first year graduate students. It is still an intriguing fact that natural protein sequences appear to be random (\cite{pmid10988023}), while most random amino acid sequences do not appear to be proteins (\cite{Keefe:2001lo}). By most statistical tests natural protein sequences could encode huge amounts of information, so it has never been obvious which signals are strictly necessary and sufficient for folding and which are simply relics of protein natural history (\cite{Zuckerkandl:1965fh}).

Guided by the ``sequence codes structure'' hypothesis, researchers have attempted to simplify the coding by replacing the apparently redundant amino-acids so that proteins can be made via a simpler palette of possibilities. Information redundancy could be experimentally traced by investigating the structure and function of simplified proteins with some of the apparent redundancy removed (\cite{Clarke:1995pr}). Simplified sequences have been made by simultaneously mutating multiple residues to a single amino acid type. When BPTI protein was mutated by replacing 29 of the total 58 amino acid residues with alanine, it was found that the sequences could still fold (\cite{Kuroda:2000sh}). The effects of a simplifying substitution depend on the position of the mutation in the sequence. Nevertheless, when properly located, simplifying a site only marginally changes overall folding stability of the domain. Most mutants were found to be thermodynamically destabilized and cumulative mutations appear to have additive effects, suggesting that the sites that were chosen to target for mutation do not strongly contribute to the cooperative folding core of this domain (\cite{Kato:2007xd}). When additional substitutions were tried, the protein became too unstable and usually a molten globule state (i.e. an ensemble of conformations) was populated. This is analogous to the situation described for the lattice model proteins where the overall destabilization of the fold relative to the disordered ensemble occurs when the variety of residue types is reduced. While the native stability often decreases because some favorable native contacts are not formed, the possibility of mis-matches of the homopolymer sections also increases. These effects can be seen as manifestations of frustration in the landscape. One key ingredient is the placing of the simplified sites where alanines were already shown to be tolerated. The point mutation A16V in an exposed loop of BPTI severely destabilizes the fold but causes no significant conformational change in the mean native structure when folding does occur. In contrast an 8 alanine substitution in the protease binding site does not severely affect stability, but does stabilize other conformations (\cite{Cierpicki:2002le}). When the effect of the substitution cannot be attributed to structural deviations, some kind of strain is usually invoked by protein engineers. Strain, tension, conflict, unsatisfaction are all terms that reflect frustration effects on protein folding. Tinkering with protein stability by multiple alanine substitutions has also revealed strain effects in the classical models of T4 lysozyme (\cite{Liu:2000wb}), and ARC repressor (\cite{Brown:1999oa}). These studies suggest that the tolerance to mutation is not evenly distributed, but instead is rather polarized in these structures, both in the native stability and contributions to the rate limiting steps (\cite{Gassner:1999xz}). A series of multi alanine substitutions carried out in single contiguous stretches show that the lysozyme fold is robust to most changes in helices, turns and loop regions but that some substitutions can lead to strong effects, revealing that local signals such as helix capping act to reduce the frustration in the natural protein (\cite{Zhang:2002jb}). When the $T_f/T_g$ ratio is large enough, the protein is robust to substitutions. As the energy gap is lowered, the conflicting effects show up, sometimes catastrophically. 

	Going beyond making substitutions of a specific and precise character, combinatorial libraries have been exploited to search for foldable sequences using reduced sets of amino acids. These libraries provide a rich source of diversity for the discovery of novel proteins. At the same time these studies begin to answer questions on the minimal set requirements while of course the practical size of the screening libraries is a concern (\cite{Bradley:2006xi}). Is the existing genetically encoded set of 20 amino acids the minimal set needed to fold all proteins ? As natural protein sequences usually derive from common ancestors and the existing fold classes are believed to be limited (\cite{Brenner:1997mq}), robust identification of fold class has been used as a constraint to test for whether reductions of coding information is possible. It has been shown that coding alphabets reduced up to 12 letters can still detect 90\% of the fold classes. Further reduction rapidly degrades the signal in homology detection (\cite{Murphy:2000wm}). The success of any reduction approach critically relies on how the residue grouping is done, thus different scoring matrices for alignments are obtained when the criteria changes (\cite{May:1999lp}). It may well be possible that the optimal grouping of the amino acids into fewer types depends on the fold class of the structures, and that the observed natural set of amino acids is the actual minimal one that would satisfy all the functional constraints. There is still no absolute standard against which to compare the groupings of amino acids, and chemical intuition has been the most common inspiration for reduced alphabets. 

\medskip
	\begin{figure}
\centering
	\includegraphics[width=0.7\textwidth]{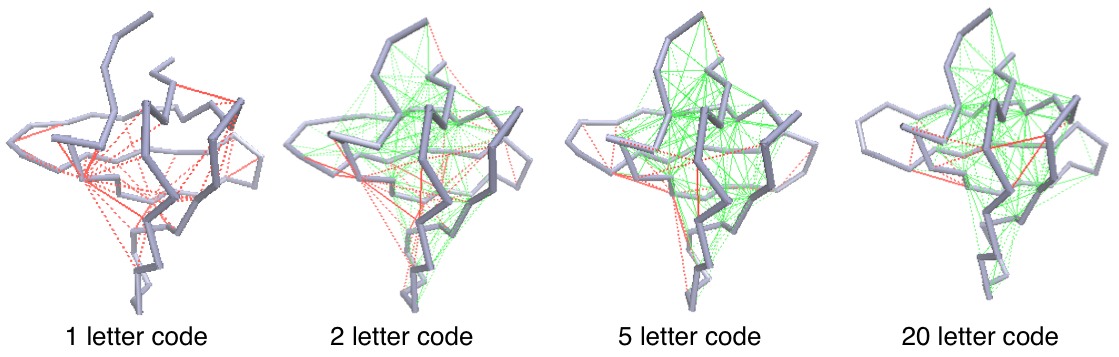}
	\caption{Is it possible to re-encode a protein structure with a reduced alphabet? Once a single overall topology is specified, successful design using oligo-letter codes becomes feasible. When the overall frustration of the starting sequence is low, competing structures are automatically destabilized by the landscape topography. The local frustration patterns of SH3 proteins with simplified sequences is shown. A natural protein is shown at the far right (pdb: 1RLQ), and next to it simplified sequences threaded on same backbone. The 5 letter code is the sequence that Riddle {\it et al.} found by selecting functional SH3 domains (\cite{Riddle:1997ta}). This sequence contains 95\% of its amino acids reduced to a 5 letter alphabet of IKEA and G. The number of highly frustrated interactions increases from 3\% to 7\% , maintaining the minimally frustrated core of interactions. Further reduction of this same sequence did not produce folded SH3 proteins. A specific structure cannot be encoded in a homopolymer. Energy landscape theory shows that increasing the alphabet allows for the possibility of decreasing the ruggedness while still maintaining the stability gap. Bigger alphabets allow for robust encoding of various specific folds. }	
	\label{fig:reduced}
\end{figure}
\medskip
	
	Energy landscape theory shows that increasing the alphabet allows for the possibility of decreasing the ruggedness while still maintaining the stability gap. So bigger alphabets allow for more robust encoding of various specific folds. Once a single overall topology is specified, successful design using few-letter codes becomes easier. When the overall frustration of the starting sequence is low, competing structures are automatically destabilized by the landscape topography. Some of the complexity of protein sequences can then be reduced, as shown in several experimental cases. Riddle et al selected a functional SH3 domain sequence that contained 95\% of its amino acids reduced to a 5 letter alphabet of IKEA and G. Physical measurements on this protein suggest that the simplified protein adopts native-like conformations (\cite{Riddle:1997ta}) and conserves the overall folding dynamics (\cite{Yi:2003bu}). Notably, this group also reported a valuable negative result: they were unable to find functional SH3 proteins when the alphabet was further reduced to the 3 letter code IKE (\cite{Riddle:1997ta}). While success has been reported by other groups in reducing $\alpha$-helical proteins to 3-letter codes (\cite{Beasley:1997wa})(see below), this apparently is not possible for the $\beta$-sheet rich SH3 topology. Frustration in tertiary interactions can be at play. Fig. \ref{fig:reduced} shows the local frustration patterns of models of this SH3 domain for various simplified sequences. It is clear that just as for the lattice model protein (\ref{fig:simplecode}), the density of minimally frustrated interactions diminishes as the complexity of the alphabet is reduced (Fig \ref{fig:reduced}). Also coding simplification leads to the appearance of more highly frustrated interactions. Frustration between alternate states limits the ability to reduce coding complexity. A specific structure cannot be encoded in a homopolymer. An extremely simple alphabet nevertheless allows for encoding the simplest, often highly symmetrical folds. Increasing the variety of interactions makes minimally frustrated structures easier to find even when the topology is intricate. Some of the apparent complexity of protein sequences thus arises presumably not just as a confounding result of historical accidents within a highly degenerate folding code, but may actually be needed in order to encode a suitably wide range of structures with sufficient discrimination in their energy landscapes (\cite{Wolynes:1997tg}). Within a rich alphabet much more varied structures become ``designable'', as many sequences can satisfy the constraints and minimize frustration.

\subsection{Frustration in artificially designed proteins}

	Protein design is a promising endeavor that still can frustrate protein engineers. The moving parts of macroscopic machines can be unambiguously catalogued and combined. This is harder to do for proteins where independent parts may be distinguishable on the larger scales such as domains and multimolecular assemblies, but even at the large scales cooperative effects often surprise us. At the lower levels of engineering single domains, combinatorial libraries of {\it de novo} amino acid sequences provide an overly rich source of diversity. Even for a relatively small protein, it is impossible to sample all possible sequences, as for a chain of 100 residues composed of the 20 amino acids, there are about 1.2 x $10^{130}$ possible polymers. This is a big number. A collection containing one molecule of each sequence would fill a volume larger than Avogadro's number of universes (\cite{Beasley:1997wa}). As randomly generated sequences rarely fold, the effective search space of protein design must be reduced. Even strong simplifications yield libraries still too vast enough to make exhaustive search impractical. The alphabet reduction approaches described in the previous section therefore become a necessary alternative. A ``binary code'' for protein design was a pioneering reduction to construct libraries where folded proteins were likely to be found (\cite{Kamtekar:1993pt}). In these libraries, encoding up to 5 polar and 6 non-polar amino acids, families of four helix bundles were found to be well ordered, and then could be successfully modified to introduce cofactor binding, catalytic activity, and other functional properties producing some of the first protein-based nanomaterials (\cite{smith2011novel}). Four helix bundles based on only three amino acids have been produced, exemplifying that symmetrical structures are more designable than asymmetric ones (\cite{pmid8962034}). 

\medskip
	\begin{figure}
\centering
	\includegraphics[width=0.8\textwidth]{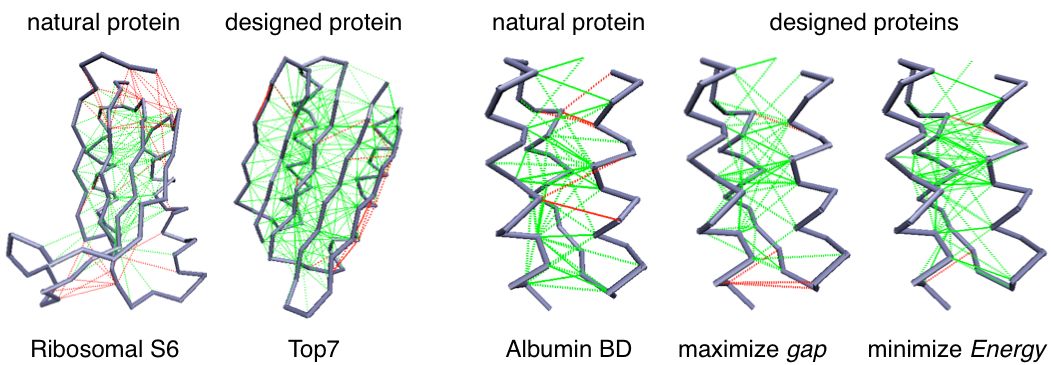}
	\caption{ Are humans too good protein designers ? Although it is still challenging for routine technological applications, protein design has had some success. Top7 is a protein with a novel fold, designed by taking a trace of a target topology and evaluating sequences that minimize the energy in that configuration. The resulting protein is unusually unfrustrated, much less than natural proteins of similar size and secondary structure such as Ribosomal S6. Folding simulations show that S6 folds smoothly in a two-state manner in contrast with Top7 that populates intermediates, even on a perfectly funneled surface (\cite{truong2013funneling}). The high stability of the fold is consistent with the original design, but competing low free energy structures appear because the sequence was not explicitly designed to avoid them. Energy landscape principles inspired the development of an automated procedure to design sequences by crafting a global funneled landscape to design out a vast number of misfolded configurations (\cite{Onuchic:1997hc}). Sequences are searched to either simply minimize the energy in the native state or to maximize the energy gap between folded and misfolded forms. An example of a natural Albumin binding domain and two redesigned sequences is shown. The small differences in frustration level impact the quality of folding (\cite{truong2013funneling}). Unlike the nicely funneled landscape of the natural sequence, landscapes of the designed proteins display complicating features, indeed related to the gap-criterion used for design. Redrawn with permission from (\cite{truong2013funneling}). }	
	\label{fig:designed}
\end{figure}
\medskip

The principle of minimal frustration provides a framework for practical engineering questions (\cite{Onuchic:1997hc}). The essence of landscape engineering is that, in its simplest form, the theory requires quantifying only a few parameters: $E_s$ , the stability gap between the ground (native, folded) state of the protein, $E_d$, the mean of the excited (misfolded) states and $\delta E$, the roughness of the energy landscape. Natural foldable proteins maximize $Tf/Tg$, suggesting a physically meaningful criterion in the design of energy functions for protein folding {\it in silico}. The essential feature of a potential function is that the energy of a sequence in the native structure measured from the mean of the misfolded states is much larger than the width of that distribution. In such a case the folded conformation will automatically have an energy lower than that of all alternative conformations. 

Inspired by these principles, the Takada laboratory developed an automated procedure to design sequences to fold to a given simple structure (\cite{Jin:2003qo}). Folding simulations of a natural (target) fold were performed and a set of misfolded structures or ``decoys" was selected. Design requires crafting a global funneled landscape by figuring out how to design out a vast number of such misfolded configurations. This can be done with statistical energy landscape theory. Sequences were searched to either simply minimize the energy in the native state or to maximize the energy gap between folded and misfolded forms as landscape theory would endorse. It was found that the sequences optimized taking into account the gap display stable secondary and tertiary interactions. The one whose design was based on simply getting a low energy native structure did not show a well dispersed NMR spectrum, suggesting that it did not have a unique fold (\cite{Jin:2003qo}). Fig. \ref{fig:designed} shows the local frustration patterns of the designed and the natural proteins. All three actually have a large fraction of minimally frustrated interactions, with the energy-optimized sequence presenting an unusually high number of them. The largest cluster of highly frustrated interactions in the natural protein coincides with the albumin binding site. This frustrated region (apparently evolved for function) is absent in the designed sequences. To explore the effects of the tertiary energetic frustration in these systems, Ha et al performed folding simulations and structural predictions using AWSEM (\cite{truong2013funneling}). They found that small differences in frustration level do impact the quality of the predictions. The fold of the natural sequence is indeed best predicted by simulated annealing, unlike the designed sequences that show lower fidelity in a structure prediction. Unlike the nicely funneled landscape of the natural sequence, landscapes of the designed proteins display complicating features. The main difference between the designed proteins and the natural ones was indeed related to the gap-criterion used for design. The natural sequences turn out to display the largest difference between the native energy and the energies of decoys (\cite{truong2013funneling}).

A very nice example showing the possible breadth of protein design was realized by the Baker laboratory in their creation of Top7, a protein with a non-natural fold (\cite{Kuhlman:2003km}). The design strategy started with a trace of the topology for which fragments of naturally occurring sequences in other proteins were selected to match the desired secondary structure elements. The sequences were then subjected to further cycles of mutation and selection with the computer scoring the energy of the sequences in the target structure. The selection criteria were based primarily on minimizing the total energy, with some few additional restrictions on the sequences in order to maintain exposed polar residues in an attempt to avoid aggregation. The resulting sequence had no significant similarity to any natural sequence but was shown to fold to a monomeric soluble form whose crystal structure matched the targeted goal (\cite{Kuhlman:2003km}). Study of the folding kinetics of designed proteins showed that Top7 folds fast relative to natural proteins of similar size. Nevertheless Top7 displayed complex multiphasic kinetics unlike its natural counterparts (\cite{Scalley-Kim:2004ez}). The effect was judged to come from slow interconversion of collapsed folding intermediates that become populated, a sign of frustration in the landscape. Apparently not all protein sequences that can be crystallized have energy landscapes as smooth as those of natural proteins. Perhaps crystal formation itself helps to separate different conformers. The source of frustration was suggested to be in the extreme regularity of the $\beta$-strands, which could favor out-of-register pairings. It is also clear that even the basic topology of Top7 makes for its noncooperative folding (\cite{zhang2010competition}, \cite{truong2013funneling}). Local frustration analysis indicates Top7 is unusually unfrustrated in an energetic sense. It has 60\% of the interactions classified as minimally frustrated and only 6\% highly frustrated. This is much less frustrated than are natural proteins of similar size and secondary structure (Fig. \ref{fig:designed}). Folding simulations of Top7 and a cognate S6 protein with a non-additive structure based model show that the topological constraints make the natural counterpart fold smoothly in a two-state manner for a fully unfrustrated landscape, while in contrast the Top7 topology allows there to be populated intermediates even on a perfectly funneled surface (Fig. \ref{fig:designed}). These kinetic topological  traps have a polarized structure, with the C-terminal fragments mostly folded with clusters of minimally frustrated interactions, unlike the transition state of the natural protein that is structurally homogeneous (Fig. \ref{fig:designed}). The high stability of the fold is consistent with the original design, but the low free energy structures appear because the sequence was not explicitly designed to avoid them. Folding simulations with the AWSEM potential also show that several non-native structures are energetically competitive, having non-native $\beta$-swapping as the main source of degeneracy, as earlier suggested. These and other results (\cite{Yadahalli:2013ov}) highlight the idea that the topology of Top7 is in itself a source for conflicts in folding, leading to a kind of ``topological frustration''.
\section{Folding kinetics and frustration}

\subsection{Kinetics on idealized funnel landscapes}

	Very often the dominant folding routes as determined via $\phi$-value analysis can be predicted from the native structure alone by assuming a uniform stabilization gain for forming any given contact (\cite{Clementi:2000gd}). In the folding transition state ensembles, many contacts usually are formed simultaneously. Therefore the law of large numbers allows the differences in energy between different specific native contacts to become averaged out. This averaging allows a uniform purely structure-based model to work. Even when such an averaged ``vanilla" funnel landscape model is inadequate, other models (slightly more nuanced) having non-uniform interaction energies work well. These heterogeneous models (\cite{Shoemaker:1999zp}, \cite{portman1998variational}, \cite{portman2001microscopic1}, \cite{portman2001microscopic}, \cite{shen2005scanning}, \cite{zong2006phi}) while nevertheless having completely minimally frustrated landscapes make improved predictions of the variations of rates. At the very least such heterogeneous contact energy models predict why for a particular system there may be an unusual sensitivity of folding route to details. Such sensitivity occurs, for example, when there are two symmetrically related routes to the folded state which compete. The successes of such pure funnel models have been reviewed several times before (\cite{pmid16780604}, \cite{Onuchic:2004wq}). 

	Folding in idealized funnels does not always progress as smoothly as might be imagined. Real proteins are not geometrically homogenous systems, so the energy gain for folding some part of the protein may not always completely compensate the chain entropy loss of fixing the residues in the native positions. Thus the free energy changes along the various routes to the native state may differ giving rise to the appearance of distinct families of macroscopic pathways. For the sake of simplicity, contrast the case in which the native interactions are made mostly between residues close in sequence (such as a $\alpha$ helical bundle) with another set of routes where the native contacts to be formed are distant in sequence space (such as a $\beta$-sheet of distant residues). In the helical bundle fortuitous interactions will be more probable between residues close in sequence, and the entropy loss will be easily compensated with the interaction energy gain. Since the entropy is already reduced, neighboring interactions will be favorably formed, leading to a zipper like folding mechanism (\cite{zimm1959theory}). In the second case of the $\beta$-sheet, the entropy loss of forming sequence distant interactions is higher, and thus more interactions will be needed to offset the corresponding entropy loss, making it less probable to form a stable enough structure. This argument is at the core of the correlation between folding rate and contact-order which is a rough proxy for loop-entropy (\cite{Plaxco:1998yj}, \cite{Clementi:2004bf}, \cite{plotkinstatmech97}). Even in simple situations, this imperfect cancellation of chain entropy by contact energy can be sufficient to make a sizable free energy barrier appear, and for the protein to behave as two-state system, populating mostly folded or unfolded configurations and not much in between. So even when energetic frustration is removed, and the contact energies between residues are taken to be the same for every pair, the mere topology of the native state gets in the way in defining the locations of the barriers and preferred routes in the folding mechanism. This type of conflict between the need for making native interactions and the necessity of preserving chain connectivity is called ``topological frustration''.

	A simple way to investigate the effects of topological frustration is to explore the folding of a protein with structure-based models (\cite{Noel:2010km}). These models use native structure as the sole input of the dynamic calculation and assign a favorable energy to every pair of residues that are ``in contact''. Contact definitions may differ, but usually a distance cutoff is sufficient to determine the most salient characteristics (\cite{Noel:2012qw}). Molecular dynamics is then performed to explore the phase-space and both thermodynamic and kinetic parameters can be determined because here there is no energetic frustration to slow down the sampling. A natural reaction coordinate here is ``how native-like the protein is'', which is closely related to the number of native contacts that can be easily counted in the snapshots of the simulation (\cite{Cho:2006vz}). Although deceptively simple, these schemes based on perfectly funneled landscapes have been proven extremely useful over the last decade. It has been shown that the structures of transition state ensembles (\cite{Ejtehadi:2004lp}), the folding rates (\cite{Chavez:2004tp}), the existence of folding intermediates (\cite{Clementi:2000gd}), dimerization mechanisms (\cite{Levy:2004ij}), and domain swapping events (\cite{Yang:2004bs}), are often very well predicted with these kind of models. 
	
	The study of topological frustration with structure-based models is not only a playground toy to understand the basics of protein folding, but it has given insights into diverse functional mechanisms. Studies on the inflammatory cytokine interleukin-1$\beta$ (IL-1$\beta$) have yielded significant insights about the modulation of the folding landscape by perturbing functional regions. Structure-based models have identified that a functionally relevant $\beta$-bulge (located between $\beta$-strands 4 and 5) lies in a region that is topologically frustrated. These simulations suggest that contacts with this region direct folding route selection (\cite{Gosavi:2008fd}). Deletion of the $\beta$-bulge maintains high affinity receptor binding but abrogates signaling activity, effectively converting the agonist into an antagonist molecule. This insight inspired experimental studies of malleability of the folding process that relate this malleability to functional properties, using proteins where the conflicting region was deleted (\cite{Capraro:2012kc}). Both the structure-based modeling and the detailed folding kinetics show that topological frustration causes premature folding of some parts that ideally would have folded later. Undoing the traps becomes a kinetic bottleneck, a phenomena termed ``backtracking'' (\cite{Gosavi:2006lg}). Occupying these traps can be inevitable and in natural environments eventually becomes a functional feature, acting as a signal facilitating export of the slow-unfolding protein, the formation of a functional excited state, etc. 
	Still, topological frustration is not trivial to detect from first principles. We generally still need to run extended simulations to localize the trapping states and characterize the kinetic bottlenecks, all of which are somehow encoded in the native topology. ``It is nice to know that the computer understands the problem. But I would like to understand it too." E. Wigner.

\subsection{Chemical frustration}

Most of the forces that give three dimensional shape to one dimensionally connected protein molecules are individually quite weak, a few $k_B$T being a typical interaction energy between amino acid side chains. Some protein chains however are partly assembled using stronger forces like covalent disulfide bonds or the coordination bonding so prevalent in metalloproteins to stabilize the three dimensional structure. Because these ``chemical" interactions are stronger than most of the  interresidue forces giving rise to the folding funnel they can lead to an especially large degree of frustration and kinetic trapping in the folding mechanism, especially when folding is studied far away from physiological conditions. While these chemical interactions are very strong they are more specific and relatively fewer in number than the other forces making it possible often to enumerate structural possibilities explicitly.

Disulfide-bonded proteins are especially interesting because of their historical place in studying protein folding. Anfinsen's famous experiment (\cite{pmid4124164}) did not directly show that the complete three dimensional fold was spontaneously formed but rather only showed that the cysteines which can pair up in multiple ways when ribonuclease is oxidatively refolded under redox buffered conditions did indeed finally find their way to their proper native pairings --a rather coarse-grained view of the structure! In the language of this review, then, the possible frustration of mispairing disulfides turned out to play a minimal role at least when the disulfide/sulfide exchange is well catalyzed by the glutathione buffer. An historically important next step was the pioneering study of the oxidative folding of bovine pancreatic trypsin inhibitor BPTI carried out by Creighton (\cite{Creighton:1984zk}). Here chemical trapping via alkylation revealed several populated species, many of which, surprisingly, exhibited non-native pairings: a sign of frustration. Indeed this laboratory observation was very much on the mind of one of us (PGW) when he first began to study frustration effects on folding in the 1980's (\cite{pmid3478708}). Of course the free energy barrier to rearranging the covalent disulfide pairings depends very much on the oxidation environment so it is not a surprise that chemical frustration can play a role when there are barriers to disulfide interchange. Rather soon after landscape ideas were introduced into folding, ironically it was found by Kim and Weissman that the non-native intermediates originally found in Creighton's study were actually not well populated when the disulfide rearrangements were better catalyzed with a different buffering system so that the ``chemical frustration" was actually minimized, leading to a dominantly funneled mechanism for folding via on-pathway intermediates in which primarily native structure was partially formed and where there was very little long lived non-native structure (\cite{Weissman:1991yg}, \cite{Weissman:1992tn}). The disulfide exchange processes, of course, are influenced by tertiary structure formation and premature burial of a disulfide can lead to an intermediate that must ``back-track", undoing and remaking disulfides to complete folding. This scenario seems to apply to hirudin folding as do some topologically frustrated proteins without disulfides (\cite{Chang:2011vp}).

Metalloproteins provide many other examples of chemical frustration. The paradigm for metalloprotein folding, again of great historical importance to the folding field is cytochrome C, which has been the subject of many experimental observations (\cite{englander2000protein}) and more recently several theoretical studies (\cite{Weinkam:2005if}, \cite{Weinkam2010book}, \cite{Weinkam:2010qc}, \cite{Weinkam:2008ud}). The intense color of the heme made studies of ultrafast folding particularly attractive. While the iron in the heme of cytochrome C is permanently ligated to a cysteine, the sixth ligand for the iron, normally a methionine, can transiently be one of many choices as a coordination partner, each of these gives a characteristic optical spectrum for the heme absorption. Under differing solution conditions then different misligated species can be populated and their concentrations monitored. Flooding the system with imidazole or going to low pH, thereby protonating the histidines, prevents such non-native histidine-heme misligation. By minimizing then this source of chemical frustration essentially by changing solution conditions, completion of folding turns out to speed up by several orders of magnitude (\cite{Weinkam:2009mw}). In contrast, chemical frustration can be increased by raising the pH. This allows many histidines and lysines to lose their protons and now be able to ligate the iron, stabilizing alternate configurations, eventually leading to an alkali denatured species. The species populated by alkaline denaturation turn out to have misligated lysines but otherwise are partially structured in a native-like fashion through the remaining minimally frustrated interactions, which of course include a dominant contribution from hydrophobic interactions of the chain with the heme. Models of perfectly funneled landscapes supplemented by just the specific strong chemically frustrated misligation interactions with lysines have been shown to describe the structural properties of the alkali denatured state quite well (\cite{Weinkam:2005if}). These models also do a good job of predicting the thermodynamics in terms of bond strengths of the coordinating ligands and the thermal unfolding characteristics of the native proteins (\cite{Weinkam:2010qc}).

\subsection{Symmetry effects}

Symmetry turns out to provide an efficient way for the minimal frustration in biomolecules to have evolved. The basis for this idea is easy to see qualitatively using the notion of the "double-or-nothing" strategy of high stakes gambling. If you are playing a game where there is a minimum threshold that must be met in order to survive so your goal is not just maximizing your expected winning, in a fair game you are very much better off to increase the size of your individual bets and to make proportionately fewer of them. For folding biomolecules the energetic threshold is set by the floor corresponding to the ground state energy of random heteropolymers, $E_g / N$ must be achieved for survival.

\medskip
	\begin{figure}
\centering
	\includegraphics[width=0.5\textwidth]{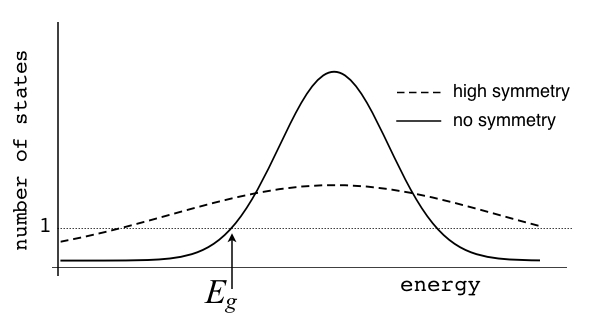}
	\caption{ Symmetry provides an efficient way for minimal frustration. In a symmetric structure the same set of interactions is used multiple times. If a strong interaction is found it will appear more often in a symmetric molecule than in the asymmetric case, broadening the energy distribution as sketched in the figure. For folding biomolecules the energetic threshold is set by the ground state energy of random heteropolymers, $E_g / N$.}	
	\label{fig:symm1}
\end{figure}
\medskip

In a symmetric structure the same set of interactions is used multiple times--this multiplicity is the meaning of symmetry, after all. So if a strong interaction is found in a symmetric molecule it will appear more often than would otherwise be expected. This corresponds to evolution's making bigger energetic bets, but again of course, fewer independent bets are made for the same size molecule. This is the double-or-nothing strategy in molecular evolution. Because the distribution of energies is thus wider for the symmetric case, low energy symmetric structures (see Fig \ref{fig:symm1}) occur much more often than might be naively expected. Symmetric structures are, indeed, prevalent, and often found by chance (symmetric homo-dimers are often found by crystallographers, even when unexpected). In addition structural symmetries are believed to have evolved from fused gene duplications even when the actual sequences found in present day examples leave no exact repetition or apparent symmetry in the sequence but only in the three dimensional structure. A rather indirect argument along these lines was made by Wolynes (\cite{pmid8962034}) and was more elegantly formalized by Wales (\cite{wales1998symmetry}). A more direct version is due to Andr\'e and Baker ( \cite{Andre:2008fb}).

\medskip
	\begin{figure}
\centering
	\includegraphics[width=0.7\textwidth]{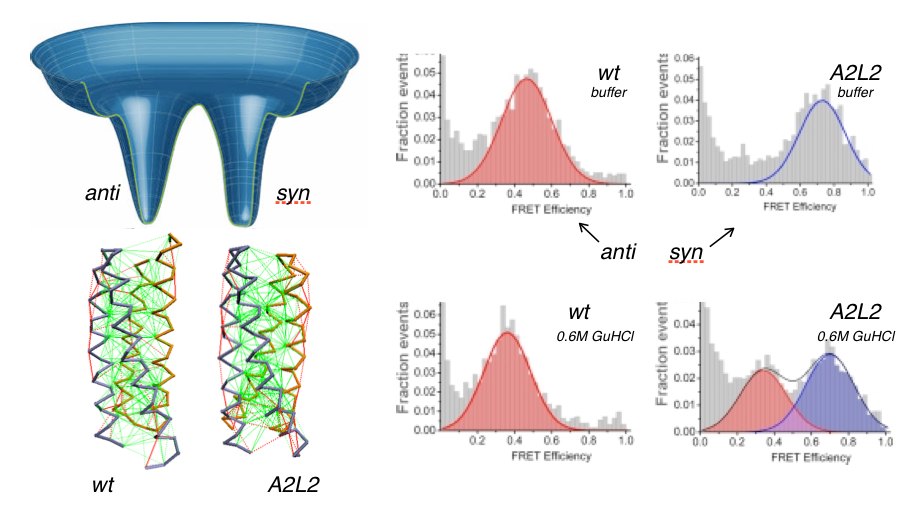}
	\caption{ Symmetric structures tend to be low in energy. ROP dimer folds as a 4-helix bundle of paired coiled coils. An alternate non-native pairing of the helices would form a structure that uses similar contact interactions and would be energetically comparable. The energy landscape appears as a dual-funnel pictured on top. The wild type (wt) protein folds in {\it anti} topology and the A2L2 mutant in a {\it syn} topology, as found by crystallography. In these structures the backbone of one monomer is colored blue and the other orange. The local frustration patterns are displayed as green lines for the minimally frustrated interactions and red lines for the highly frustrated ones. Both forms are energetically competitive. Structure based simulations predicted that the {\it syn} topology folds much faster than the native form i.e. the {\it anti} form is more topologically frustrated. The prediction that there was a structural change based on preserving basic energy landscape theory was confirmed by single molecule FRET experiments. At right, the FRET histograms for wild type protein and the A2L2 mutant obtained in native buffer (top) and in slightly denaturing conditions (bottom). The peaks at efficiencies of 0.45 and 0.75 correspond to the {\it anti} and {\it syn} conformations respectively. While the wild type protein remains stable in one form, the A2L2 mutant occupies a mixed ensemble of states at 0.6 M GuHCl. Because of the near symmetry of a macromolecule, mutations can cause a conformational switch to a nearly degenerate, yet distinct, topology or lead to a mixture of both topologies. (Redrawn with permission from \cite{pmid15701699}, \cite{Gambin:2009zl} )}	
	\label{fig:symm2}
\end{figure}
\medskip

The tendency for symmetric structures to be low in energy means they can show up, seemingly out of the blue, in folding kinetics studies as well as in the natural history of proteins. A cause celebre of this type is the folding of the ROP dimer, a 4 helix bundle made up of 2 sets of paired helical coiled coils (Fig.\ref{fig:symm2}). The usual native structure is made by an antiparallel packing of one helical pair on the partner pair of helices. The unusual kinetic anomaly noted by Regan {\it et al.} was that in some cases making alanine mutations in the individual monomers leads to a dramatic speeding up of the folding rate while at the same time leaving the equilibrium stability unchanged (\cite{Munson:1997sh}). This is not consistent with the idea of there being a unique folding funnel and was cited as a problem for landscape theory. Although a variety of alternative hypotheses were put forward (\cite{lum1999hydrophobicity}), Levy {\it et al.} conjectured that symmetry was the problem (\cite{pmid15701699}). An alternate non-native parallel pairing in the dimer structure was predicted by structure based simulations to fold much faster than would the usual native antiparallel form i.e. the parallel form is less topologically frustrated. Both parallel and antiparallel forms use similar contact interactions and are thus energetically comparable and can act as dual funnels. While originally the fast alanine mutants were thought to have the same structure since like the natural protein they bound RNA {\it in vitro}, in fact, the mutant was not active in vivo where the RNA concentration was lower. This prediction that there was a structural change based on basic energy landscape theory was confirmed by recent single molecule FRET experiments on ROP (\cite{Gambin:2009zl}). By labeling the monomers it was possible to confirm that both parallel and antiparallel packing coexist for the mutant (Fig.\ref{fig:symm2}). The {\it in vitro} RNA binding, owing to the high RNA concentration, pulls the structural ensemble over to the natural one by the law of mass action.

\subsection{Frustration in Repeat proteins}

Many natural proteins contain tandem repeats of similar amino acid stretches. These have been broadly classified in groups according to the length of the minimal repeating units. Short repeats up to five residues usually fold into fibrillar structures such as collagen or silk, while repeats longer than about 60 residues usually fold as independent globular domains (\cite{pmid21884799}). There is a class of repeat proteins that lie in between these for which the repeating units couple their folding. For these proteins unique ``domains'' are not obvious to define (\cite{Parra:2013kl}). Repeat proteins are believed to be ancient folds. Their biological activity is usually attributed to mediating specific protein-protein interactions. Their modular structure allows a versatility of recognition parallel to that of antibodies. Successful design of repeat-protein domains with novel functions based on simple sequence statistics (\cite{Binz:2003ws}, \cite{Tamaskovic:2012ti}) suggests that the folding and the functional signals present in their sequences can be partially segregated. Still, when studied in detail the coupling between folding and binding of natural proteins turns out to be intimately related to their biological function (\cite{pmid20055496}). Typical repeat-proteins are made up of tandem arrays of $\sim$20-40 similar amino acid stretches that fold up into elongated architectures of stacked repeating structural motifs. Quasi-one dimensional, these non-globular folds are stabilized only by interactions within each repeat and between neighboring adjacent repeats, with no obvious contacts between residues more distant in sequence. The internal symmetries of repeat-proteins suggest that the overall folding properties of a complete ``domain'' (the stability and cooperativity of the array) may be derived from a microscopic description of the energy balance within each folding element and its interactions with its neighbors (\cite{Aksel:2009qb}). Because of the delicate energetic balance in each subunit, subtle variations in the interactions in and between the repeats can give the impression of major changes in the folding landscape while in fact it is well funneled (\cite{Ferreiro:2007ph}). Such variations may ``decouple'' the folding of the elements, and partially folded species become populated in kinetic studies. Sufficient information about the population of these states can yield rather quantitative models of the energetic distribution along the protein (\cite{Ferreiro:2008cr}, \cite{Schafer:2012oh}). Behind the apparent simplicity of repeating amino-acid sequences, an extremely rich behavior emerges.

	Energy landscape theory argues that the three-dimensionally connected globular proteins must fold along a funneled landscape. One-dimensionality weakens this necessity (\cite{luthey1995helix}). Folding simulations based on perfectly funneled model landscapes nevertheless have revealed how finer details of the energetic contributions contribute to repeat-protein folding and are able to recapitulate the experimental results (\cite{Barrick:2008tb}). Mello and Barrick (\cite{Mello:2004nu}) have derived a funneled energy landscape from equilibrium unfolding data on the Notch seven-ankyrin-repeat domain and a simple energy function based on the folding of each repeat and the interaction with its neighbors. Werbeck and Itzhaki (\cite{Werbeck:2008cq}) have shown that the energy landscape of a large ankyrin-repeat protein is funneled yet the kinetics can be described as that for two approximately independent six-repeat subdomains. The different stabilities of these subdomains hint at unevenness in the energy landscape that results in a broad ensemble of species of similar free energies that cannot be distinguished from one another. Single point mutations cause further landscape unevenness (\cite{Werbeck:2008cq}) and give rise to multistate folding kinetics (\cite{Itzhaki:2012iq}). It is important to note that this ``unevenness'' is of a different sort than the ``roughness" discussed in landscape theory, by which is meant the energetic frustration that leads to non-native interactions. 
	
\medskip
	\begin{figure}
\centering
	\includegraphics[width=0.6\textwidth]{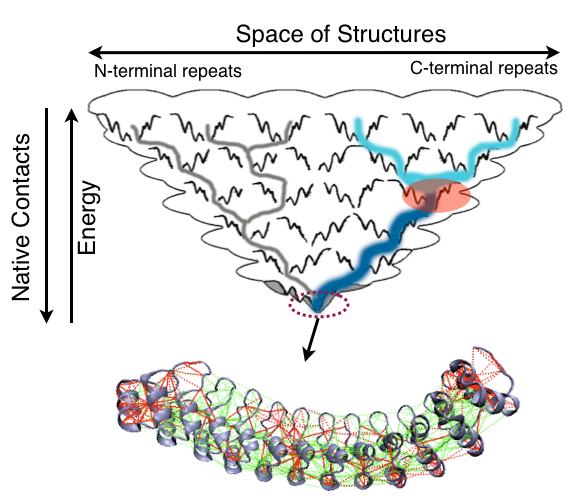}
	\caption{ The energy landscape of repeat-proteins have internal symmetries. The repeating sequence encodes similar structural elements that interact with nearest neighbors, depicted by many folding elements that merge upon interaction and comprise an overall funneled landscape (\cite{Ferreiro:2007ph}). High energy unfolded configurations are on top and the fully folded state at the bottom. In between, intermediate ensembles of partially folded states can become populated at equilibrium (red shadow). Asymmetries in the local energy distribution mold the landscape roughness that define the kinetic routes. The thick blue lines depict the preferred folding routes of the 12-ankyrin repeat protein D34, for which the intermediate ensemble is polarized towards the C-terminal repeats of the molecule (\cite{Werbeck:2008cq}). The crystal structure of D34 is pictured at the bottom, colored according to its local frustration pattern. Highly frustrated regions are located at the ends, and in a central region between repeats 5 and 6. These breaks occur where the experiments suggest the intermediate's folding boundary is. The collective influence of local interactions along the one dimensional scaffold allows sites to thermodynamically modulate each other even at considerable distance. Small perturbations, such as mutation or ligand binding, may change the interactions in or between the modules causing large effects that can be experimentally dissected. (Redrawn with permission from (\cite{Ferreiro:2008cr})}	
	\label{fig:repeats}
\end{figure}
\medskip

	Hagai et al have recently compared the folding kinetics of two repeat-protein systems that, besides having similar contact-order, display very different folding rates in both experiments and simulations. They used energetically unfrustrated models and showed that modulating the energetic strength of the interface between repeats (relative to that of the intra-repeat) can drastically change the stability, folding rate, and cooperativity of the repeating array (\cite{Hagai:2012tf}). Their study suggested that a conflict appears between forming interfaces to the left or to the right of an already folded internal repeat-unit. They showed that the formation of one interface is attenuated when the other is already formed. They argue this topological frustration will often arise as a consequence of the quasi-monodimensional geometry of repeat proteins. According to their analysis, these systems can be designed to fold faster by decreasing the strength of the interface or parts of it, resulting in less frustration in the formation of the two interfaces, and consequently in a higher probability for forming the intra repeat contacts. This reduces the conflict involved in forming the two interfaces of a given repeat unit and in turn gives rise to a less cooperative behaviour (\cite{Hagai:2012tf}).
	
	Unlike globular domains, solenoid repeat architectures can be severely truncated and nevertheless still retain cooperative folding properties. Javadi and Main (\cite{Javadi:2009pt}) have described the folding kinetics of a series of designed TPR-repeat proteins assembled with identical repeats, from 2 to 10 repeats, and have described how the energy landscape changes with the addition of repeat units. Consistent with theoretical predictions (\cite{Ferreiro:2008cr}), they observed that the longer proteins display more complex multi-state relaxations, Moreover, although the initial folding event of all these proteins involves a nucleus with similar solvent accessibility, the subsequent folding of the other regions depends directly on the repeat number, with the longest protein populating an apparent off-pathway intermediate (\cite{Javadi:2009pt}). Other repeat protein topologies show congruent folding characteristics. Long arrays of Ankyrin and Heat repeats display complex denaturant dependent relaxations, indicative of sequential transition states and parallel pathways (\cite{Itzhaki:2012iq}). Kinetic studies on site-directed mutants suggest that the sequential transition states arise from uneven unzipping of the repeat array with local trapping, and parallel pathways result from the inherent symmetry of repeat-protein structures that initiate folding in either or both N-terminal to C-terminal, depending on the local stability of the initially nucleating repeats. Moreover, the addition of terminal stabilizing repeats can shift the transition state ensemble toward those regions, rerouting the folding of the whole repeat array (\cite{Tripp:2008mw}). Local energetics are thus crucial in determining the routes actually followed and the traps populated in these quasi one-dimensional proteins. Experimentally tracing the kinetics of repeat-protein folding further details the richness of mechanisms that can be gained by the molding traps. As mentioned, these traps may arise purely topologically or as an effects of local energetic frustration. 
	
	Energetic frustration seems to be at work in determining the traps in the folding of the ankyrin repeat protein D34. This protein unfolds via a fast process that yields an intermediate that can also be detected at equilibrium (\cite{Werbeck:2007mb}). The barrier separating the native from the intermediate states appears ``broad," consistent with the picture of neighboring repeats unfolding sequentially (Fig \ref{fig:repeats}). A slower unfolding phase corresponds to unravelling the C-terminal subdomain, where mutations affect both the folding rate and its urea dependence in a manner consistent with unfolding by two parallel routes. How does this change in folding regime appear? Local frustration analysis points out that there are three regions enriched in frustrated interactions in an otherwise strongly cross-linked web of minimally frustrated contacts (Fig \ref{fig:repeats}). One of the highly frustrated regions colocates with the central region where the experiments suggest the intermediate's folding boundary is found (\cite{Ferreiro:2008rq}). The highly frustrated interactions are apparently responsible for the roughness that gives rise to a long-lived folding intermediate. Modification of this interface so that it is minimally frustrated should destabilize the trap, making the D34 landscape appear much smoother. Experiments redesigning repeat-proteins so as to conform to minimally frustrated ``consensus'' sequences show that indeed traps are destabilized and kinetic barriers are considerably reduced (\cite{Kramer:2010or}, \cite{Street:2009um}). The overall picture emerging from repeat-protein folding studies is that the differences in folding rates are related to escape from local traps. These traps appear as manifestations of frustration, either topological, energetic or both. There is a fine balance between these signals that can be at play at the secondary structure and/or the tertiary contact levels. A local effect is felt globally because the near neighbor interactions are extraordinarily important in stabilizing the repeats, and weak biases can tip the balance to complete folding. Single substitutions can affect local biases (such as helix propensity), and thus exert profound effects on the overall folding of these domains. The larger repeating arrays are more likely to tolerate ``cracks'' and the folding at the repeats ends may become anticorrelated (\cite{pmid18483553}). Such sensitivity allows for specific encoding of traps by means of making few sequence or environmental modifications. This may be used in intracellular targeting and sorting. Owing to the symmetry of the repeating array, long arrays can be fine tuned to populate partially folded states, and these ensembles can be coopted in functional mechanisms. Changing the stability of a single repeating element (by posttranslational modifications or binding of other macromolecules) may affect the behavior at a distant site, providing a coupling mechanism that can transmit allosteric signals to long distances within a single repeating geometry.

\subsection{Anomalous $\phi$ values}

\medskip
	\begin{figure}
\centering
	\includegraphics[width=0.9\textwidth]{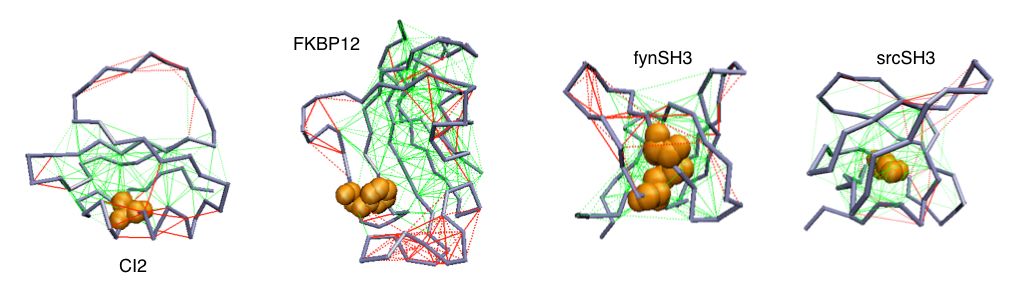}
	\caption{ Anomalous $\phi$-values often involve frustrated sites. In a minimally frustrated landscape there is a relation between the energetic contribution of residues in the native state with the energetic contribution to the transition state ensembles along the primary folding routes. These can be experimentally measured by analyzing the change in stability and folding speed upon single-site mutation by $\phi$-value analysis. Consistent with the notion of widespread minimal frustration, more than 1000 reliable measurements (\cite{Naganathan:2010ea}) lie in the range 0 $< \phi <$ 1 as expected on a perfectly funneled landscape. The figure shows the frustratograms for proteins in which deviations have been confidently measured, with the anomalous residues highlighted in orange. CI2 mutant A16G ($\phi$ = 1.12) ; FKBP12 mutant F36A ($\phi$ = -0.1); fynSH3 mutants F26I ($\phi$ = -0.4) A39V ($\phi$ = 2.34) A39F ($\phi$ = -0.16); srcSH3 mutant I34A ($\phi$ = 2.01). }	
	\label{fig:phivalues}
\end{figure}
\medskip

In the laboratory, $\phi$-values are determined by measuring the changes in both overall stability and folding speed of a protein and single point mutant(s) of it and comparing them (\cite{Fersht:1992pi}). The stability change is $\Delta \Delta G^{eq}_{mut-wt} = \Delta G^{U-F_{mut}}_{mut} - \Delta G^{U-F_{wt}}_{wt}$ ; and the change in folding rate usually can be ascribed to the effect of the mutation on the activation free energy $\Delta \Delta G^{TSE}_{mut-wt} = -RT ln (k_{mut} / k_{wt})$. This analysis assumes that the transmission coefficient, related to configurational diffusion and internal friction, remains constant upon mutation, which may not always be the case. The $\phi$-value is defined as $\Delta \Delta G^{TSE} / \Delta \Delta G^{eq} $. For minimally frustrated proteins, realistic pictures of the rate-limiting steps in folding can be constructed by measuring the effect of mutations not just at the folding midpoint but over a range of thermodynamic conditions. In contrast to the situation for a minimally frustrated protein, on a very rugged landscape the $\phi$-values become very difficult if not impossible to interpret in terms of the native folded protein structure alone. This is because the kinetics of re-organization is determined by the stability of unknown non-native contacts in the transition state ensemble. Since these need not be present in the final native structure for highly frustrated systems, their energetic change would be unrelated to stability changes of the protein as a whole. The globally unfrustrated character of the energy landscape of natural proteins indicates that there will be relatively few thermally accessible states that can be associated with such folding ``mistakes'' at the folding temperature ($T_f$) . When $T_f / T_g$ is large, the structures will have mostly native interactions formed and only a few weak non-native interactions are anticipated. Partially folded forms of the protein may become populated, but highly misfolded alternate structures are not expected to occur in general. This simplifying feature of a minimally frustrated heteropolymers' landscape is what allows one to infer the existence of a relation between the energetic contribution a residue pair makes in the native state with the energetic contribution to the populated intermediate or transition state ensembles in the folding routes. The thermodynamic contribution of an interaction will be simply scaled by the fraction of the time that the pair contact is actually formed in the ensemble, and the folding rates will largely follow native state stability giving rise to ``rate equilibrium free energy relations'' (\cite{Leffler:1953am}) . This regularity is predicted by a funneled landscape and provides the ultimate theoretical basis of the analysis of protein folding kinetics using $\phi$-values (\cite{Matouschek:1989ap}). Strong deviations from this simple behavior then are likely manifestations of localized frustration in the energy landscape.

Many natural proteins with simple kinetic behavior have been subjected to $\phi$-value determination and analysis. Almost 1000 mutations have been independently evaluated and the general behavior is in accordance with the expectations outlined above (\cite{Naganathan:2010ea}). Most of the $\phi$-values indeed lie in the range 0 $< \phi <$ 1 , i.e. they are consistent with an ensemble averaged fractional occupancy of local native structure (reviewed elsewhere (\cite{Onuchic:1997hc})). In general, the $\phi$-value increases linearly with the stability perturbation and, in conditions of isostability, where the unfolded and folded forms are equally populated, most of the $\phi$-values cluster around $\phi = 0.36 \pm 0.11$ (\cite{Naganathan:2010ea}). This average value is anticipated from both mean field and capillarity theories on perfectly funneled landscapes if the activation free energy barrier originates from an imperfect cancellation of the entropy loss by the energy gain upon folding, as we have discussed previously. $\phi$-values are not completely uniform, however. Site-specific variations of $\phi$-values can be roughly predicted by knowing the topology of native structure under the assumption of a minimally frustrated energy landscape (\cite{Munoz:1999tz}, \cite{Clementi:2000gd}). These predictions are pretty good when the conformations belonging to the $TSE$ have a large connected web of native contacts established. Sometimes, specially for $\alpha$-helical proteins, inhomogeneities of the energies do make the vanilla predictions less accurate (\cite{Cho:2009mb}). If the interactions around the probed site are sparse, mutations can change the location of the TSE dramatically and the perturbations can no longer be treated as linear free energy perturbations (\cite{Sanchez:2003hz}). Very large deviations from structure-based predictions or $\phi$-values outside the expected 0-1 range are signatures that energetic frustration could be at play. Sometimes, the deviations can be directly attributed to highly frustrated interactions that one can identify in a native structure. Fig. \ref{fig:phivalues} show the local frustration pattern of four example proteins where specific mutations perturb the folding rate more than would be expected by the native stability change ($\phi > 1$, Fig \ref{fig:phivalues}a) or where the mutations have opposite effects on the $TSE$ and the native state ($\phi < 0$ Fig \ref{fig:phivalues}b). In both CI2 and FKBP12 the sites indicated by these anomalous $\phi$-values map to regions that are enriched in highly frustrated interactions in the native state (Fig \ref{fig:phivalues}). However, in the cases of the fynSH3 and srcSH3 domains, the mutations for which anomalous $\phi$-values were measured do not coincide with frustrated interactions for tertiary contacts (Fig \ref{fig:phivalues} c,d). For these sites the anomaly must be attributed to other sources of frustration. One major component of the stabilization energy that is not visible in the frustratograms of Fig. \ref{fig:phivalues} is the secondary structure energy term. Strongly stabilizing tertiary interactions may locally conflict with the secondary structure propensity of a backbone that will otherwise occupy other conformations if unrestrained. Conflict between secondary structure formation and tertiary interactions has been commonly observed in the SH3 fold class (\cite{Klimov:2002sv}). Notice that in these cases the regions of anomalous $\phi$-values are enriched in minimally frustrated tertiary interactions, perhaps compensating for the secondary/tertiary conflict (Fig \ref{fig:phivalues} c,d).

\subsection{Non-native intermediates}

\medskip
	\begin{figure}
\centering
	\includegraphics[width=0.4\textwidth]{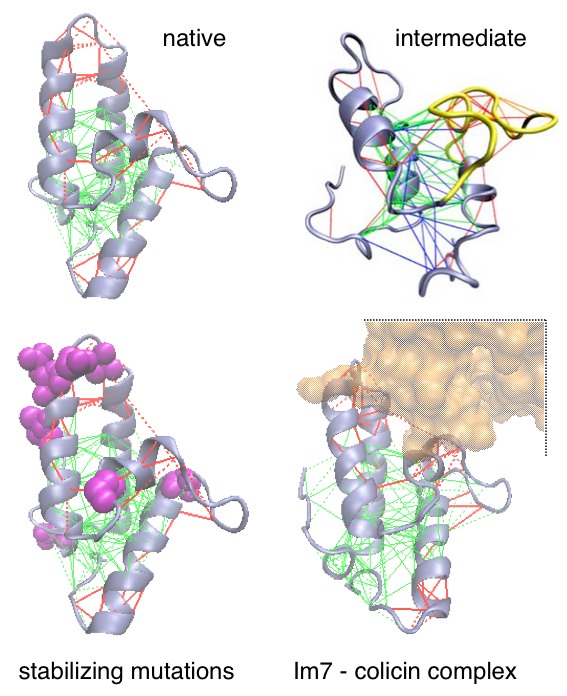}
	\caption{ Non-native intermediates highlight frustrated sites. Functional constraints for specific binding constrain the polypeptide sequences and can give rise to energetic conflicts with the rest of the polymer. Im7 is an example in which smooth folding is compromised by function, leading to local frustration. In the top panel is shown the average structure of the native protein where highly frustrated interactions are marked with red lines. Traps with non-native structure give rise to a folding intermediate, which is observed both in experiment and simulations. The intermediate is stabilized by favorable non-native interactions formed by residues that, in the native form, cluster on the surface and form a locally frustrated region. Minimally frustrated contacts present in the native structure (green) are distinguished from those that are nonnative (blue). A distinct nonnative cluster can be observed involving interactions between helix IV and the helix IÐII region. Native (red) and nonnative (orange) frustrated contacts surround the core. Random mutagenesis identified sites where substitutions increase the thermodynamic stability (purple). Many of the residues identified as forming non-native contacts during the early stages of folding of Im7 lie in regions that have a functional role: they are involved in the recognition and inactivation of colicin toxins forming the interaction interface with them. The observation of clusters of frustrated interactions in the native states points the way to a general mutational strategy to reduce the ruggedness of real protein folding landscapes.}	
	\label{fig:im7}
\end{figure}
\medskip

Immunity Protein 7 (Im7) is an 86-residue-long protein that in vitro folds through an on-pathway intermediate state. Kinetic studies suggested that this intermediate contains some non-native interactions of three of the four helices around a specific hydrophobic core (\cite{Capaldi:2002fk}). This exception to the behavior expected for minimally frustrated heteropolymers hints that a functional constraint may be conflicting with rapid folding. Does frustration change the way the Im7 intermediate is formed? Sutto et al explored this question using simulations that include native bias between residues close in sequence but that allowed for non-native (and possibly frustrated) interactions between residues farther apart (\cite{Sutto:2007dq}). These simulations predict an intermediate state stabilized by favorable non-native interactions formed by residues that, in the native form, cluster on the surface to form a locally frustrated region (Fig \ref{fig:im7}). Further mapping of the structural transitions using protein engineering combined with kinetic folding experiments and restrained molecular dynamics characterized more completely the details of the basins and bottlenecks, and confirm that indeed the energy landscape of Im7 is unusually rugged (\cite{Friel:2009kx}). Rather than folding by progressively increasing the number of native contacts, the sequence of Im7 contains other signals that constrain efficient folding. Many of the residues identified as forming non-native contacts during the early stages of folding of Im7 lie in regions that have a functional role: they are involved in the recognition and inactivation of colicin toxins forming the interaction interface with them. By coupling protease susceptibility and antibiotic resistance traits, Foit et al developed a novel assay to screen for protein stability {\it in vivo} (\cite{Foit:2009vn}). When random mutagenesis and selection for antibiotic resistance was conducted on the Im7 protein, most of the mutations that were found to enhance thermodynamic stability map again to the predicted binding surface as would be expected from the local frustration analysis (Fig \ref{fig:im7}). It appears that the functional constraint for specific binding restricts the sequences and gives rise to energetic conflicts with the rest of the polymer. Smooth folding of this small domain is compromised by frustration and in the case of Im7, traps with non-native structure give rise to an intermediate, which is observed both in experiment and simulations (\cite{Figueiredo:2013uq}, \cite{Whittaker:2011ys}). The observation of clusters of frustrated interactions in the native state points the way to a general mutational strategy to reduce the ruggedness of folding landscapes. The most effective approach to destabilize the intermediate of Im7 should be to stabilize the native state by alleviating the energetic conflicts in this region (\cite{Sutto:2007dq}). When redesigned mutants were simulated, the intermediate is no longer populated when residual native-state frustration is reduced. The redesign strategy based on minimizing native state frustration is more effective than specific negative design of the intermediate. A mutant specifically redesigned to destabilize the wild-type intermediate did not fold without an intermediate but rather retained an intermediate having a different structural ensemble, but nevertheless involving some non-native contacts (\cite{Sutto:2007dq}).
\subsection{Friction and roughness effects on rates}

\begin{figure}
\centering
	\includegraphics[width=0.9\textwidth]{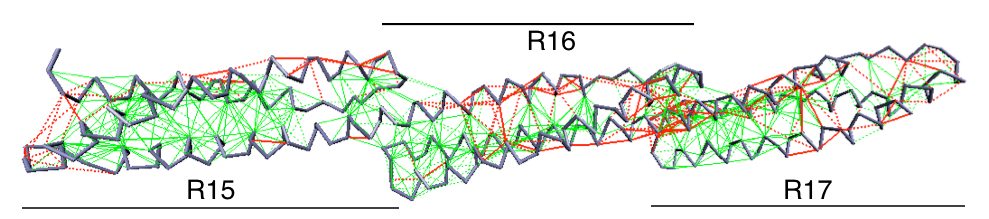}
	\caption{ Frustration can affect proteins' internal friction. Search in a rough energy landscape is affected by the population of local traps. When there are many traps these can be thought of as a source of friction on the diffusive folding. The structurally similar Spectrin repeats were shown to fold with very different rates (\cite{Wensley:2010ly}). The frustratograms of a natural protein that include these repeats is shown. R16 and R17 have a central cluster of highly frustrated residues and fold three orders of magnitude slower than does R15, which is more uniformly minimally frustrated. This central cluster of frustrated residues is involved in the main crossing from an initial transition state for folding to a later exit transition state for folding (\cite{Borgia:2012zr}). The frictional effect can be revealed by the viscosity dependence of the folding reactions, which can be studied by varying the solvent conditions and keeping the relative stabilities unchanged. For R15 the folding rate scales inversely to the solvent viscosity while for the frustrated systems R16 and R17 the folding rates are independent of solvent viscosity (\cite{Wensley:2010ly}).}	
	\label{fig:spectrins}
\end{figure}

The Im7 example shows that one of the effects of frustration is to introduce kinetic traps or non-native intermediates into the folding mechanism, a possibility first pointed out by Bryngelson and Wolynes (\cite{pmid3478708}).  If the system is small the number of trapping states will be small and can be described as arising from specific intermediates. The structural uniqueness of these intermediates is however in each case an open issue. As we saw in the Im7 example destabilizing one particular misfolded structure, generally just gives rise to an alternate trap, if frustration still remains in the native structure. What is the effect of these ensemble traps on the folding kinetics? In general searching through the traps slows folding down. Bryngelson and Wolynes showed how this slowing effect, especially where there are many traps, could be thought of as a source of ``friction'' on the diffusive progress of the protein as it is organized by the remaining guiding forces of the funnel. A very nice experimental system that seems to illustrate this idea is the extensive study of the folding of spectrin domains by Clarke (\cite{Wensley:2010ly}. The experiments of Clarke and coworkers have shown that structurally highly homologous members of the spectrin domain family fold with very different rates. The frustratograms of the three natural systems that have been studied are shown in Fig \ref{fig:spectrins}. R16 and R17 which have a central cluster of frustrated residues fold three orders of magnitude slower than does R15 which is more uniformly minimally frustrated. This central cluster of frustrated residues has been shown to be involved in the main step crossing from an initial transition state for folding to a later exit transition state for folding (\cite{Borgia:2012zr}, \cite{Wensley:2012ve}). As in other quasi-one dimensional proteins such as the repeat-proteins, this extended transition region entails what may be described either as a broad malleable transition state region or a high energy intermediate (\cite{Sanchez:2003qf}).

\medskip
\medskip

One of the effects of having localized frustration in the part of the protein that is being studied at this key kinetic bottleneck is thermodynamic--the transition state region is destabilized by the frustrated contacts and, by itself, this effect diminishes the folding rate. This thermodynamic effect might be said, with more precision, to be an effect of energetic heterogeneity rather than frustration per se since no alternate structure is involved. Yet a residual search or friction effect contributes to the slowing as well. This frictional effect of frustration is revealed by the viscosity dependence of the folding reactions which can be studied by varying the solvent, while keeping the stability unchanged. For the unfrustrated R15 the rate scales inversely to the solvent viscosity while for the frustrated systems R16 and R17 the folding rates are independent of solvent viscosity (\cite{Wensley:2010ly}).

Rather general arguments suggest that solvent viscosity effects depend on how much chain motion occurs as the transition state is traversed (\cite{Frauenfelder:1985bh}, \cite{portman2001microscopic}). On the smooth R15 landscape, large amounts of chain motion occur but when frustrated traps are present, as in R16 and R17, many individual smaller steps must be made searching through the misfolds. These smaller reconfiguration events displace little solvent. The idea that several misfolded configurations are involved in the frustrated system is supported by atomistic simulations of the R16 system that show explicitly helix sliding events coming from mismatches in the registry of the helices in this frustrated region (\cite{Best:2012cr}).

\section{Functional consequences of Frustration}

\subsection{Binding sites are often frustrated}

\begin{figure}
\centering
	\includegraphics[width=1\textwidth]{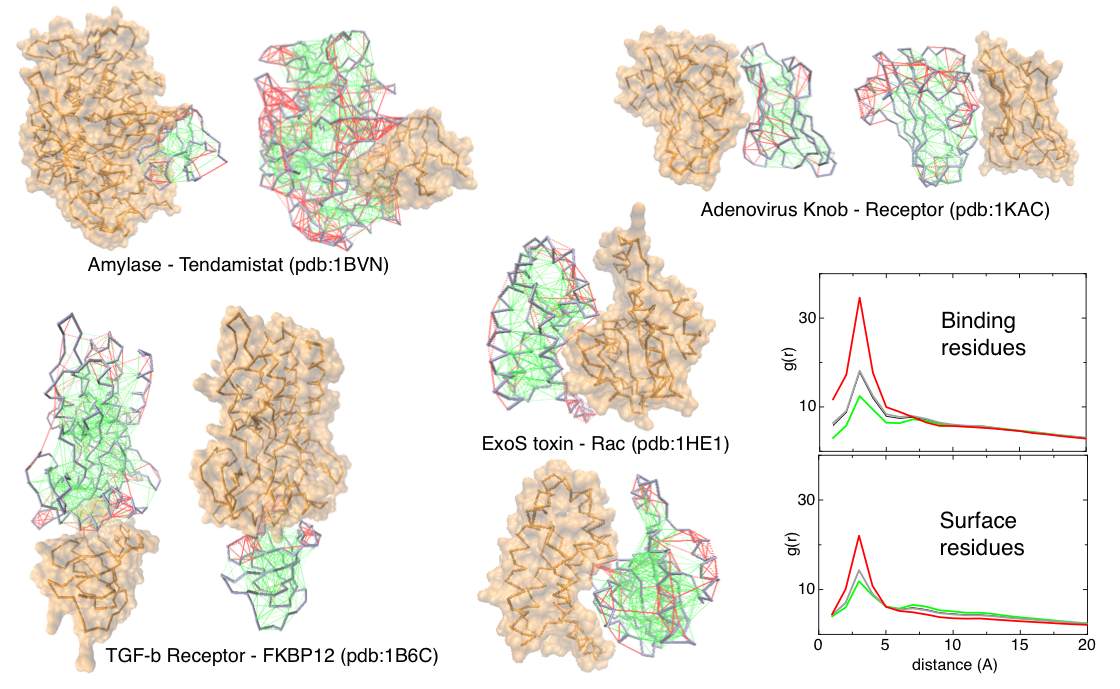}
	\caption{ Frustration guides protein-protein association. The structural coordinates of natural protein complexes were taken the pdb, and local frustration was calculated on the unbound monomers. The configurational frustration pattern of one ligand is shown together with the surface of the other. Minimally frustrated contacts are green, highly frustrated contacts in red, water-mediated interactions with dotted lines. A statistical survey was performed with a non-redundant sample of 85 complexes (\cite{Ferreiro:2007bh}). The pair distribution functions between the centers of mass of the contacts and the C$\alpha$ of either binding or non-binding surface residues is shown. Whereas many of the highly frustrated interactions are close to the binding residues as defined in the co-crystal structures, there are additional patches of frustrated interactions on the protein surfaces, suggesting there may be other functional reasons to retain these conflicting residues. }	
	\label{fig:binding1}
\end{figure}

Landscape theory approaches have already proven useful for describing protein -- protein interactions (\cite{Zheng:2012ys}). From the point of view of the localized frustration index, a natural question is whether and how often the residues that can conflict with folding of a single domain are there specifically to allow protein-protein association. Ferreiro {\it et al.} analyzed the distribution of the frustration index in protein assemblies on a curated set of 85 nonredundant hetero-dimeric protein complexes (\cite{Ferreiro:2007bh}). Fig \ref{fig:binding1} shows the frustration indices of the contacts and residues in some examples from the database. When the calculations of local frustration are performed in the unbound monomers, indeed one sees that near the binding sites the proteins are enriched in highly frustrated interactions (Fig \ref{fig:binding1}). The statistics over the whole database showed that the highly frustrated contacts both cluster and tend to be closer to the binding residues than they are to other surface residues. Minimally frustrated interactions are also particularly excluded from the vicinity of binding residues and neutral interactions follow the native topological distribution (\cite{Ferreiro:2007bh}). Interestingly, when the non-binding surface is examined, some clusters of highly frustrated interactions can still be detected (Fig \ref{fig:binding1}). Whereas many of the highly frustrated interactions are close to the binding residues as defined in the co-crystal structures, these additional frustrated patches on the protein surface suggest there may also be other functional reasons to retain conflicting residues. These reasons may be binding to other unknown molecular partners or, perhaps, these frustrated regions are sites relevant for allostery ({\it vide infra}). A corollary to this observation is that if the binding surfaces are too big with respect to the area of the domain, they are expected to enrich the total amount of frustrating interactions, compromising the folding of the unbound domain. It is interesting to note that coupling between folding and binding is more prevalent when the interaction surface is large (\cite{Papoian:2003jl}). 

\medskip
\medskip

	Changes in frustration upon binding may help guide specific association. A survey of the change in local frustration upon binding was performed by calculating the frustration indices for proteins in the complexes and comparing these to the frustration indices of the unbound monomers making up those complexes (\cite{Ferreiro:2007bh}). When calculated over a non-redundant set of hetero-oligomer complexes, the newly formed interactions at the binding interface have on average a local frustration distribution similar to that for contacts in single domains. This implies that specific association can as well be accounted by the principle of minimal frustration, as suggested earlier (\cite{Papoian:2003jl}). Most of the contacts strictly within the monomers do not change their frustration index upon binding, but approximately 25\% of the them do become less frustrated upon association. A minor fraction about 7\%, change in the opposite direction, becoming more frustrated. The decrement in frustration index is nearly constant and mostly comes from the change in the solvent accessibility upon binding, captured by the burial term and the contact type in the AMW energy function. In summary, the frustration index of the regions close to the binding site changes upon association, usually becoming less frustrated in the complex, and the newly formed interfaces have a similar distribution of local frustration to the interiors of unbound domains.
\subsection{Frustration in allostery and local conformational changes}

If the minimal frustration principle were to be satisfied everywhere, and the temperature sufficiently low, the protein molecule becomes a beautiful sculpture that would tend to move as a rigid body. For real proteins this {\it rigor mortis} cannot be the whole story, as the basic functions of proteins usually involve responding to changes in their environment. The detection and logical control capabilities of proteins rely on their ability to amplify the effect of a relatively small environmental change so as to produce a large discrete transition in a different chemical realm. Monod emphasized how the apparent gratuity of such ``transductions'' provide the essential computational capabilities of protein molecules (\cite{monod1973hasard}). We can thus expect that the minimal frustration principle will be at least locally violated by natural functional proteins. Local violations open possibilities for more complex energy landscapes, such as those needed for allostery and large-scale conformational changes. 

\begin{figure}
\centering
	\includegraphics[width=0.8\textwidth]{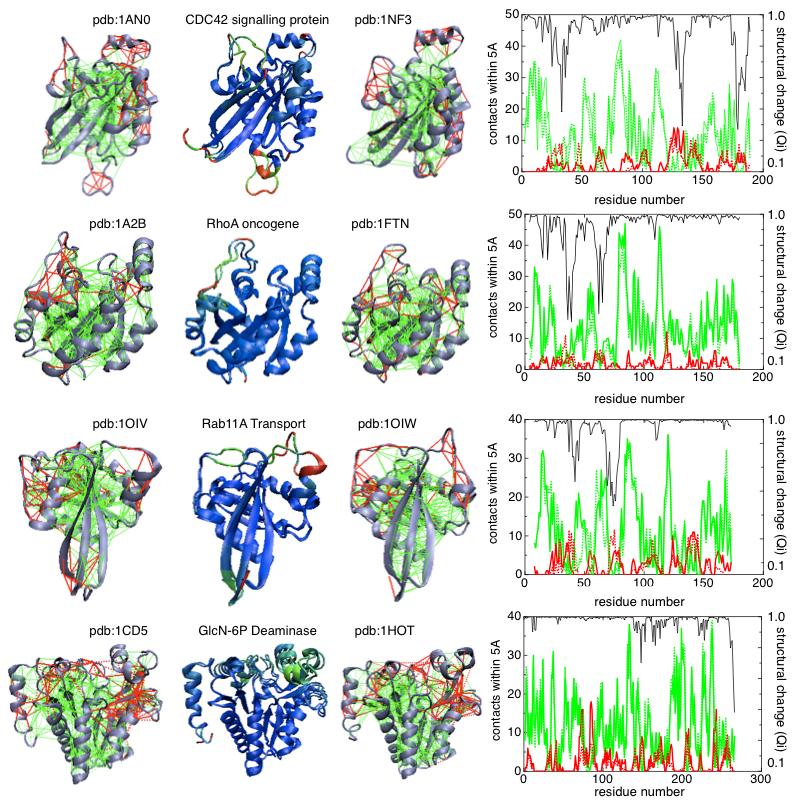}
	\caption{ Local frustration in allosteric domains. Proteins can change their structural forms responding to signals in their environment; some have been crystallized in two different conformations. Examples of structural alignment of experimentally determined conformations is shown above, colored according to the structural deviation (blue low, red high). The frustratograms for the individual conformations are shown at the sides, with the minimally frustrated interactions in green lines, the highly frustrated interactions in red lines. At far right a quantification of the local frustration projected on the linear sequence of the protein, minimally frustrated interactions (green) or highly frustrated interactions (red) in the vicinity of each residue, in either forms (solid or dashed). The structural deviation between both states is quantified by a local $Q_i$ score. Clusters of high local frustration colocate with residues whose local environment shifts between the two structures. Redrawn with permission from (\cite{Ferreiro:2011nx}) }	
	\label{fig:allostery1}
\end{figure}

A statistical survey of the local frustration patterns of allosteric protein domains indeed does show that the regions that reconfigure are often enriched in patches of highly frustrated interactions (\cite{Ferreiro:2011nx}). In contrast, the more rigid parts of the proteins, which are locally structurally superimposable in both forms, are connected by a dense web of minimally frustrated interactions (Fig \ref{fig:allostery1}). Regions that are highly frustrated typically are where the action is and are seen to reconfigure locally between the two different forms. In some cases, frustrated regions display rather extensive reconfigurations of compact regions, while in others, the frustrated clusters localize around apparent pivot points joining the more rigid elements. In general, clusters of high local frustration colocate with residues whose local environment shifts between the two structures (see Fig \ref{fig:allostery1}). It is apparent that the major structural ``core" is conserved between the large scale taxonomic conformational substates, and differences between them tend to localize on the protein surface. The changing frustrated regions often include the binding site of an allosteric effector. While conformational changes enter, keeping around 80\% of the pairwise interactions, 10\% of the interactions are exclusively found in one conformational substate or the other. The distribution of local frustration for contacts that are common to both structures is similar to the distribution for contacts exclusively found in only one substate. The interactions that are found in both substates do not dramatically change, but instead new contacts are formed in one substate that do not exist in the other. Thus the conformational change is brought about by a balanced exchange of local frustration upon switching substates. In surveying several allosteric domains there are examples in which only one structure of a pair has a relevant frustrated cluster but where the frustration is relieved in the other conformation (\cite{Ferreiro:2011nx}). In other cases, both structures display highly frustrated interactions in the same region, and these residues interchange the interactions amongst each other. There are no examples of structure pairs of monomeric proteins where the locally displaced regions are completely minimally frustrated in both configurations. 

\medskip
\medskip

\begin{figure}
\centering
	\includegraphics[width=0.4\textwidth]{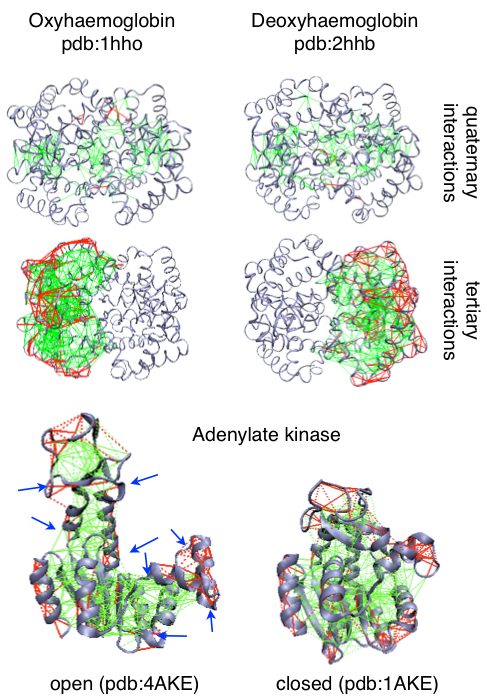}
	\caption{ Frustration in prototypic allosteric proteins. Haemoglobin is a tetrameric protein in which near rigid-body rotation allows symmetrical packing of subunits to occur. In this multimeric protein the interface between the subunits is not frustrated in either form. Highly frustrated interactions do occur internally in each subunit near the heme binding site. This is a classical example of a Wyman-Monod view of symmetry leading to near degeneracy, for which allostery does not depend on frustration. Below, the case of adenylate kinase, a protein that undergoes a large-scale conformational transition upon binding substrates. Steered molecular dynamics identified locations where ``hinge'' motions are believed to occur, shown here with blue arrows (\cite{Henzler-Wildman:2007kl}). An extensive minimally frustrated network of contacts rigidifies the molecule in the closed form, and highly frustrated regions co-locate with the hinges. Motion of adenylate kinase along the low-frequency normal modes contributing to the closure accumulates stress in some regions (\cite{Miyashita:2003tg}). A high-stress region can ``crack" or locally unfold releasing the strain and catalyzing the motion. This region is highly frustrated in both forms. The presence of interactions that conflict with folding an enzyme is a general theme in the realization of effective catalysts. Redrawn with permission from (\cite{Ferreiro:2011nx})}
	\label{fig:allostery2}
\end{figure}

The term allostery was in fact first used only to describe the regulation of protein activity through changes in quaternary structure of multisubunit complexes (\cite{Changeux:2013nx}). Landscape theory underscores how preserving symmetry is an easy path for achieving low energy near degenerate conformations, an idea taken as axiomatic by Monod, Wyman and Changeux. Using the same counting argument that makes symmetric structures individually more easily designable, pairs of symmetric structures provide a statistically favored way of achieving a switch-like behavior. The local frustration distribution in the multimeric forms of the allosteric proteins show that, on average, interfacial (quaternary) interactions are less frustrated than the monomeric (tertiary) interactions (\cite{Ferreiro:2011nx}). The interfaces in multimers appear in fact thus to be depleted of highly frustrated interactions, suggesting that the symmetry is reflected in the local frustration patterns when alternate configurations are used as decoys. The primal example of allosterism, hemoglobin is shown in Fig. \ref{fig:allostery2}. Here the interface between the subunits is not frustrated in either form, rather a rigid-body rotation allows symmetrically related interactions to occur. These distant structural forms involve a different knob-in-hole packing of the helices of one monomer on the others, and both packings appear to be minimally frustrated (Fig \ref{fig:allostery2}). Multiple funnels to structurally distinct low-free-energy states can be achieved with nearly rigid subunits packing in a number of symmetrically equivalent ways. In this sense, the natural ``engineering'' of frustration may be not essential for allostery. The classic Wyman-Monod view of symmetry of multimeric assemblies leading to near degeneracy does not depend on frustration at all. Because this usually involves only a few $k_b T$ of free energy, weakly assembled complexes may accidentally exhibit allostery without any strong frustration, as recently argued by Kuriyan and Eisenberg (\cite{Kuriyan:2007oq}).

\medskip
\medskip

	The precise mechanisms by which localized frustration allows allostery clearly deserves further investigation. Local frustration may simply allow a small discrete set of configurations interconverting by local motions. The allosteric dynamical transition may then involve specifically defined paths of transformation much like a small molecule or a macroscopic machine with hinges. Alternatively, the local frustration may destabilize a part of the protein in favor of an ensemble of rather high entropy. In other words, a local region can locally unfold or ``crack''. This possibility has been entertained previously (\cite{Miyashita:2003tg}) and elegantly explains the observations of the way denaturants can catalyze conformational changes (\cite{Zhang:1997hc}). Resolving experimentally the issue of hinges versus cracks requires an analysis of the effect that local mutations exert on the conformational kinetics (\cite{Whitford:2008qa}). 

Adenylate kinase is celebrated example of large-scale conformational change related to a functional transition. The gross opening and closing of this protein requires at least two reaction coordinates to be taken into account. Steered molecular dynamics calculations have identified several locations where ``hinge'' motions are believed to occur (\cite{Henzler-Wildman:2007kl}). The frustratograms of both substates are shown in Fig. \ref{fig:allostery2}. Nearly rigid-body motions of the lid and the core are involved in the closure. An extensive minimally frustrated network of contacts rigidifies the molecule in the closed form, and the places identified as hinges indeed co-locate with highly frustrated regions. Miyashita {\it et al.} have examined the motion of adenylate kinase along the low-frequency normal modes contributing to closure and concluded that there were regions where high stress accumulates at the transition state for conformational change (\cite{Miyashita:2003tg}). This high-stress region can however crack or locally unfold releasing the strain and catalyzing the motion. This high-strain region is highly frustrated in both forms. The presence of these interactions that conflict with folding the enzyme may be a general theme in the design of effective catalysts.

Protein phosphorylation is a paramount example of allosteric control of enzymatic activity. The activation of protein kinases usually occurs in a sharp switch-like fashion upon phosphorylation, making the ``kinome'' a rich source of computational cellular complexity where recurring networks can be constructed. These days particular attention has come to some members of the multitude of kinases, because mutations of kinases have been found to be associated with pathological cell growth. In some cancers the activating oncogene involves substitutions on ABL and EGFR kinases quite often. Oncogenic mutations perturb the minimally frustrated core of these proteins, while the locally frustrated clusters appear related to highly dynamic regions that are involved in allosteric activation (\cite{Dixit:2011ij}). In particular, the cancer-related mutations were found to both amplify dynamics of the inactive state and relieve local frustration in the active state (\cite{Dixit:2011ij}). Thus, the redistribution of local frustration may facilitate conformational transitions and at the same time change the relative stability of the substates.  

\subsection{Frustration in metastable and multistable proteins}

The universe of conformational substates of any protein molecule is huge. Therefore the complete energy landscape of proteins is complex, yet many experiments at physiological conditions can be interpreted using a simpler free energy spectrum of states, which can be considered as ensembles of low temperature substates (\cite{pmid19436496}). The low free energy part of the spectrum of excitations corresponds with partially folded ensembles, that get naturally populated as a consequence of the hierarchical nature of the mostly funneled landscape. Often the properties of these excited states can be predicted by native topology based models (see above). As we have seen, frustration, the conflict between inconsistent stabilizing interactions evolved for other purposes than optimizing folding, offers another mechanism for forming low free energy excitations. If these states have large structural differences, then the interactions that stabilize one form are expected to destabilize the other, making natural proteins metastable in certain conditions. Given that minimal frustration corresponds to folding stability, does high local frustration correspond to metastability? 

\medskip
	\begin{figure}
\centering
	\includegraphics[width=0.6\textwidth]{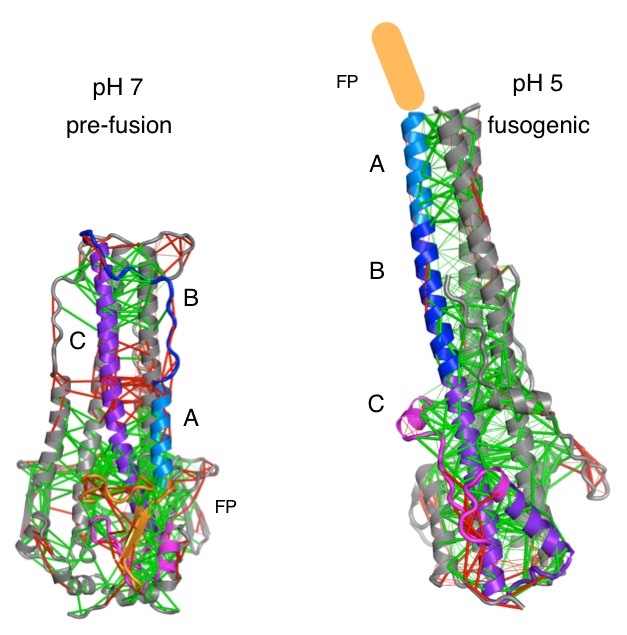}
	\caption{ Local frustration in metastable proteins. Hemagglutinin (HA) is a viral protein that undergoes a dramatic conformational change in response to an environmental pH change. The figure shows the local frustration patterns of the HA molecule crystallized in different conditions. The protein is a trimer that protrudes from the membrane of coated viruses, rearranging so that parts of it move by as much as 100 $\AA$, labeled here A,B,C and colored on a monomer to ease visualization. Several regions of high frustration are identified at pH7 (red lines). The local frustration of these region changes upon rearrangement and is diminished overall in the structure at pH 5. The regions involved in the major conformational change are already destabilized in the pre-fusion state and are more likely to ``crack''. There is also a region of high local frustration of the fusion peptide (FP, orange) in the pH 7 structure that conflicts with the core of the HA trimer. This peptide interacts with the membrane of the host endosome upon triggering by pH. }	
	\label{fig:haemaglut}
\end{figure}
\medskip

	Hemagglutinin (HA) is a fascinating example of metastability. The process of viral infection starts with the interaction of multiple HA molecules on the infecting virion interacting with multiple sialic acids on the cell to be infected (\cite{Luo:2012bs}). This interaction triggers endocytosis, and the virion is transported to the endosome where the pH is 1.5 - 2 pH units lower than outside the cell. At this lower pH, the HA molecule undergoes an irreversible and dramatic conformational change that was first mapped-out by Wiley and co-workers in the early 1990's (\cite{Bullough:1994fv}). The molecule rearranges so that parts of it move by as much as 100 $\AA$  from one structure to the other. The local frustration patterns of the HA molecule crystallized at pH 7 show several regions of high frustration (Fig. \ref{fig:haemaglut}). The frustration of these region changes upon rearrangement and is diminished in the structure crystallized at pH 5.5. Thus, the regions that are involved in the conformational change are already destabilized by local interactions, and are more likely to ``crack'', as in the cases of allostery describe above. Another interesting feature of these frustratograms is the high local frustration of the ``fusion peptide'' in the pH 7 structure (Fig. \ref{fig:haemaglut}). This peptide is poised to interact with the membrane of the endosome upon triggering, and promote the fusion of the virion membrane with that of the infected cell's endosome. It is evident that when the peptide is occluded in the pre-transition structure, it conflicts with the core of the HA trimer. Presumably this structure can be held in place by the minimally frustrated network of interactions that the other residues of the core contribute. 
	
\begin{figure}
\centering
	\includegraphics[width=0.6\textwidth]{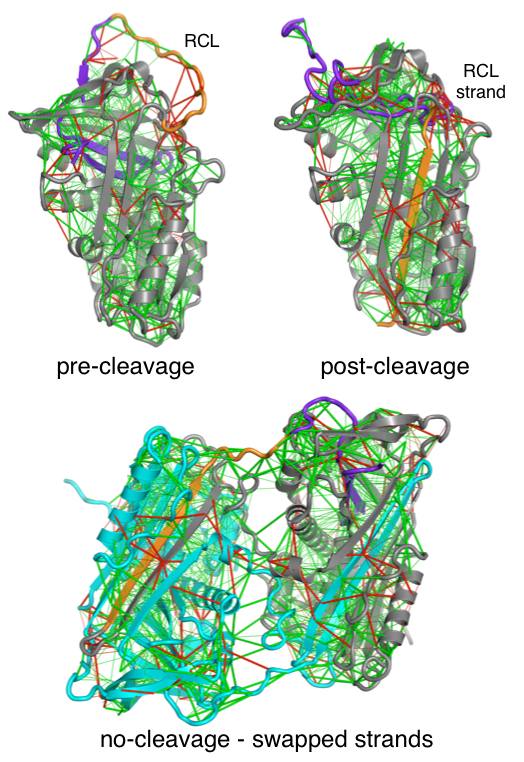}
	\caption{ Frustration in Serpin strand swapping. Metastable proteins can undergo large conformational changes upon backbone connectivity changes. When a protease cleaves the $\alpha$1-antitrypsin molecule in the RCL loop (top left, purple and orange) the whole molecule rearranges and this part of the polypeptide is found as a $\beta$-strand inserted in the central $\beta$-sheet (top right, orange). The local frustration patterns shows that the RCL is ``strained'' in the pre-cleavage form, where local destabilization promotes conformational change. Accommodating the new $\beta$-strand requires sub-optimal interactions among residues in the $\beta$-sheet region of the protein, which is also locally frustrated. Some serpins can spontaneously undergo a conformational rearrangement without RCL cleavage. The structure of antithrombin III (AT-III) strand-swapped dimer is shown at the bottom, with one monomer colored cyan and the other gray. In a classic domain-swap fashion, the polypeptides exchange part of the $\beta$-sheet with one another. The strand that gets swapped corresponds to the RCL region that can get cleaved. Regions of high local frustration in the dimer correspond with the regions that undergo rearrangements, the base of the swapped $\beta$-strand in addition to the $\alpha$-helix that is dislodge from the face of the $\beta$-sheet for strand insertion.}	
	\label{fig:serpin1}
\end{figure}

	Serpins are another large group of metastable proteins that undergo a dramatic conformational change when cleaved by target proteases (\cite{Huntington:2000dz}). The protease is itself strongly inhibited by the cleaved serpin. The cleavage occurs on an exposed amino-acid sequence in the serpin molecule (called the reactive-center loop (RCL), and the resulting cleaved loop becomes an additional $\beta$-strand in the central $\beta$-sheet (colored orange in Fig. \ref{fig:serpin1}). The uncleaved serpin is thought to be more stable than the cleaved form (\cite{Im:2000fu}). Part of the energy difference comes from the cleavage of a peptide bond in the RCL, but other energies are also implicated. The ``strain" of the metastable form of the serpin is evident in the frustatograms. The RCL itself is enriched in highly frustrated interactions. High mutational frustration is observed both before and after cleavage in the region of the structure where the conformational change occurs, again highlighting the role of local destabilization promoting conformational change. 	
	The structural ``openness" required to accommodate a newly inserted $\beta$-strand apparently requires sub-optimal interactions among residues in that region of the protein. There is also high local frustration in the $\alpha$-helix that stacks over the face of the central $\beta$-sheet (Figure \ref{fig:serpin1}). These regions have unusually rapid hydrogen/deuterium exchange dynamics (\cite{Tsutsui:2008kl}) and may remain metastable even in the cleaved form. Also it has been suggested that the frustrated helix may be easy to detach from the face of the $\beta$-sheet so as to initiate serpin aggregation. 
	
\medskip
\medskip

	Certain serpins (for example, plasminogen activator inhibitor-1 and antithrombin III (AT-III)) can spontaneously undergo a conformational rearrangement without RCL cleavage to form an inactive latent protein. Sometimes, ``beads-on-a-string-type'' structures have been observed in electron micrographs, suggesting that this latent form is an ordered oligomer.  In addition, mutations in some serpins promote the formation of highly stable, inactive polymers in the endoplasmic reticulum of liver cells. It was originally thought that the mutation might subvert the conformational flexibility required for the inhibitory activity of serpins, and that polymerization occurred through the insertion of the RCL of one serpin molecule into the $\beta$-sheet of another, and so on. However, Yamasaki {\it et al.} recently solved the structure of an antithrombin domain-swapped dimer, the structure of which revealed how any serpin adopting a similar configuration could form long-chain polymers. The domain swap involves a third of the  $\beta$-sheet, which forms an equivalent region in another molecule (Fig. \ref{fig:serpin1}). In Yamasaki and colleagues' dimer structure, the two molecules simply exchange part of the $\beta$-sheet with one another (a classic domain swap). The strand that swaps is the same strand that inserts during serpin maturation after cleavage (colored orange in Fig. \ref{fig:serpin1}). Examination of the frustration in the AT-III monomer both before and after thrombin cleavage reveals that as was seen in the $\alpha$1-antitrypsin, the base of the $\beta$-strand that swaps is frustrated (Fig. \ref{fig:serpin1}). In addition, the helix that is thought to have to dislodge from the face of the $\beta$-sheet in order to allow strand insertion and/or swapping is highly frustrated.  Comparison of the ATIII monomer before and after cleavage reveals that the observed frustration is present both before and after strand insertion hinting that this part of the protein can readily unfold even after strand insertion. A few highly frustrated contacts are observed in the center of the swapped strand, but it is not obvious why the strand swap is favored. While the ubiquity of domain swapping is a consequence of minimal frustration (\cite{Yang:2004bs}), a frustrated region of a protein can nevertheless unfold more readily, encouraging partial domain swapping if the protein contains frustrated elements. To form a polymer capable of incorporating more than two monomers, extensive domain swapping among multiple Serpin monomers is required, but it is readily apparent how this can occur from the structure (\cite{Yamasaki:2008qa}). The fact that the overlaying helix remains frustrated in all of the serpin forms suggests that domain swapping may occur even after serpin activation. 
	
\subsection{Frustration, dynamics, and catalysis} 

We have reviewed here how natural sequences are shaped via constraints imposed by the interaction of the protein with its immediate environment. These create conflicts in the folding of a polypeptide that can even compromise its foldability, become multistable or, in the extreme case, appear unfoldable, ``intrinsically disordered''. On the other hand, these frustrating interactions facilitate excursions out of the thermodynamic minima so that the polypeptide can explore low-lying excited states that can become fundamental for function. For enzymes, an obvious requirement for activity is, in addition to foldability, catalytic power, which in the cellular context must be jointly met with the requirement for regulation. Usually the regulation comes about through an allosteric site, allowing otherwise disconnected activities to become coupled, gratuitously, in functional space (\cite{monod1973hasard}). 

\medskip
	\begin{figure}
\centering
	\includegraphics[width=0.9\textwidth]{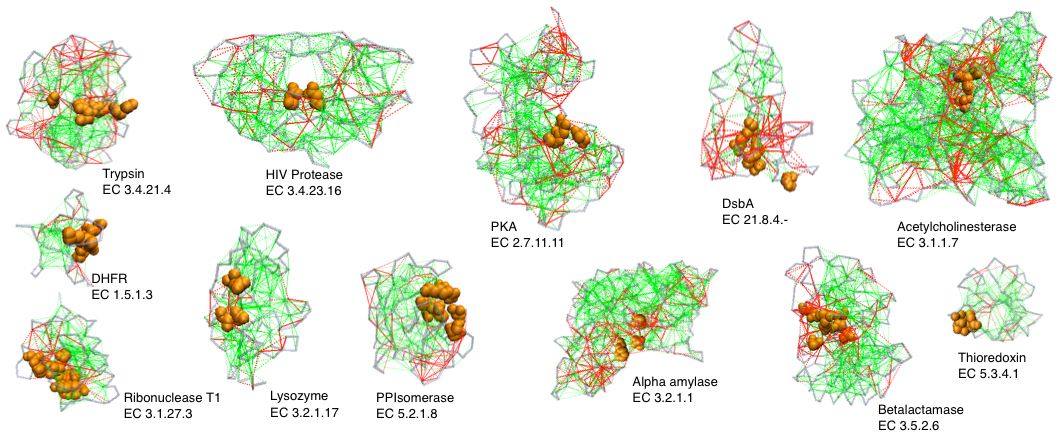}
	\caption{ Local frustration in enzymes. Catalytic sites are stringent, usually bringing together in space residues that would otherwise adopt different interactions. Examples of local frustration patterns in prototypic enzymes is shown. Orange spheres mark the catalytic residues annotated in the Catalytic Site Atlas (\cite{Porter:2004mi}). The catalytic residues are typically involved in highly frustrated interactions. Additionally the catalytic sites are often surrounded by a dense minimally frustrated network of interactions. It is apparent that large enzymes also contain other surface patches of highly frustrated interactions that could mark other functional sites, such as the binding of an allosteric effector. These general aspects of frustration in enzymes appear to hold irrespective of enzyme class or catalytic mechanism, as classified by the E.C. numbers.}	
	\label{fig:catsites}
\end{figure}
\medskip

Catalysis requires detailed chemistry. Reaction rates strongly depend on the precise arrangement of reacting groups at sub-$\AA$ resolution. Because of this, catalytic sites are generally stringently conserved, bringing together in space residues that would otherwise adopt different interactions. Thus in principle we would expect natural enzymes to have locally frustrated catalytic sites. Some examples of frustration patterns of enzymes are shown in Fig. \ref{fig:catsites}. In most cases the catalytic residues are involved in highly frustrated interactions either between them or with the rest of the molecule. Additionally however these sites are often surrounded by a dense minimally frustrated network of interactions, that could in principle be related to the requirement of second-shell interactions to be favorable enough to compensate for the frustrating catalytic groups keeping it in place. It is apparent that large enzymes also contain other surface patches of highly frustrated interactions that mark other functional sites, such as the binding of an allosteric effector (Fig. \ref{fig:catsites}). These general aspects of frustration in enzymes appear to hold irrespective of enzyme class or catalytic mechanism. This observation is consistent with the fact that mutations at catalytic sites often make the protein fold more stable (\cite{Shoichet:1995tw}, \cite{Meiering:1992pi}). Conversely, mutations that decrease the stability of folds are often associated with increased catalytic power, usually referred to as to stability-function trade-offs (\cite{Tokuriki:2008ye}). Many studies have noted that the evolution of new enzymatic activities is accompanied by a loss of protein stability. Also mutations that modulate activity are as destabilizing as any other random mutation (\cite{Tokuriki:2012ff}). On the other hand, substitutions in the substrate binding pockets compromise stability to a larger extent than surface mutations that underline neutral, non-adaptive evolutionary changes. It is suspected that new enzymatic activities arise with the help of compensatory stabilization from otherwise silent mutations (\cite{Tokuriki:2009lh}). These compensatory changes can be located in regions of the protein that were neutral to the stability of the original enzymes. The core of the protein would therefore act as a buffer for the accumulation of mutations in other regions. Enzymes are polypeptidic foldable systems for which many functional requirements simultaneously need to be met: bind substrate/s, catalyze, release product/s, bind effectors, switch on/off, evolve. Each of these activities may require compromise with robust folding, resulting in sites of higher local frustration. In the cases of the enzymes shown in Fig. \ref{fig:catsites} it is still not possible to deconvolute the contribution of each of these effects to the observed frustration patterns. Many of the activities in the catalytic mechanism overlap: the site where the substrate is bound is necessarily close to the catalytic residues. Thus we cannot distinguish whether the local frustration comes about just for thermodynamic binding, for kinetic catalysis or for both. Experimentally these differing causes could be distinguished by mapping perturbations that differentially affect catalysis and substrate recognition. Perhaps these effects can be computationally distinguished by analyzing the changes in energy distributions upon binding, transition state formation and conversion. Because of the sensitivity of chemistry to structural details, such a study requires fine-grained energy functions along with extensive sampling of conformational substates.

\begin{figure}
\centering
	\includegraphics[width=0.6\textwidth]{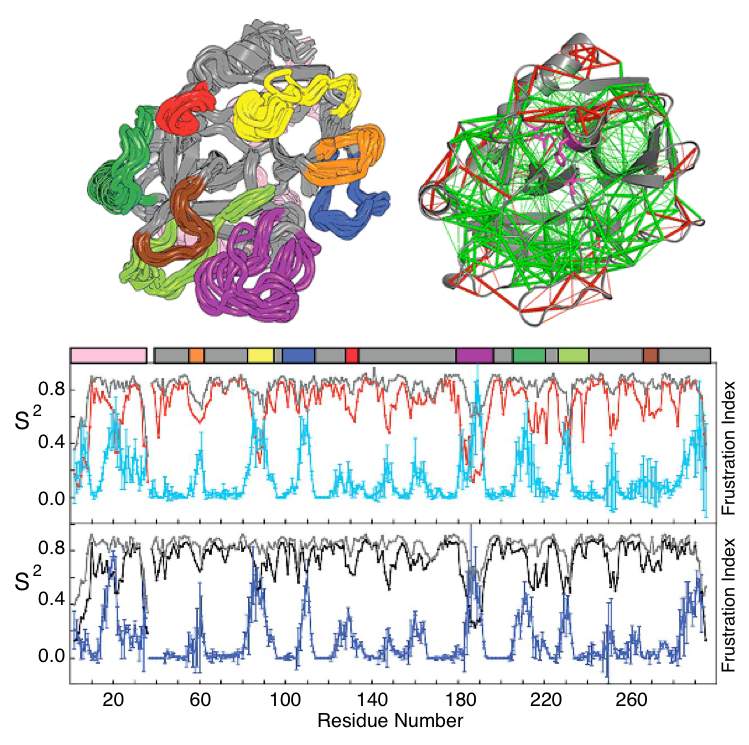}
	\caption{ Slow conformational dynamics and local frustration. Catalytic mechanisms require motion. The dynamics of the protease Thrombin change upon ligand binding. In particular the active site loops display motions over several orders of magnitude - from picoseconds to milliseconds time scales. The ensembles of structures can be visualized with experimentally calibrated molecular dynamics. An ensemble of the representative structures of this protein is shown at top left, with dynamic loops identified with different colors. The local frustration pattern of the lowest-energy structure from the ensemble is shown on the right. Minimally frustrated contacts are shown in green, highly frustrated ones in red, thin lines for water-mediated contacts. Below, a quantitative comparison of the order parameters $S^2$ reflecting nanosecond time scale motions $S^2_{ns}$ (gray) and longer time scale motions $S_{AMD}^2$ (red, black), with the local frustration distribution (cyan, blue). The average per residue fraction of highly frustrated contacts in the three lowest-energy structures is shown with error bars corresponding to 1 s.d. The results from the inhibitor-bound ensemble are shown in the lower panel, the structural ensemble of the free protein in the top panel. Redrawn with permission from \cite{Fuglestad:2013jl}.}	
	\label{fig:thrombin}
\end{figure}

A common denominator to any catalytic mechanism is the requirement for motion. The details of how enzyme dynamics affect the activity has aroused much controversy (\cite{Olsson:2006il}, \cite{Karplus:1983zt}). The regulation of enzymatic activity is perhaps less controversial than the catalytic acts themselves. An interesting example is Thrombin. Research on the coagulation protease Thrombin, has suggested that its activity is allosterically regulated by changes in its dynamics. The core serine protease domain structure typified by trypsin and chymotrypsin is highly conserved in thrombin. It consists of two interacting $\beta$-barrel structures. Compared to trypsin and chymotrypsin, however, thrombin contains insertions at several of the loops connecting the $\beta$-strands. These loop insertions create a bipartite substrate (fibrinogen) binding site where fibrinogen is bound at the catalytic site as well as at a second site on the backside of the protease domain termed anion binding exosite-1. This exosite is also the binding site of the allosteric regulator protein, thrombomodulin, and thrombomodulin binding alters the catalytic activity of thrombin towards a different substrate, protein C. Binding studies have demonstrated thermodynamic linkage between these two sites, but no differences are seen in the various bound structures determined by x-ray crystallography.  Evidently the large change in enzymatic activity comes about from changes in the dynamics of the enzyme, which arise from a rather malleable ensemble of conformations, a situation that has been termed ``entropic allostery''. Recently, a combination of NMR and MD studies on thrombin has revealed the large extent of changes in motions on binding, particularly in the active site loops, spanning time scales from picoseconds to milliseconds (\cite{Fuglestad:2013jl}). To capture the longer timescale motions, accelerated molecular dynamics protocols were implemented, from which the average residual dipolar coupling constants for the backbone NH groups of the Boltzmann-reweighted ensembles could be back calculated and compared with NMR observables. By ascertaining which acceleration level produced the ensemble that best matched the NMR data, it was possible to explore excited states that were otherwise inaccessible but in fact occur in solution. Accelerated-MD of thrombin has identified that the changes upon active site occupation occur in groups of residues linked through networks of correlated motions and physical contacts (\cite{Fuglestad:2013jl}). Upon active site ligation, loop motions are quenched, but new ones appear connecting the active site with distal sites where allosteric regulators bind. The analysis of local frustration revealed that all of these loops are highly frustrated in the mean native structure of thrombin (Fig \ref{fig:thrombin}). 

\medskip
\medskip

We can use the computed structural models of the excited states of thrombin to look for the changes in frustration along the ensembles. Both the inhibitor bound and apo thrombin have regions of high frustration in the surface loops (Fig \ref{fig:thrombin}). The regions of high frustration map to dynamic regions, as seen with NMR dynamic measurements obtained at different time-scales. On the one hand, the order parameters derived from conventional MD simulations agree with order parameters derived from NMR relaxations experiments proving motions in the nanosecond time regime. On the other hand, order parameters derived from the NMR-calibrated accelerated MD identify disorder resulting from motions on longer time scales, typically microsecond to milliseconds. The fast dynamics mapped in this system do not correspond well to the regions of high local frustration while the correspondence with the dynamics on longer timescales is remarkable (Fig \ref{fig:thrombin}). These results highlight the notion that regions of high local frustration are evolved features of the proteins that allow control of the motions on the slower time scales, which can occur with relative energetic ease. Energetically unfavorable interactions within the surface loops facilitate large amplitude motions, smoothing conversion between the ensembles of states needed for catalysis.

\subsection{Frustration and the initiation of aggregation}

Most of our biochemical illustrations of the frustration concept have focused on frustration within a single biomolecule or functional association, but what about frustration between biomolecules in the proteome that are not supposed to interact? Biomolecules are never alone during their life in the cell. Proteins are not even always born singly, but in prokaryotes especially are born in a litter -- they are synthesized in  "bursts" when several ribosomes gather to translate a single messenger RNA. During this burst the local protein concentration can be quite high. After being born and becoming first folded they will occasionally unfold and have to re-fold, again in the presence of neighbors, some of whom might be unfolded themselves. How do these unfolded proteins then avoid binding to the wrong partners? 
The principle of minimal frustration helps explain why even within the cell re-folding can be robust. (By the way it is obviously not always the case under all thermodynamic conditions-we can scramble eggs and do not expect them to unscramble when they cool!) The lack of frustration in finding partners is just what would be expected if achieving minimal frustration is a result of evolution. While the native contacts within a protein are quite stable, random contacts between two different proteins will not be as stable, so it will be rare for proteins with distinct sequences to be able to interact strongly with each other. Transient aggregates can unfold and eventually settle in their proper natively folded states. Clearly the dangerous partners that a protein must avoid are other copies of itself: native-like interactions between distinct partners can lead to association. This is called "domain swapping" and if it goes on sequentially indeed an aggregate can form, as we have seen in the Serpin example (Fig \ref{fig:serpin1}).

\medskip
	\begin{figure}
\centering
	\includegraphics[width=0.7\textwidth]{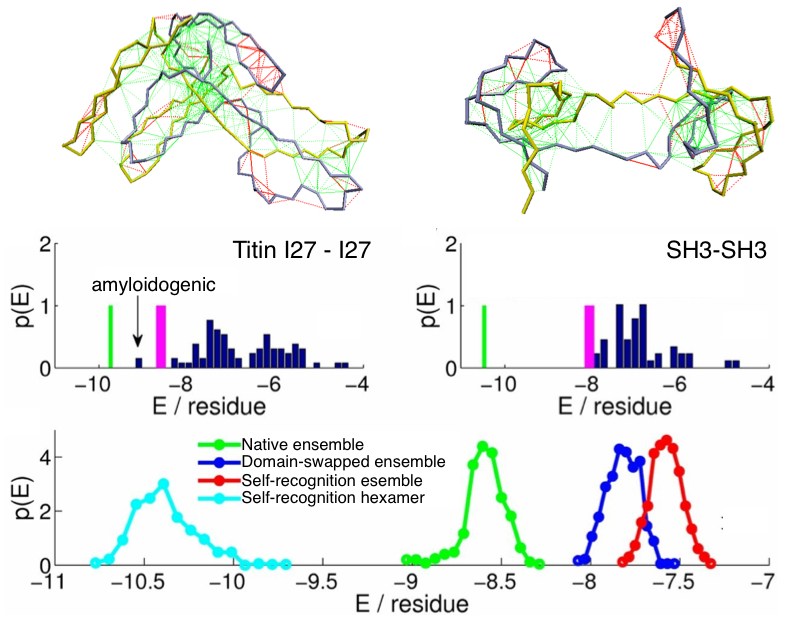}
	\caption{ Frustration and the initiation of aggregation. Self-recognition of short sequences in two copies of the same protein sequence can lead to native-like interactions, resembling the first steps in the formation of an amyloid. The folding of tandem copies of TitinI27 or SH3 were simulated using AWSEM (\cite{Zheng:2013zr}). Besides native-like structures, misfolded structures with a significant amount of self-recognition interactions were found. The different levels of frustration in the tertiary contacts is illustrated in two representative structures for I27 dimer (left), SH3 dimer (right), with the backbone of the sub-domains in blue and yellow. The swapped contacts formed at the interface of SH3-SH3 are minimally frustrated, as expected from the principle of minimal frustration. The self-recognition contacts formed at the domain interface of I27-I27 are also minimally frustrated, indicating that these contacts are stronger than random contacts. Below, comparisons of the stability (energy per residue) among various structures and ensembles of thermally sampled structures of I27 (left) and SH3 (right). The stability for the native monomeric structure in shown by a green vertical line, the strongest non-native hexapeptide pairing in magenta. Misfolded configurations are typically less stable than the native structure, indicating that misfolding by inappropriate pairing of strands will be unlikely in folding of the monomers for both I27 and SH3. Blue bars show the distribution of the stability of all of the self-recognition hexapeptide pairs, calculated with the AWSEM-Amylometer based on the energies of a $\beta$-sheet formed from the hexapeptide (\cite{Zheng:2013zr}). If the stability of the strongest self hexapeptides pair is competitive with the native structure, as in the case of I27-I27, the particular self-pair is responsible for the misfolding of the fused protein and can trigger further aggregation. For SH3-SH3, the fused protein folds as well as the monomer, because all self hexapeptides pairings are weaker than the most stable nonnative hexapeptides pairing in the monomer. At low temperature, folding to the native state is robust, but the amyloid-like structures can act as kinetic traps. At higher temperature their disordered parts give them an entropic stability advantage. They can act in the initial states for further aggregation. Frustration can be manifested more strongly in partially disordered ensembles and may not always be apparent in the final crystallographic structure. These examples show that looking at pair-level frustration as in typical frustratograms may not reveal all situations where energetic frustration can show up. Redrawn with permission from \cite{Zheng:2013zr}.}	
	\label{fig:aggregation}
\end{figure}
\medskip

Domain swapping indeed does occur and seems to play a role in the initiation of some aggregation processes. The possibility of misfolding via domain swapping is a symmetry effect, the possibility of which cannot be easily eliminated by evolution. Simply symmetrizing the interactions of a structure-based pure funnel model for a monomer allows one to predict possible domain swapped dimers (\cite{Yang:2004bs}). Different dimers differ in stability primarily through topological effects that determine the fraction of contacts that are prevented from being made in the more crowded dimer situation. Transient domain swapping seems to be a universal phenomenon when a protein is present at high concentrations but transient domain swapping resolves itself, suggesting the monomeric native form still has at least an entropic advantage at the concentrations found in most cells. A problem can occur if the local concentration is high. One of the worst cases of high concentration is in multi-domain tandem repeat proteins, where the covalent connection means two identical domains can never get far away from each other. Evolution does seem to take notice. Han {\it et al.} have shown by surveying multi-domain sequences in several genomes that in such multidomain repeat proteins the nearest-neighboring domains, even if structurally similar, are very distinct in sequence, while more sequentially distant domains often can still have pretty high sequence identity (\cite{Han:2007gb}).

Domain swapping is likely to be a major reason aggregates formed in disease states usually involve a single protein. Zheng {\it et al.} computationally studied the misfolding of dimers of domains in titin (\cite{Zheng:2013zr}). This system had been experimentally studied in the laboratory by Clarke and provided support for the ideas about domain swapping found in the bioinformatic survey (\cite{Wright:2005mb}). This simulation study showed there was another source of misfolding--the "self-recognition" of short sequences in two copies of the same protein. Fig. \ref{fig:aggregation} shows both the domain swapped dimer and a misfolded dimer with the self-recognizing segments paired up. This self-recognition resembles the first steps in the formation of an amyloid. Indeed the self-recognizing fragment is recognized by software trained to recognize amyloidogenic fragments bioinformatically. In Fig. \ref{fig:aggregation} we show histograms of the energies per residue of hexapeptide fragments in the I27 fused dimer system along with the native energy. The amyloid fragment comes perilously close to competing with the native structure locally. But it does not compete globally. A histogram of the global energies of native structures, self-recognizing structures and domain swapped structures are shown in Fig.\ref{fig:aggregation}. Here the native structure clearly wins. At low temperature folding should be fine but the amyloid-like structures can act as kinetic traps at higher temperature due to their disordered parts giving them entropic stability (\cite{Zheng:2013zr}). Globally the energy landscape is a funnel but still misfolding can give a kinetic trap, which can act as the initial state of further aggregation. We see that frustration can manifest itself more strongly in partially disordered ensembles even when it may not be always apparent in the final crystal structure. This example shows that looking at pair-level frustration as in typical frustratograms does not reveal all situations where energetic frustration can show up.

Pair level frustration in the monomer can however encourage aggregation. An example is insulin. The metastability of insulin is revealed in its natural tendency to self-assemble into oligomers, a feature that was first appreciated when the structure was determined in the 1960s ({\cite{adams1969structure}, \cite{Baker:1988cq}) . The insulin monomer contains two polypeptide chains, an A chain of 21 amino acids and a B chain of 30 amino acids, which contain helices between residues A1 and A8, A13 and A19, and B9 and B19. The two chains are covalently linked by disulfide bridges between A7 and B7 and A20 and B19. Identical hydrophobic surfaces of two monomers come together forming an antiparallel beta-strand between residues B40 and B46 of each monomer. In the crystal, insulin is a hexamer consisting of three dimers aggregated around two zinc ions forming a soluble, globular protein structure (\cite{Baker:1988cq}). The hexamer has been seen to adopt one of three conformational states known as T6, T3R3, and R6, depending on the arrangement of the first eight residues of the B chain. In the T state (Fig. \ref{fig:insulin}, PDB code 3P2X; residues B9-B12 are colored orange and residues B16-B19 are colored yellow), residues B9-B12 adopt an elongated conformation, while in the R state, PDB code: 1MPJ; same coloring scheme), they form part of the B-chain helix. Residues B9-B12 do not appear to be making many contacts with the rest of the insulin molecule, as has been suggested by NMR studies on the T-state (\cite{Hua:2008kh}). The A-chain is highly frustrated in all structures, and one might surmise that it is mainly held-together by the disulfide bonds it makes with the B-chain.
	Researchers have identified a single residue, a proline at position B28, which is largely responsible for making favorable contacts with residues B20-B23 of the adjacent monomer. Experiments with B chain C terminus substitutions and truncations have shown that residues B26-B30 are essential for dimer formation, and single substitutions at residue B28 markedly increase the tendence to form monomeric units upon dilution (\cite{Brems:1992rq}). We explored the residual local frustration in the B28 Pro -- Asp mutant, which crystallized in an R6-like conformation (PDB code: 1ZEH). Plots of the fraction of contacts that are either minimally or maximally frustrated show that the region near the site of mutation is highly frustrated ((Fig. \ref{fig:insulin}b). Close examination of this region reveals that the Asp makes inter-chain electrostatic contacts that form an extended backbone structure, however, these contacts are frustrated. The increased local frustration is likely the cause of the increased conformational flexibility in the B chain C terminus observed in this mutant (\cite{Whittingham:1998fc}). It is interesting to speculate that the frustration might partly be ameliorated by the binding of small aromatic molecules such as phenol or cresol, which bind in this region of the protein, and may provide stabilization of the R6 conformation. 

\medskip
	\begin{figure}
\centering
	\includegraphics[width=0.8\textwidth]{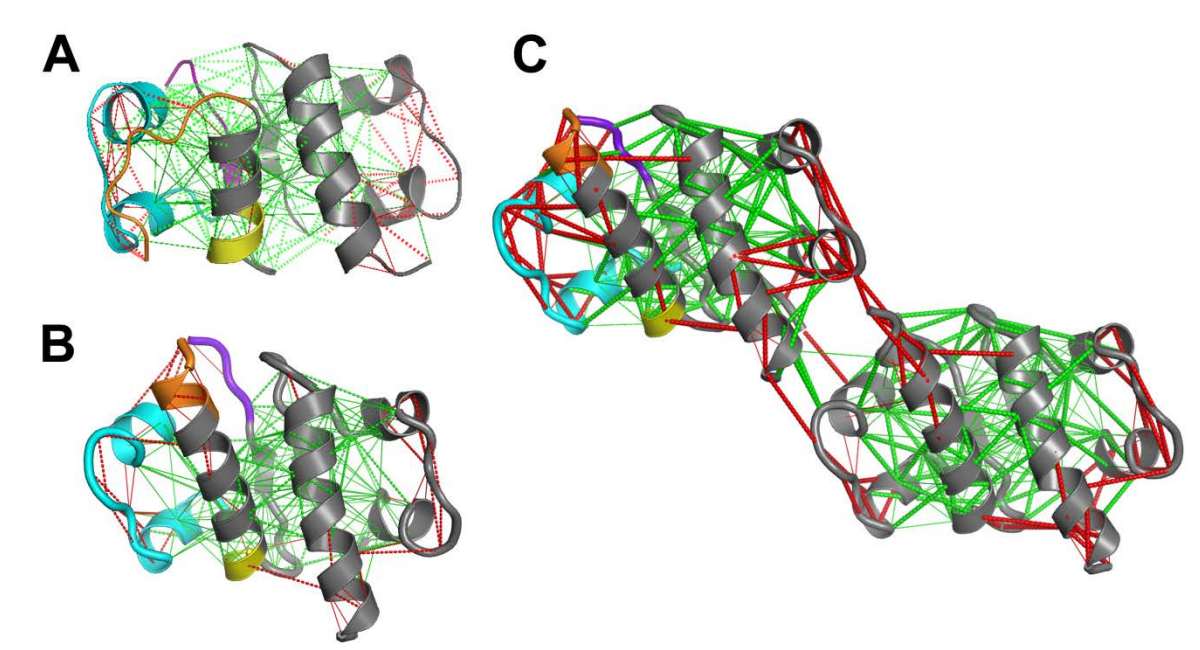}
	\caption{Frustration in a monomer can encourage aggregation. Insulin crystallizes in either the (A) T-state (PDB code 2P2X), exemplified by the loss helical structure in the N-terminal residues of the B-chain helix (colored orange) as compared to (B) the R-state (PDB code 1ZEH), which is stabilized by the Pro(B28) Ð Asp mutation and by binding of phenol or cresol molecules to pocket formed between the N-terminal residues of the B-chain helix and the the C-terminal residues of the B-chain (purple). Such structural maleability is typically indicative of frustration, and indeed insulin is highly frustrated. Insulin can, on occasion cause an amyloid disease and Weiss and colleagues managed to crystallize the fibrillar form (C) (\cite{Weiss:2013ss}). Interestingly, the molecular interactions observed in the crystal structure showed the participation of the A-chains (cyan in all structures), the most frustrated part of the insulin structure. The fibrillar form, however, appears to remain frustrated.}	
	\label{fig:insulin}
\end{figure}
\medskip

	Insulin can, on occasion cause an amyloid disease. This phenomenon was investigated recently by Weiss and colleagues who managed to crystallize the fibrillar form of insulin (\cite{Weiss:2013ss}). Interestingly, the molecular interactions observed in the crystal structure showed the participation of the A-chains, the most frustrated part of the insulin structure. No obvious alteration in the amount of minimally frustrated contacts were observed in the fibrillar form, however (Fig. \ref{fig:insulin}C).

\subsection{Frustration in nucleic acids and protein complexes}

	We have reviewed here why proteins are such a special kind of polymers: minimally frustrated globule-like structures can be encoded within an amino acid sequence. The richness of fold types is constrained at every level by the variety of the alphabet and the environmental conditions where the correct decoding occurs. Energy landscape theory shows that in the case of polypeptides the present 20-mer alphabet is more than sufficient to allow robust folding of a vast collection of topological families. But proteins are not the only polymers found in living organisms. How does the concept of frustration apply to other biological polymers? 
	
	In terms of dry mass, polysaccharides abound in cells. Some of them are simple linear concatenations of the same monosaccharide units (cellulose), others display branched covalent structures (glycogen, starch), and even more complex varieties are still being characterized. Admittedly more needs to be done with respect to their understanding. Yet it seems that most polysaccharides do not spontaneously fold to specific structures, but rather form disordered colloids. Although the variety of saccharide units is very high due to local modifications, the chemical ligation properties of the individual units is sufficiently uniform along the chain, so that the interactions between any two parts is as nearly (de)stabilizing as any other. In energy landscape terms, the energetic variance of alternative structures $\Delta E^2$ is small but the gap to the ground state appears to be nearly negligible, and a glass transition is expected to be predominant (Honey is a good example !). Here frustration is the rule that plagues the landscape with local minima. Not surprisingly specific synthesis and handling of even the most simple biological polysaccharides is challenging, and much could be gained if this were not the case (imagine making paper out of plain sugar !).
	
	Yet there is a special sort of polysaccharide that does spontaneously fold. Nucleic acids are essentially linear phospho-pentoses in which each unit is modified with a nitrogen-rich base, typically one out of four possibilities. In these polymers the unit-unit interaction modes are restricted, with electrostatics dominating repulsion and hydrophobic base-base interactions favoring association. The most stable structures are periodic packings of bases between two or more chains forming helices. Ever since the proposal of the antiparallel double-helix (\cite{watson1953molecular}), it has been evident that this structure is to a certain extent nearly independent of the sequence of the monomers, as long as specific base-pairing is maintained. This makes nucleic acids the prime molecular species where faithful templated polymerization can occur. Spontaneous formation of the double-helical structure is at the core of genetic engineering technologies, yet our present understanding of the folding landscape of DNA is limited. The melting temperature of short chains can be well approximated with simple empirical rules that take into account the unit composition and the saltiness of the solution (\cite{Schildkraut:1965ud}). This corresponds with a landscape in which the ``native'' structure can be reached with little competition of alternative states, with a large gap dominating the energetics. However longer chains display more complicated behaviors, with marked hysteresis, history and concentration dependent kinetics, all hints of glassiness of the landscape. It should be kept in mind that the differences in stabilization energy of specific base-pairing interactions is low compared to the mere stacking of the bases, and thus the discrimination between correct and incorrect base-pairing is expected to become more difficult as the chain grows longer. In contrast to proteins, nucleic acid folding is dominated by the formation of continuous secondary structure stretches, with small relative contribution of long-range tertiary interactions, which under ordinary conditions are discouraged by electrostatics. The apparent simplicity of the short oligonucleotide folding allows for engineering large assemblies by adjusting local rules (\cite{Rothemund:2006bd}). Indeed statistical mechanical theories of nucleic acid structure formation suggest that due to its quasi-one dimensional character the landscape is naturally hierarchical (\cite{Fernandez:1990kx}). 
	
	The most common source of frustration in nucleic acids is the competition between base-pairing continuous stretches of a single chain and between similar chains. The formation of alternative, metastable stem-loop structures was early suggested by Crick as a plausible mechanism for gene regulation (\cite{crick1971general}). In recent years single molecule spectroscopy has allowed for detailed characterization of the overall timescales in which such transitions occur, providing direct evidence of the existence of many collapsed intermediate structures (\cite{Ma:2007qc}). The free energy landscapes drawn out from these and other experiments portray pictures of overall smooth surfaces with discrete metastable traps (\cite{Woodside:2006mw}).The location of the traps and energy barriers separating them can be readily manipulated by adjusting the hairpin sequence, such that the population of folding intermediates can be partially controlled introducing mismatches and changing loop length. Frustration between these structures provides the grounds for engineering nucleic acid polymers that can act as switches, with ingenious applications envisioned (\cite{Viasnoff:2006if}).
	
	The natural occurrence of functional aspects of frustration in nucleic acids is best exemplified by RNA, not DNA. Small stereochemical restrictions in the dihedral angles in the backbone make the small alphabet of ribonucleotide sequences generally not sufficient to specify a set of secondary and tertiary interactions. Studies on the folding of catalytic RNAs (ribozymes) have highlighted the peculiar propensity of RNA to adopt alternate conformations in solution. This conformational heterogeneity is attributed to misfolded structures or long-lived folding intermediates and it is frequently observed in natural isolates. The high probability of even random RNA sequences to have fairly stable stretches of base-pairs makes the number of alternate structures significant. Still, the RNA folding complexity is reduced by its more hierarchical nature as compared to proteins. Typically secondary structures form fast and last for minutes to hours at physiological temperatures, while tertiary interactions are comparatively weak, interconverting in the timescale of milliseconds to seconds (\cite{Solomatin:2010sw}). This marked difference in stability and kinetics constrains the accessible tertiary structures. Conversely, isolated secondary structural elements of larger RNAs often fold properly in absence of tertiary interactions. This phenomenon underlies the successful prediction of RNA structure by two-dimensional mixing and matching of fragments, and also gives rise to deep folding intermediates separated by large barriers. These long-lived alternatives can be coopted to perform specific functions and should be treated as alternative ``native'' states. Such functional alternatives have been described in the autocatalytic Tetrahymena Group I intron, sarcin-ricin loop, the ribosome, and in vitro selected aptamers (\cite{Marek:2011kc}). Furthermore, the stability of these frustrated states can depend on the differential binding to other molecules, as exemplified in riboswitches (\cite{Serganov:2013lo}). In these examples, binding of a small ligand to a segment of mRNA is sufficient to affect a structural transition in the mRNA that in turn regulates the translation rate, making them effective modulators of protein abundance. 

	Frustration has an important functional role when proteins meet up with DNA. Some proteins display the remarkable ability to differentially recognize a specific stretch of DNA out of a vast excess of structurally similar sites. Transcription factors locate their cognate site in a genome with rates that appear to break the diffusion speed limit. Delbr$\ddot{u}$ck first envisioned a mechanism to explain this puzzle proposing that proteins may bind to any site on a linear chain of DNA and then slide along it, reducing the three dimensional search problem to one dimension. Since then the existence of such ``facilitated diffusion'' mechanisms has been confirmed in a variety of systems (\cite{Hippel:1989fh}). This mechanism requires the protein to have a relatively small variance of the sequence-dependent energy landscape of interaction, so that fast sliding can occur. Gerland and Hwa have analyzed this with the same mathematics employed by Bryngelson and Wolynes for folding proteins (\cite{Gerland:2002pr}). At the same time however a DNA binding protein must also display a relatively large thermodynamic stability at its target site. The requirement of both fast search and stability of the cognate protein-DNA complex impose conflicting constraints on the landscape: whereas rapid search requires a fairly smooth landscape, the stability of the protein-DNA complex requires low energy at the target site. Slutsky and Mirny (\cite{Slutsky:2004hs}) showed that this apparent paradox can be ameliorated if the protein occupies (at least) two distinct conformational states that differ in the variance of the binding energy distribution. In one state, the protein binds DNA sites with a small variance, thus it cannot strongly discriminate different sequences but allows sliding to occur rapidly. The other state is characterized by a large variance with greater differences in affinity for the various sequences. The differences in the free energy and the switching rate between the protein states defines an energy gap. In this model, the search time to stably bind a cognate site is optimized when the protein spends most of the time in the state that favors sliding but can undergo fast conformational transitions to the other state. In this scenario frustration in protein-DNA binding effectively creates a landscape with only a little ruggedness and a few alternate sites with low energies (\cite{Slutsky:2004hs}).

	Biochemical studies have highlighted the fact that DNA-binding proteins bind non-specifically to any DNA stretch (or any polycationic surface for that matter), with electrostatic interactions predominating the energetics. The surfaces of DNA binding proteins usually display positively charged patches that complement the negatively charged DNA backbone, providing a rationale for the lack of sequence specificity. On the other hand, the structures of several specific protein-DNA complexes show that these are stabilized by intimate contacts between residues and bases that could discriminate the subtle structural differences between different DNA sequences, mostly present at the bottom of the major groove (\cite{Norambuena:2010sh}). The differential ``'binding modes'' may thus be related to the formation of distinct interaction regions. A recent survey of the structures of DNA-binding proteins has found that some binding proteins utilize one interface to specifically recognize a cognate sequence while another region is employed to establish nonspecific associations, resembling two distinct active sites. In contrast, sometimes these two patches overlap (\cite{Marcovitz:2011xd}). The system now has to choose between these alternatives, frustrating the protein-DNA interaction. The free energy barrier between the binding modes can in principle be related to the overlap of these binding surfaces. MD simulations of the sliding/binding process shows that as this local frustration increases, the rate of switching between sliding and binding decreases (\cite{Marcovitz:2013le}). On the one hand high frustration within the protein smoothes the landscape for sliding, but it makes the switching to the specific binding mode slower, which slows down the overall target location rate. On the other hand, if frustration is lowered too much switching becomes rare and the protein often misses the cognate site. The situation encountered is reminiscent to the effect frustration has in folding a polypeptide chain: a little is helpful, lots can be disastrous. 
	
	The formation of a specific protein-DNA complex may be coupled to the folding of the protein. While protein flexibility is involved in induced-fit recognition and to a larger extent in intrinsically unstructured DNA binding proteins, the electrostatic field of the DNA can induce flexibility and a partial unfolding of the protein (\cite{Levy:2007kl}). Protein disorder facilitates the three dimensional diffusive search through the ``fly-casting'' mechanism where the protein binds DNA nonspecifically in a partially folded state and completes its folding when it binds the specific site (\cite{Shoemaker:2000rp}). Indeed, several DNA binding proteins are known to be partially unstructured in the unbound state that only fold upon binding to the target (\cite{Spolar:1994oa}). The interplay between fly-casting and electrostatics is observed even for weak electrostatic forces and is expected to vary with the electrostatic screening due to salt and the intrinsic folding barrier, both of which can be modulated. Increasing the strength of the electrostatic forces destabilizes the isolated protein due to frustration between general electrostatic interactions and the specific native interactions. In the presence of DNA, a larger destabilization effect is seen as a result of competition between folding and binding (\cite{Levy:2007kl}). This latter frustration reflects a ``tidal force'' from electrostatics tending to pull the molecule apart. Under strong electrostatic forces, the negatively charged DNA pulls the protein apart, and the protein-DNA association can therefore advance protein folding.
	
	Papillomavirus E2c protein is a remarkable model in which the effects of frustration in protein-DNA association have been experimentally studied. When confronted with a small double-stranded oligonucleotide the E2c protein finds its specific site using two parallel routes (\cite{Sanchez:2011xz}). In one of them the process is well described as a two-state reaction with a small activation barrier of entropic origin. The effect of single point mutations on the DNA-contacting surface to the kinetics and thermodynamics indicated that the formation of direct contacts between residues and bases constitute the bottleneck of complex formation (\cite{Ferreiro:2008jb}). Few nonspecific ionic interactions mainly stabilize the final consolidated complex. The binding landscape along this route is smooth, consistent with an hydrated protein-DNA interface. In contrast, binding along the other route involves several steps. Attaining an initial, diffuse, transition state ensemble of structures with some nonnative contacts is followed by formation of extensive nonnative interactions that drive the system into a kinetic trap (\cite{pmid20375284}). These nonnative contacts then slowly rearrange into native-like interactions. Dissimilar protein-DNA interfaces are formed in the intermediate and the final complex, as anticipated from the computational studies described above. Frustration between the binding modes creates a degeneracy of interactions that drives the complex into long-lived non-native species (\cite{pmid20375284}). It has been speculated that this slowing down of the binding reaction allows for extra functionality, as this metastable species exposes different surfaces that could be linked with other binding activities, as usually seen in eukaryotic protein-nucleic acid complexes.

\section{Frustration is a fundamental aspect of biochemistry}

Some biologists take for granted that proteins just 'do-their-job' as if they were robots made by some blind adaptive designer. Physicists are often amazed by the intricacies of how this peculiar kind of chemical scum can organize itself at the temperatures the biosphere experiences. We have explored the foundations of the concept of {\it Frustration} and how it bridges these fields in the study of biomolecules. The impossibility of simultaneously satisfying constraints leads to degeneracy of states. Finite degeneracy allows for captivating emergent behaviors where apparently unrelated parts of a system respond to subtle changes in local energetics. Too much frustration and degeneracy leads to qualitatively different dynamics where searching through the many states cannot be carried out in a meaningful time and the system freezes out in a glass. On the other hand, systems that lack any frustration at all are well behaved, display sharp phase transitions, get ordered spontaneously in the right conditions and, much like crystals, are beautiful but rather dull. As in many other realms of natural processes, interesting things happen within a balance between orderly and chaotic regimes (\cite{pmid12861080}, \cite{kauffman2013beyond}). 

The principle of minimal frustration has lead to an understanding of the general aspects of protein folding. It has fruitfully lead to testable hypotheses, successful interpretations of a diverse range of experiments and to the development of useful computational tools. As with any good theory, the deviations from its expectations mark situations in which discoveries await. Frustration opens up possibilities for degeneracy, recognizing proteins as metastable systems that can compute and modify their structures according to environmental inputs. In turn, the fact that these computations are robust to small perturbations make proteins evolvable, such that the overall behavior can be finely tuned, and trained, over time (\cite{Bray:1995xi}). Present day protein molecules contain information both about historical chances and physical necessities imprinted in their structures. It is still challenging for humans to distinguish the relevant information from the inevitable noise that overwhelms our current intellectual capabilities.

Studying the aspects of frustration averaged over many proteins provides ways to infer energy functions for structure prediction and has allowed us to appreciate the apparent complexity of the protein folding code. Frustration affects laboratory folding kinetics and deviations from perfectly funneled landscapes reflect previously unrecognized functional constraints. We have reviewed here how a large part of the functions of proteins is related to subtle local frustration effects and how frustration influences the appearance of metastable states, the nature of binding processes, catalysis and allosteric transitions. ``Doing something'' is the prime characteristic of protein molecules. The teleonomic, apparent purposefulness of biological function often involves several ``activities'' coming together in a coordinated way (\cite{monod1973hasard}). At any scale it is challenging to disentangle what are the meaningful parts of living things and what are the relevant interactions between them (\cite{simondon2005individuation}). The apparent individuality of the polypeptides we can purify from present-day organisms is restricted by our skewed view and becomes highlighted when a theory based on statistics is available. The study of life beyond the scale of single proteins has to treat basic aspects of the stability and dynamics of finite states, and as such we envision that frustration is a fundamental concept that will help (often frustrated) researchers for many years.

The huge number of proteins in a cell requires that proteins must interact with a relatively high degree of specificity, otherwise most of cellular matter would be tied up in a mess of inappropriate interactions, instead of cells we would have scrambled egg cells. This functional specificity has been achieved by the coevolution of sequences that also lead to little scrambling even within each individual protein molecule. Low free energy excited state ensembles for a biomolecule can come about in several ways: harmonic motions of large length scale or special harmonic motions utilizing weakly connected links, local folding/unfolding on a funneled landscape, frustrated local interactions giving energetically degenerate structures and sequence symmetry that gives rise to alternate structures. Various combinations of such motions with low free energy costs occurring at the same time will also give low free energy excited states and kinetically dominant reaction paths. An example of such a combination mode is ``cracking'' in which the system makes excursions along a low frequency normal mode (not costly in energy), during this excursion the molecule locally unfolds (not too costly in free energy) and then refolds. If the region that unfolds is frustrated, so much the better: its unfolding will be still faster and cheaper in free energy terms. Cracking seems to be a likely mechanism for many physiologically relevant biochemical processes.

The problem of self-organization for large nanoscale objects is not stability but the kinetics of getting to a stable structure. Kinetic barriers can be avoided if sufficient entropy remains to fluidize the motion. Up to a point, robust self-assembly in a cell therefore often relies on folding upon binding - a sort of a just-in-time policy for macromolecular economy. Nevertheless, kinetic barriers for folding (even for folding upon binding) tend to scale with the size of the assembled system. The larger barriers occur for frustrated systems but also may grow with size even when the minimal frustration principle is respected, as in the capillarity picture. At some size scale the assembly rates generally become too slow and far-from-equilibrium assembly takes over. This crossover already seems to occur when the largest proteins are folded {\it in vivo}. Rubisco, a very large protein, requires chaperones in order to fold. It is likely the chaperoned catalysis of proper folding is actually a kind of kinetic proofreading, in which misfolded protein molecules are ripped apart by unfolding machines and given a chance to try again (\cite{Gulukota:1994mq}, \cite{Todd:1996wm}). The role of the ATP-induced motions of a chaperone can be thought of as increasing the effective temperature for a limited fraction of the degrees of freedom of the protein.

Cells are not formless bags of molecules. Organization persists even at the cellular scale and changes its nature at this supramolecular or ultra structural level. We are only beginning to probe the atomically weak but colloidally strong forces that act on this larger scale length. It is also clear that in living things, motion on this size scale is not just equilibrium Brownian motion. Instead, chemical energy is flowing through the system: various molecular motors carry parts around in a hurly-burly reminiscent of a large city or ant colony (\cite{pmid21876141}). This motorization probably allows the problem of large barriers and high glass transition temperatures to be overcome much as in kinetic proofreading by chaperones where ATP is used. By comparing responses to fluctuations on the colloidal scale (\cite{Lau:2003lp}), an effective temperature of the cytoskeleton has been measured to be 10,000 K. Obviously, this high temperature applies to only a minute fraction of the degrees of freedom. Studying the organization of living matter on this longer length scale will require new tools, both experimental and theoretical . We can anticipate, however, that the strong foundation of energy landscape concepts learned in the study of molecular folding and function, will be of great help in finding the required conceptual framework for this far-from-equilibrium regime.

\section{Acknowledgments}
We thank Weihua Zheng, Nick Schafer, Bobby Kim, Ha Truong and Roc\'io Espada for stimulating discussions and for providing help with figures. Supported by Grants R01 GM44557 and P01 GM071862 from the National Institute of General Medical Sciences. Additional support was also provided by the D. R. Bullard-Welch Chair at Rice University, the Agencia Nacional de Promoci\'on Cient\'ifica y Tecnol\'ogica (ANPyCT) and the Consejo Nacional de Investigaciones Cient\'ificas y T\'ecnicas de Argentina (CONICET).

\addcontentsline{toc}{section}{References}  %

\printbibliography

\end{document}